\documentclass{aa}
\usepackage[varg]{txfonts}
\usepackage[]{natbib}
\usepackage{appendix}
\usepackage{comment}
\usepackage{array}
\usepackage{supertabular}
\usepackage{multicol}
\usepackage{amsmath}

\newcommand{\kms}{km\,s$^{-1}$}
\newcommand{\vsini}{$v\sin i$}
\newcommand{\vrad}{$v_{\rm rad}$}
\newcommand{\vmic}{$v_{\rm mic}$}
\newcommand{\vmac}{$v_{\rm mac}$}
\newcommand{\msun}{M$_{\odot}$}

\newcommand{\bell}{$B_{\ell}$}

\newcommand{\teff}{$T_{\rm eff}$}
\newcommand{\logg}{$\log{g}$}

\newcommand{\rstar}{$R_{\star}$}
\newcommand{\mstar}{$M_{\star}$}
\newcommand{\blmax}{$|B_{\ell}|_{\rm max}$}

\begin{document}

\title{The magnetic fields of intermediate-mass T Tauri stars : \\ I. Magnetic detections and fundamental stellar parameters}
\titlerunning{Magnetic properties of intermediate-mass T Tauri stars}

\author{F.~Villebrun\inst{1}
  \and E.~Alecian\inst{1}
  \and G.~Hussain\inst{2,3}
  \and J.~Bouvier\inst{1}
  \and C.~P.~Folsom\inst{3}
  \and Y.~Lebreton\inst{4,5}
  \and L.~Amard\inst{10}
  \and C.~Charbonnel\inst{3,6}
  \and F.~Gallet\inst{1,6}
  \and L.~Haemmerlé\inst{6}
  \and T.~B\"ohm\inst{3}
  \and C.~Johns-Krull\inst{7}
  \and O.~Kochukhov\inst{8}
  \and S.~C.~Marsden\inst{9}
  \and J.~Morin\inst{11}
  \and P.~Petit\inst{3}
     }

\offprints{F. Villebrun, \email{florent.villebrun@univ-grenoble-alpes.fr}}

\institute{Universit\'{e} Grenoble Alpes, CNRS, IPAG, F-38000 Grenoble, France, e-mail: florent.villebrun@univ-grenoble-alpes.fr
  \and European Southern Observatory, Karl-Schwarzschild-Str. 2, D-85748 Garching bei M\"unchen, Germany
  \and IRAP, Universite de Toulouse, CNRS, CNES, UPS, F-31400 Toulouse, France
  \and LESIA, Observatoire de Paris, PSL Research University, CNRS UMR 8109, 5 Place Jules Janssen, F-92195 Meudon, France
  \and Institut de Physique de Rennes, Universite de Rennes 1, CNRS UMR 6251, F-35042 Rennes, France
  \and Department of Astronomy, University of Geneva, Chemin des Maillettes 51, 1290 Versoix, Switzerland
  \and Department of Physics \& Astronomy, Rice University, 6100 Main Street, MS-108, Houston, TX 77005, USA
  \and Department of Physics and Astronomy, Uppsala University, Box 516, 75120 Uppsala, Sweden
  \and University of Southern Queensland, Centre for Astrophysics, West Street, Toowoomba, QLD 4350, Australia
  \and University of Exeter, Department of Physics \& Astronomy, Stoker Road, Devon, Exeter, EX4 4QL, UK
  \and LUPM, Universit\'{e} de Montpellier \& CNRS, Montpellier, Cedex 05, France
     }

\date{Received xxxxxxxxx 2018}

\abstract
{The origin of the fossil magnetic fields detected in 5 to 10\% of intermediate-mass main sequence stars is still highly debated.} 
{We want to bring observational constraints to a large population of intermediate-mass pre-main sequence (PMS) stars in order to test the theory that convective-dynamo fields generated during the PMS phases of stellar evolution can occasionally relax into fossil fields on the main sequence.} 
{Using distance estimations, photometric measurements, and spectropolarimetric data from HARPSpol and ESPaDOnS of 38 intermediate-mass PMS stars, we determined fundamental stellar parameters ($T_{\rm eff}$, $L$ and $v \sin i$) and measured surface magnetic field characteristics (including detection limits for non-detections, and longitudinal fields and basic topologies for positive detections). Using PMS evolutionary models, we determined the mass, radius, and internal structure of these stars. We compared different PMS models to check that our determinations were not model-dependant. We then compared the magnetic characteristics of our sample accounting for their stellar parameters and internal structures.}
{We detect magnetic fields in about half of our sample. About 90\% of the magnetic stars have outer convective envelopes larger than $\sim$25\% of the stellar radii, and heavier than $\sim$2\% of the stellar mass. Going to higher mass, we find that the magnetic incidence in intermediate-mass stars drops very quickly, within a time-scale of the order of few times 0.1 Myr. Finally, we propose that intermediate-mass T Tauri stars with large convective envelopes, close to the fully convective limit, have complex fields and that their dipole component strengths may decrease as the sizes of their convective envelopes decrease, similar to lower-mass T Tauri stars.}
{}

\maketitle

\section{Introduction}

A very peculiar type of magnetic field can be found in 5 to 10\% of Herbig Ae/Be (HAeBe) and A/B stars: a very strong field (300 - 30kG), stable over many years and even decades, and that tends to be dominated by low-order largely dipolar large-scale fields \citep{Alecian2013,Donati2009}. The other 90 - 95\% of HAeBe and A/B stars do not host magnetic fields, or at least they are not detected. As these stars do not have a convective envelope, the dynamo process (that usually explains the presence and the characteristics of magnetic fields in cooler stars) cannot be invoked to explain either the presence or the characteristics of the fields observed in HAeBe and A/B stars. In addition, the dynamo field formed in the core of intermediate mass stars on or very near the main sequence would take too long to rise to the surface \citep[e.g.][]{Moss2001} raising the need for an alternative explanation for the presence of such magnetic fields in intermediate-mass stars. 

The underlying mechanisms causing these fields in non-convective stars is unclear and the debate centres around two theories, both of which implying that the magnetic fields come from earlier evolutionary stages of the stars, leading people to call these "fossil fields". The first theory consists of frozen-in magnetic fields originally present in the interstellar medium during the collapse of the pre-stellar cloud \citep{Moss2001}, while the other consists of relaxing the dynamo field that is created during the pre-main sequence (PMS) convective phase in stars that meet specific conditions related to their rotation \citep{Duez2010,Braithwaite2004,Emeriau2015,Gaurat2015}. In this paper, we will not focus on the first scenario, and will instead derive new observational constraints to the second scenario by measuring the rotational and magnetic properties of HAeBe and A/B stars' progenitors: the intermediate-mass T Tauri stars (IMTTS). Because IMTTS correspond to the very last convective stage of intermediate-mass stars (i.e. before they lose their convective envelopes to become fully radiative), they are the best objects to study to test the relaxation scenario.

T Tauri stars (TTS) are PMS stars with spectral types between late F and M that are gravitationally contracting while evolving towards the main sequence \citep{Herbig1962}. Some of them possess circumstellar disks, and many are still actively accreting. Intermediate-mass T Tauri stars (1.2 to 4 M$_{\odot}$) are precursors to HAeBe stars, and later to A/B stars. Studies of 16 low-mass T Tauri stars (LMTTS) have revealed that - unlike in HAeBe stars - magnetic fields are ubiquitous in these systems and of dynamo type \citep[][also see \citealt{Hill2017} for a summary of all the results to date]{Donati2008,Donati2010,Donati2011a}. \citet{Hussain2009} detected and reconstructed the large-scale surface magnetic fields in 2 IMTTS (CV Cha and CR Cha), but aside from this work, no other surface magnetic field maps of IMTTS have been published. Our knowledge of IMTTS magnetic properties is therefore very limited. 

To test the relaxation theory we need to determine the magnetic topologies and strenghts of a sample of IMTTS well distributed among the PMS phase from their fully convective to fully radiative evolutionary stages. In particular, we need to determine the ratio between poloidal and toroidal fields \citep[][Mathis private communication]{Emeriau2015}, which is only accessible with ZDI (Zeeman Doppler Imaging) magnetic mapping \citep[e.g.][]{Hussain2009}. As this technique requires continuous monitoring, evenly sampling at least one stellar rotation period, a considerable investment of telescope time is needed for each star. We therefore need to filter out the most suitable targets for mapping, i.e. those with strong large-scale magnetic fields. To this end, we selected a set of targets aiming to cover the part of the HR diagram that corresponds to IMTTS as thoroughly as possible. Future observations to obtain full ZDI maps for a small subset of the sample with the strongest magnetic fields will be reported in a subsequent paper.

Here, we report the initial step in our programme, an observational campaign of a large sample of IMTTS using ESPaDOnS (at the Canada France Hawaii Telescope, CFHT) and HARPSPol (at ESO La Silla) to acquire a few high-resolution spectropolarimetric observations per star (Section~2). We used these data to detect magnetic fields, and to determine the effective temperature of our sample consistently and as accurately as possible (Section~3). For the first time, we have therefore been able to place accurately these stars in the HR diagram, and compare their positions with PMS evolutionary tracks computed using different models, to predict their internal structure (Section~3). We discuss our results and present our conclusions in Sections~4 and 5, respectively.

\section{Observations}

We based this study on 92 spectra {\bf for} 38 IMTTS we obtained in 2012. For each star, we acquired up to 4 spectra. These spectra have been obtained with either the ESPaDOnS or HARPSpol instruments. ESPaDOnS is a high-resolution spectropolarimeter, and has been used in the polarimetric Stokes $V$ mode \citep{Donati2003}. It is located at the CFHT and has a spectral resolution $R=65000$ (its wavelength range covers between 369\,nm and 1048\,nm). HARPSpol is the polarimetric module of the HARPS spectrograph \citep{Piskunov2011} located at La Silla observatory, Chile (ESO 3.6m), which has a spectral resolution $R= 115000$ (its wavelength range covers between 378\,nm and 691\,nm). For both instruments, the circular polarisation state of the stellar light is obtained by acquiring 4 successive spectra at different phase retarder configurations. The data have been reduced using Libre-ESpRIT when obtained with ESPaDOnS, and the REDUCE {package} adapted for HARPSpol \citep[e.g.][]{Alecian2011}. Libre-ESpRIT is based on the earlier ESpRIT \citep{Donati1997} and uses the same basic reduction methods. Some spectra from HARPSpol were also reduced by Libre-ESPRIT to double-check the reduced spectra. From this data reduction, we get one Stokes $I$ and one Stokes $V$ spectrum per observation, as well as one diagnostic spectrum \citep[null-spectrum or N-spectrum, see][]{Donati1997} where the data are combined in such a way to cancel stellar polarization signal. This null-spectrum enables us to make sure the polarized light we acquire does not come from spurious origins (e.g., instrumental or observing conditions). The median signal-to-noise ratio (SNR) of our data is 170 at peak per CCD pixel for one observation (for fainter sources, one observation could consist of several successive Stokes $V$ sequences). A log of the observations can be found in Table B.1.

\longtab{
\tiny
{\renewcommand{\arraystretch}{1.2}
\begin{longtable}{ccccccc}
\caption{Log of the observations.}\\
\hline\hline
Target & Coordinates & Instrument & HJD (2450000+) & $t_{exp}$ (s) & SNR (per ccd pxl) & SNR (LSD) \\
\hline
BN Ori & 05 36 29.347 +06 50 02.17 & ESPaDOnS & 6196.01169 & 2160 & 340 & 18832 \\
- &  &  & 6272.05218 & 2160 & 290 & 15171 \\
- &  &  & 6286.08442 & 2160 & 315 & 16392 \\
- &  &  & 6293.85675 & 2160 & 325 & 17661 \\
CHX 22 & 11 12 42.689 -77 22 23.05 & HARPSpol & 6140.48253 & 4800 & 55 & 3769 \\
CO Ori & 05 27 38.335 +11 25 38.92 & ESPaDOnS & 6197.00549 & 1560 & 170 & 7064 \\
- &  &  & 6284.80619 & 1560 & 145 & 5639 \\
COUP 1350 & 05 35 26.202 -05 27 36.67 & ESPaDOnS & 6285.04068 & 4000 & 200 & 11450 \\
- &  &  & 6288.86765 & 4000 & 205 & 13249 \\
- &  &  & 6292.84744 & 4000 & 225 & 14480 \\
CPD-43 7188 & 15 38 43.068 -44 11 47.42 & HARPSpol & 6128.60140 & 7200 & 90 & 10425 \\
CQ Tau & 05 35 58.467 +24 44 54.09 & ESPaDOnS & 6284.02976 & 6000 & 445 & 19472 \\
- &  &  & 6287.06958 & 6000 & 350 & 4382 \\
- &  &  & 6289.89715 & 6000 & 425 & 19086 \\
- &  &  & 6290.84611 & 6000 & 465 & 21871 \\
GW Ori & 05 29 08.391 +11 52 12.67 & ESPaDOnS & 6195.99311 & 480 & 200 & 11701 \\
- &  &  & 6258.12749 & 960 & 145 & 6981 \\
- &  &  & 6272.07160 & 480 & 165 & 8975 \\
HBC 741 & 23 20 52.118 +74 14 07.10 & ESPaDOnS & 6100.12768 & 880 & 220 & 14587 \\
- &  &  & 6102.12400 & 880 & 205 & 13583 \\
- &  &  & 6104.11588 & 1760 & 305 & 19938 \\
HD 133938 & 15 08 38.499 -44 00 51.97 & HARPSpol & 6140.56344 & 7200 & 75 & 7072 \\
HD 135127 & 15 14 39.583 -34 45 41.31 & HARPSpol & 6131.52313 & 3152 & 110 & 10272 \\
HD 137059 & 15 25 17.010 -38 45 25.88 & HARPSpol & 6129.49163 & 5600 & 50 & 4926 \\
- &  &  & 6131.57721 & 5600 & 70 & 7878 \\
HD 143978 & 16 04 57.074 -38 57 15.71 & HARPSpol & 6129.70610 & 4000 & 115 & 12340 \\
- &  &  & 6130.55291 & 2800 & 95 & 10037 \\
- &  &  & 6130.63407 & 2800 & 60 & 6411 \\
HD 147048 & 16 21 12.193 -40 30 20.59 & HARPSpol & 6131.65409 & 6704 & 50 & 4914 \\
IRAS 22144 & 22 15 41.908 +69 38 56.69 & ESPaDOnS & 6100.10110 & 2640 & 175 & 10716 \\
- &  &  & 6101.08220 & 2640 & 145 & 9021 \\
- &  &  & 6103.02717 & 2640 & 165 & 10562 \\
IRAS 22152 & 22 16 31.105 +70 02 39.35 & ESPaDOnS & 6100.10110 & 2640 & 175 & 9174 \\
- &  &  & 6101.08220 & 2640 & 145 & 7448 \\
- &  &  & 6103.02717 & 2640 & 165 & 8230 \\
NGC 2264 108 & 06 40 51.185 +09 44 46.12 & ESPaDOnS & 6270.15205 & 3520 & 105 & 6008 \\
- &  &  & 6285.13122 & 3520 & 85 & 4670 \\
- &  &  & 6289.03847 & 3520 & 140 & 9062 \\
NGC 2264 121 & 06 40 56.507 +09 54 10.42 & ESPaDOnS & 6267.16535 & 800 & 30 & 1066 \\
- &  &  & 6269.13658 & 800 & 55 & 2843 \\
- &  &  & 6286.11356 & 800 & 60 & 3465 \\
- &  &  & 6288.03579 & 800 & 50 & 2510 \\
NGC 2264 84 & 06 40 42.183 +09 33 37.44 & ESPaDOnS & 6201.13037 & 2840 & 120 & 7375 \\
- &  &  & 6288.98198 & 5680 & 170 & 10222 \\
NGC 6530 SCB7 & 18 03 22.59 -24 22 04.7 & ESPaDOnS & 6108.00466 & 1280 & 215 & 14153 \\
- &  &  & 6110.84930 & 1280 & 215 & 14131 \\
- &  &  & 6111.85859 & 1280 & 225 & 14845 \\
NGC 6530 SCB739 & 18 04 36.103 -24 26 44.80 & HARPSpol & 6129.68867 & 8800 & 230 & 16375 \\
- &  &  & 6131.75909 & 2400 & 85 & 6062 \\
PAR 102 & 05 29 11.440 -06 08 05.40 & ESPaDOnS & 6200.03445 & 440 & 105 & 5936 \\
- &  &  & 6284.82240 & 440 & 100 & 5777 \\
- &  &  & 6288.76243 & 440 & 95 & 5373 \\
PAR 1391 & 05 34 15.196 -05 11 49.44 & HARPSpol & 6265.58358 & 5520 & 75 & 6545 \\
- &  &  & 6267.76109 & 6000 & 90 & 8617 \\
PAR 1394 & 05 34 14.163 -05 36 54.25 & ESPaDOnS & 6284.94365 & 3120 & 325 & 18347 \\
- &  &  & 6288.79597 & 3120 & 275 & 21118 \\
- &  &  & 6200.08117 & 3120 & 285 & 16998 \\
PAR 1414 & 05 34 21.377 -04 18 38.71 & HARPSpol & 6266.71310 & 12800 & 70 & 6768 \\
- &  &  & 6268.84148 & 7600 & 45 & 4302 \\
PAR 1455 & 05 34 24.961 -05 22 05.53 & HARPSpol & 6265.64831 & 4800 & 75 & 6766 \\
- &  &  & 6267.63691 & 14400 & 95 & 8797 \\
PAR 1646 & 05 34 55.208 -04 20 38.87 & HARPSpol & 6265.86174 & 3400 & 95 & 8997 \\
- &  &  & 6268.58229 & 4800 & 75 & 6568 \\
- &  &  & 6266.86804 & 3400 & 80 & 8031 \\
PAR 1736 & 05 35 05.195 -05 14 50.34 & HARPSpol & 6265.72010 & 7200 & 100 & 6536 \\
- &  &  & 6268.65637 & 7200 & 70 & 4160 \\
PAR 2441 & 05 36 51.270 -04 25 39.97 & ESPaDOnS & 6286.88901 & 2080 & 195 & 14347 \\
- &  &  & 6288.82923 & 2080 & 195 & 14536 \\
- &  &  & 6260.12791 & 2080 & 200 & 14457 \\
PX Vul & 19 26 40.250 +23 53 50.72 & ESPaDOnS & 6104.04050 & 2920 & 200 & 6679 \\
- &  &  & 6108.07777 & 2920 & 220 & 7356 \\
- &  &  & 6110.87961 & 2920 & 215 & 7322 \\
RY Ori & 05 32 09.942 -02 49 46.77 & ESPaDOnS & 6283.91923 & 5600 & 180 & 8169 \\
- &  &  & 6284.88690 & 5600 & 205 & 9706 \\
- &  &  & 6290.76770 & 5600 & 230 & 11095 \\
V1000 Sco & 16 11 08.908 -19 04 46.86 & ESPaDOnS & 5969.16094 & 2400 & 200 & 13827 \\
- &  &  & 6103.75536 & 2400 & 205 & 11784 \\
V1001 Sco &  16 11 59.272 -19 06 53.36 & ESPaDOnS & 5968.12625 & 1700 & 160 & 10412 \\
- &  &  & 5972.11674 & 1200 & 115 & 6970 \\
- &  &  & 6103.78273 & 1700 & 170 & 9340 \\
V1002 Sco & 16 12 40.516 -18 59 28.27 & ESPaDOnS & 5969.10633 & 6000 & 420 & 31570 \\
- &  &  & 6109.77816 & 6000 & 525 & 35521 \\
V1044 Ori & 05 34 16.462 -05 36 45.57 & ESPaDOnS & 6202.00591 & 760 & 100 & 6839 \\
- &  &  & 6271.11957 & 760 & 75 & 4660 \\
- &  &  & 6290.93247 & 760 & 90 & 6085 \\
V1149 Sco & 15 58 36.913 -22 57 15.22 & ESPaDOnS & 5967.09378 & 1000 & 180 & 13096 \\
- &  &  & 6099.78064 & 1000 & 225 & 14354 \\
- &  &  & 6101.78171 & 1000 & 240 & 15598 \\
V1152 Sco & 16 01 25.630 -22 40 40.29 & ESPaDOnS & 5968.16100 & 3040 & 190 & 15537 \\
- &  &  & 6102.75795 & 3040 & 210 & 15119 \\
V1156 Sco & 16 04 47.752 -19 30 22.92 & ESPaDOnS & 5968.09636 & 2800 & 260 & 21648 \\
- &  &  & 5972.09027 & 2800 & 200 & 15268 \\
\hline
\end{longtable}}
}

The normalisation provided by the data reduction pipelines being insufficient for our work, we used a polynomial renormalisation routine \citep[][]{Folsom2008,Folsom2012}. This routine determines continuum points in the stellar spectrum order by order, and then fits a polynomial function based on these continuum points. The original spectrum is then divided by the fitted polynomial function, making sure that the spectrum is properly normalised.

In order to increase the SNR of our data, we built the mean Least-Squares Deconvolved (LSD) profiles of our spectra as described in \citet{Donati1997}. For all of the stars, the LSD profiles have been normalized using a mean wavelength of 500 nm and a mean Land\'e factor of 1.2 \citep{Kochukhov2010}. For each star, a tailored list of atomic lines parameters, called a line-mask, is compiled from a synthetic spectrum created with the effective temperature, and \logg\ (determined in Section~3). We used solar abundances which are appropriate for our sample (see Section~3). We used the Vienna Atomic Line Database \citep[VALD3 version,][]{Ryabchikova2015} to build the line-masks, and excluded from these masks spectral regions affected by Balmer lines, emission lines, telluric lines, and regions with a poor SNR.

\section{Results}

\subsection{Stellar properties}

We want to determine stellar properties of our sample to better constrain their position in the HR diagram and better understand their magnetic behaviour in terms of their effective temperature and internal structure.

\subsubsection{Effective temperatures and rotational velocities}

To derive the effective temperature, $T_{\rm eff}$, and projected rotational velocities, $v \sin i$, of these stars, we used the spectral fitting method: for each observed spectrum, we created an associated synthetic spectrum using the ZEEMAN code \citep{Landstreet1988,Wade2001,Folsom2012} that, according to a $\chi^{2}$-fitting procedure, best fits the observed spectrum. We used a grid of stellar atmospheric models from the MARCS code \citep{Gustafsson2008} using global solar abundances from \citet{Asplund2009}. Model atmospheres were interpolated between grid points, using linear interpolation of the log of the model quantities, to obtain exact \teff\ and \logg\ combinations. Atomic data were extracted from the VALD3 database using "extract stellar" requests with parameters corresponding to the range of stellar parameters of interest. 

We first needed initial estimates of the star's fundamental parameters (\vrad\ , $T_{\rm eff}$, $v \sin i$ and $v_{\rm mac}$) which we determined by eye using the IDL visualisation script BINMAG2\footnote{see the webpage http://www.astro.uu.se/$\sim$oleg/binmag.html} \citep[see][]{Alecian2016}. These first estimates of fundamental stellar parameters are then used by the LMA semi-automatic routine \citep{Folsom2012} as a starting point for a $\chi^{2}$ minimisation using the Levenberg-Marquardt technique. Synthetic spectra are created by the ZEEMAN spectrum synthesis code and compared to the observed spectrum until all free parameters converge to their optimal solution. The free parameters in this fitting procedure are: \teff, \vsini, the microturbulence velocity \vmic, and the radial velocity \vrad. We fixed \logg\ to 4.0 as our data were of variable quality -- the ability to determine \logg\ depends on both \vsini\ and the SNR of the spectrum --  and so this quantity could not be consistently determined across the entire sample. \logg\ = 4.0 is a reasonable value for the part of the Hertzsprung-Russell diagram (HRD) covered by our sample. We used global solar metallicity from \citet{Asplund2009} and found that they reproduce our data well: there was therefore no need to fit the abundances as well. We also fixed the macro-turbulent velocity, \vmac, to 2.0\,\kms\ as it reproduces well our data for low \vsini\ stars, and could not be constrained for high \vsini\ stars (as the rotation dominates line broadening).

We took into account the broadest possible spectral range, but had to discard several wavelength windows that were strongly affected by telluric absorption lines, or night sky or YSO emission lines. We also had to discard the blue part of the spectra (i.e. $\lambda$ < 400\,nm) because noise is high and blending with Balmer lines becomes severe. As a result of this filtering, we fitted 12 wavelength windows for ESPaDOnS spectra ($\sim$ 90\,nm per window, totaling 1085\,nm) and 10 wavelength windows for HARPSpol spectra (totaling 885\,nm). For each window, we produced an estimate of $T_{\rm eff}$ and $v \sin i$. We then averaged all the values of $T_{\rm eff}$ and $v \sin i$ for all the windows of a single spectrum in order to obtain a preliminary estimate of $T_{\rm eff}$ and $v \sin i$ for the entire spectrum. To discard potential bad fits, we excluded the windows that deviated from the mean value by 1$\sigma$, excluded them, and re-calculated a new average. We did this operation only once: this 1$\sigma$-clipping resulted in the exclusion of 1 or 2 windows in most cases, and up to 4 or 5 windows at the most (for specra that were noisier than the rest and / or for fast rotating stars). Following this procedure, we derived $T_{\rm eff}$ and $v \sin i$ with standard deviations of the mean of 100-150\,K and 1-2\,\kms\ respectively (450\,K and 20\,\kms\ for the most challenging case, i.e. for spectra that were noisier than the rest and / or for fast rotating stars). All these results are compiled in Table B.2. 

\longtab{
{\renewcommand{\arraystretch}{1.6}
\begin{longtable}{ccccccccc}
\caption{Table compiling stellar properties of our sample: effective temperature, projected rotational velocity, total radius of the star, mass, radius of the radiative core, convective turnover time at Hp/2 (at half a pressure scale height above the base of the convection zone), and accretion type. Error bars on \teff\ and \vsini\ are detailled in Section 3.1.1., and error bars on the other parameters come from interpolations of the CESAM grids.}\\
\hline\hline
Identifier & $T_{\rm eff}$ & $v \sin i$ & Radius & Mass & $R_{\rm rad. core}$ & log($\tau_{c}$) & Type & References for accretion type \\
 & (K) & (km/s) & (R$_{\odot}$) & (M$_{\odot}$) & ($R_{\odot}$) & (s) &  &  \\
\hline
BN Ori & 7020 $\pm$ 450 & 214 $\pm$ 24 & 3.250$^{+0.476}_{-0.416}$ & 2.149$^{+0.130}_{-0.112}$ & 3.221$^{+0.427}_{-0.407}$ & 4.304$^{+0.673}_{-0.573}$ &  &  \\
CHX 22 & 5260 $\pm$ 200 & 9.87 $\pm$ 0.14 & 2.204$^{+0.300}_{-0.264}$ & 1.772$^{+0.168}_{-0.175}$ & 1.378$^{+0.141}_{-0.121}$ & 6.690$^{+0.168}_{-0.175}$ & wTTS & \citet{Spangler2001} \\
CO Ori & 6290 $\pm$ 220 & 69 $\pm$ 6 & 6.174$^{+0.543}_{-0.499}$ & 3.225$^{+0.175}_{-0.158}$ & 5.958$^{+0.355}_{-0.391}$ & 4.434$^{+0.760}_{-0.409}$ & cTTS & \citet{Astraatmadja2016} \\
COUP 1350 & 5590 $\pm$ 130 & 61.8 $\pm$ 1.0 & 2.934$^{+0.201}_{-0.188}$ & 2.099$^{+0.144}_{-0.143}$ & 2.099$^{+0.112}_{-0.107}$ & 6.415$^{+0.152}_{-0.165}$ &  &  \\
CPD -43 7188 & 5360 $\pm$ 120 & 25.4 $\pm$ 0.5 & 1.320$^{+0.107}_{-0.099}$ & 1.181$^{+0.082}_{-0.074}$ & 0.883$^{+0.056}_{-0.055}$ & 6.467$^{+0.080}_{-0.083}$ &  &  \\
CQ Tau & 6800 $\pm$ 290 & 94 $\pm$ 6 & 2.491$^{+0.278}_{-0.250}$ & 1.747$^{+0.101}_{-0.074}$ & 2.451$^{+0.191}_{-0.229}$ & 3.697$^{+0.965}_{-0.180}$ &  &  \\
CR Cha & 4800 $\pm$ 230 & 34.5 & 3.113$^{+0.366}_{-0.327}$ & 1.657$^{+0.341}_{-0.413}$ & 0.953$^{+0.525}_{-0.953}$ & 7.134$^{+0.002}_{-4.133}$ & cTTS & \citet{Hussain2009} \\
GW Ori & 5700 $\pm$ 150 & 46.0 $\pm$ 1.6 & 7.651$^{+0.568}_{-0.529}$ & 3.751$^{+0.189}_{-0.186}$ & 6.316$^{+0.350}_{-0.331}$ & 5.991$^{+0.247}_{-0.294}$ & cTTS & \citet{Astraatmadja2016} \\
HBC 741 & 5470 $\pm$ 70 & 40.0 $\pm$ 0.8 & 2.658$^{+0.197}_{-0.183}$ & 2.026$^{+0.117}_{-0.105}$ & 1.718$^{+0.122}_{-0.108}$ & 6.651$^{+0.068}_{-0.069}$ & cTTS  & \citet{Ducourant2005} \\
HD 133938 & 5290 $\pm$ 110 & 20.40 $\pm$ 0.23 & 1.714$^{+0.154}_{-0.141}$ & 1.459$^{+0.108}_{-0.099}$ & 1.106$^{+0.079}_{-0.073}$ & 6.592$^{+0.085}_{-0.083}$ & wTTS & \citet{Xing2010} \\
HD 135127 & 6740 $\pm$ 120 & 108 $\pm$ 5 & 1.401$^{+0.113}_{-0.085}$ & 1.403$^{+0.023}_{-0.095}$ & 1.316$^{+0.119}_{-0.095}$ & 4.656$^{+0.023}_{-0.095}$ & wTTS & \citet{Krautter1997} \\
HD 137059  & 5840 $\pm$ 70 & 27.9 $\pm$ 1.7 & 2.786$^{+0.243}_{-0.223}$ & 1.913$^{+0.125}_{-0.116}$ & 2.171$^{+0.201}_{-0.182}$ & 6.077$^{+0.099}_{-0.109}$ & wTTS & \citet{Xing2010} \\
HD 143978  & 6010 $\pm$ 120 & 36.1 $\pm$ 0.8 & 1.171$^{+0.077}_{-0.072}$ & 1.138$^{+0.024}_{-0.107}$ & 0.929$^{+0.049}_{-0.123}$ & 5.946$^{+0.122}_{-0.112}$ & wTTS & \citet{Xing2010} \\
HD 147048  & 5410 $\pm$ 120 & 21.54 $\pm$ 0.41 & 1.433$^{+0.231}_{-0.199}$ & 1.256$^{+0.164}_{-0.141}$ & 0.962$^{+0.141}_{-0.120}$ & 6.461$^{+0.096}_{-0.092}$ &  &\\
IRAS 22144 + 6923 & 5720 $\pm$ 90 & 34.9 $\pm$ 1.0 & 2.541$^{+0.316}_{-0.281}$ & 1.842$^{+0.173}_{-0.159}$ & 1.883$^{+0.248}_{-0.217}$ & 6.261$^{+0.111}_{-0.119}$ &  &\\
IRAS 22152 + 6947 & 6620 $\pm$ 130 & 43.7 $\pm$ 2.4 & 1.744$^{+0.231}_{-0.204}$ & 1.423$^{+0.071}_{-0.071}$ & 1.625$^{+0.216}_{-0.216}$ & 4.749$^{+0.306}_{-0.337}$ &  &\\
NGC 2264 108 & 6130 $\pm$ 200 & 52.6 $\pm$ 1.8 & 3.387$^{+0.675}_{-0.563}$ & 2.108$^{+0.296}_{-0.253}$ & 2.939$^{+0.607}_{-0.500}$ & 5.509$^{+0.348}_{-0.467}$ & cTTS & \citet{Sousa2016} \\
NGC 2264 121 & 5450 $\pm$ 260 & 144 $\pm$ 12 & 3.922$^{+0.841}_{-0.693}$ & 2.628$^{+0.373}_{-0.405}$ & 2.697$^{+0.559}_{-0.454}$ & 6.581$^{+0.321}_{-0.369}$ &  & \\
NGC 2264 84 & 6160 $\pm$ 140 & 36.3 $\pm$ 0.7 & 3.407$^{+0.652}_{-0.548}$ & 2.113$^{+0.281}_{-0.243}$ & 2.986$^{+0.620}_{-0.510}$ & 5.433$^{+0.287}_{-0.363}$ & cTTS & \citet{Sousa2016} \\
NGC 6530 SCB 7 & 5490 $\pm$ 70 & 5.00 $\pm$ 0.22 &  &  &  &  &  &\\
NGC 6530 SCB 739 & 6400 $\pm$ 100 & 4.28 $\pm$ 0.42 &  &  &  &  &  &\\
Par 102 & 6160 $\pm$ 100 & 49.3 $\pm$ 1.5 & 2.894$^{+0.334}_{-0.300}$ & 1.889$^{+0.154}_{-0.136}$ & 2.495$^{+0.304}_{-0.269}$ & 5.518$^{+0.185}_{-0.221}$ &  &\\
Par 1391 & 6350 $\pm$ 160 & 14.99 $\pm$ 0.47 & 2.293$^{+0.248}_{-0.223}$ & 1.607$^{+0.115}_{-0.095}$ & 2.045$^{+0.210}_{-0.191}$ & 5.215$^{+0.318}_{-0.391}$ &  &\\
Par 1394 & 6260 $\pm$ 140 & 62.9 $\pm$ 1.7 & 3.236$^{+0.258}_{-0.239}$ & 2.032$^{+0.119}_{-0.106}$ & 2.906$^{+0.213}_{-0.199}$ & 5.204$^{+0.312}_{-0.384}$ &  &\\
Par 1414 & 5920 $\pm$ 120 & 33.9 $\pm$ 1.0 & 1.830$^{+0.233}_{-0.207}$ & 1.416$^{+0.131}_{-0.113}$ & 1.415$^{+0.181}_{-0.158}$ & 6.021$^{+0.147}_{-0.163}$ &  &\\
Par 1455 & 5890 $\pm$ 120 & 23.5 $\pm$ 0.6 & 2.781$^{+0.215}_{-0.199}$ & 1.895$^{+0.128}_{-0.113}$ & 2.201$^{+0.161}_{-0.150}$ & 6.006$^{+0.165}_{-0.180}$ &  &\\
Par 1646 & 6310 $\pm$ 80 & 15.67 $\pm$ 0.39 & 3.174$^{+ 0.227}_{-0.211}$ & 2.001$^{+0.099}_{-0.092}$ & 2.886$^{+0.210}_{-0.198}$ & 5.096$^{+0.190}_{-0.236}$ &  &\\
Par 1736 & 6180 $\pm$ 110 & 55.3 $\pm$ 2.6 & 6.861$^{+0.416}_{-0.392}$ & 3.459$^{+0.135}_{-0.134}$ & 6.504$^{+0.363}_{-0.333}$ & 4.775$^{+0.375}_{-0.423}$ &  &\\
Par 2441 & 5780 $\pm$ 110 & 13.40 $\pm$ 0.39 & 2.780$^{+0.172}_{-0.162}$ & 1.936$^{+0.116}_{-0.106}$ & 2.119$^{+0.114}_{-0.108}$ & 6.171$^{+0.140}_{-0.155}$ &  & \\
PX Vul & 7860 $\pm$ 410 & 81 $\pm$ 5 & 3.556$^{+0.632}_{-0.536}$ & 2.494$^{+0.257}_{-0.223}$ & 3.515$^{+0.624}_{-0.529}$ & 5.330$^{+0.127}_{-0.160}$ &  &\\
RY Ori & 6120 $\pm$ 110 & 49.7 $\pm$ 1.7 & 2.562$^{+0.368}_{-0.322}$ & 1.746$^{+0.172}_{-0.157}$ & 2.160$^{+0.329}_{-0.283}$ & 5.640$^{+0.189}_{-0.223}$ & cTTS & \citet{Richards2012} \\
V 1000 Sco & 4830 $\pm$ 200 & 25.8 $\pm$ 1.9 & 2.369$^{+0.343}_{-0.299}$ & 1.606$^{+0.168}_{-0.303}$ & 1.075$^{+0.193}_{-0.566}$ & 7.023$^{+0.099}_{-0.188}$ & wTTS & \citet{Luhman2012} \\
V 1001 Sco & 5180 $\pm$ 130 & 24.6 $\pm$ 0.8 & 1.782$^{+0.223}_{-0.198}$ & 1.512$^{+0.136}_{-0.135}$ & 1.115$^{+0.108}_{-0.099}$ & 6.671$^{+0.106}_{-0.103}$ & wTTS & \citet{Luhman2012} \\
V 1002 Sco & 4950 $\pm$ 220 & 76.7 $\pm$ 3.3 & 2.260$^{+0.244}_{-0.220}$ & 1.697$^{+0.073}_{-0.221}$ & 1.195$^{+0.100}_{-0.311}$ & 6.917$^{+0.186}_{-0.185}$ &  & \\
V 1044 Ori & 5500 $\pm$ 140 & 26.7 $\pm$ 0.8 & 2.804$^{+0.202}_{-0.188}$ & 2.072$^{+0.138}_{-0.149}$ & 1.927$^{+0.101}_{-0.094}$ & 6.520$^{+0.149}_{-0.166}$ & cTTS & \citet{DaRio2009} \\
V 1149 Sco & 5740 $\pm$ 90 & 12.06 $\pm$ 0.23 & 2.172$^{+0.185}_{-0.170}$ & 1.641$^{+0.111}_{-0.102}$ & 1.602$^{+0.136}_{-0.124}$ & 6.241$^{+0.102}_{-0.107}$ & cTTS & \citet{Luhman2012} \\
V 1152 Sco & 4800 $\pm$ 120 & 13.1 $\pm$ 0.7 & 1.373$^{+0.119}_{-0.110}$ & 1.206$^{+0.058}_{-0.066}$ & 0.816$^{+0.038}_{-0.039}$ & 6.756$^{+0.096}_{-0.088}$ & wTTS & \citet{Prugniel2011} \\
V 1156 Sco & 4820 $\pm$ 180 & 48.3 $\pm$ 3.2 & 1.816$^{+0.176}_{-0.160}$ & 1.438$^{+0.057}_{-0.107}$ & 0.972$^{+0.062}_{-0.136}$ & 6.893$^{+0.152}_{-0.139}$ &  & \\
\hline
\end{longtable}}
}

Some stars of our sample may still be strongly accreting. We therefore checked for the presence of veiling by comparing the measured $T_{\rm eff}$ at high and low wavelengths (i.e. in the red and blue parts of the spectra respectively). Because accretion shocks emit more radiations (relative to photospheres) in the blue / UV part of the spectra, the calculated $T_{\rm eff}$ may be over-estimated at shorter wavelengths: the lines are shallower due to an extra continuum contribution from accretion hot spots at or near the stellar surface. Except for 3 stars, no significant veiling was found in our sample. For the three affected by veiling (V1000~Sco, V1152~Sco and V1156~Sco), the \teff\ was consistently higher in the blue by 400-500\,K. In these cases, it is more cautious to ignore the calculated $T_{\rm eff}$ in the blue part, and only consider the calculated $T_{\rm eff}$ in the red part (as it is less affected by veiling). Thanks to the 1$\sigma$-clipping, these abnormally high $T_{\rm eff}$ estimates have been excluded, and the final temperatures of these 3 stars are largely based on the temperatures found in their red spectral windows (>580\,nm). We thus are confident that veiling did not significantly skew our $T_{\rm eff}$ determination.

\subsubsection{Luminosities}

To position our sample of IMTTS in the HR diagram, we needed to compute their luminosities, which first requires the determination of the extinction. We based our determination of A$_j$ on the ($V-J$) color, ($B-V$) being more affected by accretion and circumstellar extinction. We looked for the most consistent photometric measurements available in the literature. In most cases, we used the $J$ magnitudes from the 2MASS survey \citep{Cutri2003} and $V$ magnitudes from \citet{Kharchenko2001} or the NOMAD catalogue \citep{Zacharias2004}. Using our determination of effective temperature, we derived the associated empirical ($V-J$)$_{\rm o}$ of 5-30 Myr intermediate-mass stars from \citet{Pecaut2013} to compute the color excesses E($V-J$) and extinctions A$_{J}$. The total to selective extinction R$_{J}$=0.437 has been determined from the color excesses and extinctions found in \citet{Casagrande2010} following the relationship:
\begin{equation}
\frac{\textrm{R}_{J}}{\textrm{R}_{V}} \times \frac{\textrm{E}(V-J)}{\textrm{E}(B-V)} = \frac{\textrm{A}_{J}}{\textrm{A}_{V}}
\end{equation}

We used the distance from Tycho-Gaia Astrometric Solution \citep[TGAS,][]{GAIA2016} whenever available in order to determine the absolute magnitudes M$_{J}$. However, for half of the stars, TGAS distances were not yet available: we thus used the best distance estimate of their associated star-forming region or cluster found in the literature (see Table B.3 for distance estimates). We then used bolometric corrections (BC)$_{J}$ from \citet{Pecaut2013} to get their bolometric magnitudes, and finally determined the bolometric luminosities using a solar bolometric absolute magnitude of 4.755, also from \citet{Pecaut2013}. All the informations about the magnitudes and luminosities of our sample are compiled in Table B.4.
 
\longtab{
{\renewcommand{\arraystretch}{1.3}
\begin{longtable}{ccccc}
\caption{Table compiling the 38 IMTTS identifiers, the associations they belong to, and the distances estimations we found in the literature (sources are mentionned : Gaia-TGAS paralax whenever possible, otherwise the associated star forming region ro cluster distance).}\\
\hline\hline
Main identifier & 2MASS identifier & Association & Distance (pc) & Source \\
\hline
BN Ori & J 05362935 + 0650020 & RSF1 (OriB) & 388 $\pm$ 5 & \citet{Kounkel2016} \\
CHX 22 & J 11124268 - 7722230 & Cha1 & 160 $\pm$ 15 & \citet{Whittet1997} \\
CO Ori & J 05273833 + 1125389 & IC1 & 388 $\pm$ 5 & \citet{Kounkel2016} \\
COUP 1350 & J 05352620 - 0527366 & Ori Neb Clus & 388 $\pm$ 5 & \citet{Kounkel2016} \\
CPD -43 7188 & J 15384306 - 4411474 & Lupus & 127 $\pm$ 6 & GAIA\\
CQ Tau & J 05355845 + 2444542 & Tau & 160 $\pm$ 7 & GAIA\\
CR Cha & J 10590699 - 7701404 & Cha & 188 $\pm$ 8 & GAIA\\
GW Ori & J 05290838 + 1152126 & Ori & 388 $\pm$ 5 & \citet{Kounkel2016} \\
HBC 741 & J 23205208 + 7414071 & Cep & 188 $\pm$ 9 & GAIA\\
HD 133938 & J 15083849 - 4400519 & Lup & 166 $\pm$ 10 & GAIA\\
HD 135127 & J 15143959 - 3445412 & Lup & 132 $\pm$ 7 & GAIA \\
HD 137059 & J 15251700 - 3845261 & Lup & 150 $\pm$ 10 & \citet{Crawford2000} \\
HD 143978 & J 16045707 - 3857157 & Lup & 96 $\pm$ 2 & GAIA \\
HD 147048 & J 16211219 - 4030204 & Lup & 147 $\pm$ 20 & GAIA\\
IRAS 22144 + 6923 & J 22154189 + 6938566 & Cep (L1219) & 392 $\pm$ 41 & GAIA\\
IRAS 22152 + 6947 & J 22163111 + 7002393 & Cep (L1219) & 376 $\pm$ 41 & GAIA \\
NGC 2264 108 & J 06405118 + 0944461 & NGC 2264 & 913 $\pm$ 150 & \citet{Baxter2009} \\
NGC 2264 121 & J 06405650 + 0954104 & NGC 2264 & 913 $\pm$ 150 & \citet{Baxter2009} \\
NGC 2264 84 & J 06404218 + 0933374  & NGC 2264 & 913 $\pm$ 150 & \citet{Baxter2009} \\
NGC 6530 SCB 7 & J 18032258 - 2422046 & NGC 6530 & 1800 $\pm$ 100 & \citet{Sung2000} \\
NGC 6530 SCB 739 & J 18043607 - 2426447 & NGC 6530 & 1800 $\pm$ 100 & \citet{Sung2000} \\
Par 102 & J 05291144 - 0608054 & Ori & 342 $\pm$ 32 & GAIA \\
Par 1391 & J 05341519 - 0511494 & Ori & 388 $\pm$ 5 & \citet{Kounkel2016} \\
Par 1394 & J 05341416 - 0536542 & Ori Neb Clus & 388 $\pm$ 5 & \citet{Kounkel2016} \\
Par 1414 & J 05342137 - 0418387 & NGC 1981 & 380 $\pm$ 17 & \citet{Maia2010} \\
Par 1455 & J 05342495 - 0522055 & Ori & 388 $\pm$ 5 & GAIA \\
Par 1646 & J 05345520 - 0420389 & NGC 1981 & 388 $\pm$ 5 & \citet{Kounkel2016} \\
Par 1736 & J 05350519 - 0514503 & Ori & 388 $\pm$ 5 & \citet{Kounkel2016} \\
Par 2441 &  & Ori OB1c & 388 $\pm$ 5 & \citet{Kounkel2016} \\
PX Vul & J 19264025 + 2353508 & R Vul R2 & 420 $\pm$ 50 & \citet{Manoj2006} \\
RY Ori & J 05320993 - 0249467 & Ori & 362 $\pm$ 44 & GAIA \\
V 1000 Sco & J 16110890 - 1904468 & Upper Sco & 145 $\pm$ 14 & \citet{DeZeeuw1999} \\
V 1001 Sco &  & Sco-Cen & 145 $\pm$ 14 & \citet{DeZeeuw1999} \\
V 1002 Sco & J 16124051 - 1859282 & Upper Sco & 131 $\pm$ 4 & GAIA \\
V 1044 Ori & J 05341646 - 0536455 & Ori & 388 $\pm$ 5 & \citet{Kounkel2016} \\
V 1149 Sco & J 15583692 - 2257153 & Sco & 166 $\pm$ 10 & GAIA \\
V 1152 Sco & J 16012563 - 2240403 & Upper Sco & 141 $\pm$ 7 & GAIA \\
V 1156 Sco & J 16044776 - 1930230 & Upper Sco & 140 $\pm$ 5 & GAIA \\
\hline
\end{longtable}}
}

\longtab{
{\renewcommand{\arraystretch}{1.2}
\begin{longtable}{cccccccc}
\caption{Table compiling photometric measurements found in the literature : magnitudes V B and J, bolometric corrections in J band, theoretical (V-J)$_{o}$ and A$_{J}$ from \citet{Pecaut2013}, and the corresponding log(L/L$_{\odot}$) with their associated uncertainties.}\\
\hline\hline
Identifier & V & B & J & (BC)$_{J}$ & (V-J)$_{o}$ & A$_{V}$ & log(L/L$_{\odot}$) \\
\hline
BN Ori & 8.80 & 10.04 & 8.604 & 0.67 & 0.68 & 0.000 & 1.364 $\pm$ 0.042 \\
CHX 22 & 11.103 & 12.02 & 8.645 & 1.23 & 1.48 & 1.377 & 0.525 $\pm$ 0.091\\
CO Ori & 10.30 & 12.33 & 7.983 & 0.95 & 1.00 & 1.858 & 1.731 $\pm$ 0.042\\
COUP 1350 & 11.78 & 12.46 & 9.703 & 1.13 & 1.28 & 1.123 & 0.880 $\pm$ 0.042\\
CPD -43 7188 & 10.28 & 11.03 & 8.805 & 1.21 & 1.40 & 0.106 & 0.111 $\pm$ 0.057\\
CQ Tau & 10.0 & 10.78 & 7.926 & 0.73 & 0.73 & 1.894 & 1.077 $\pm$ 0.055\\
CR Cha & 11.0 & 12.15 & 8.462 & 1.34 & 1.72 & 1.152 & 0.667 $\pm$ 0.054\\
GW Ori & 10.10 & 10.83 & 7.698 & 1.12 & 1.26 & 1.610 & 1.746 $\pm$ 0.042\\
HBC 741 & 10.22 & 11.17 & 8.308 & 1.21 & 1.42 & 0.694 & 0.723 $\pm$ 0.058\\
HD 133938 & 10.54 & 11.17 & 8.950 & 1.19 & 1.39 & 0.281 & 0.315 $\pm$ 0.066\\
HD 135127 & 9.15 & 9.64 & 8.235 & 0.76 & 0.77 & 0.203 & 0.564 $\pm$ 0.061\\
HD 137059 & 8.75 & 9.42 & 7.410 & 1.03 & 1.12 & 0.310 & 0.911 $\pm$ 0.070\\
HD 143978 & 9.20 & 9.70 & 8.143 & 0.99 & 1.06 & 0.000 & 0.207 $\pm$ 0.044\\
HD 147048 & 10.482 & 11.45 & 8.964 & 1.18 & 1.37 & 0.210 & 0.199 $\pm$ 0.125\\
IRAS 22144 + 6923 & 11.60 & 12.40 & 9.875 & 1.08 & 1.19 & 0.755 & 0.794 $\pm$ 0.099\\
IRAS 22152 + 6947 & 11.39 & 11.77 & 10.196 & 0.79 & 0.80 & 0.555 & 0.721 $\pm$ 0.103\\
NGC 2264 108 & 11.88 & 12.42 & 10.744 & 0.95 & 1.00 & 0.190 & 1.163 $\pm$ 0.148 \\
NGC 2264 121 & 12.47 & 13.33 & 10.783 & 1.18 & 1.365 & 0.455 & 1.088 $\pm$ 0.148 \\
NGC 2264 84 & 12.01 & 12.57 & 10.759 & 0.95 & 1.00 & 0.355 & 1.177 $\pm$ 0.148\\
NGC 6530 SCB 7 & 10.583 & 11.578 & 8.721 & 1.16 & 1.33 & 0.748 & 2.547 $\pm$ 0.063\\
NGC 6530 SCB 739 & 8.644 & 9.234 & 7.428 & 0.85 & 0.87 & 0.487 & 3.156 $\pm$ 0.063\\
Par 102 & 10.39 & 11.02 & 9.030 & 0.95 & 1.00 & 0.506 & 1.035 $\pm$ 0.091\\
Par 1391 & 10.698 & 11.2 & 9.65 & 0.88 & 0.905 & 0.200 & 0.887 $\pm$ 0.078\\
Par 1394 & 10.219 & 10.697 & 8.993 & 0.91 & 0.95 & 0.390 & 1.161 $\pm$ 0.056\\
Par 1414 & 11.48 & 12.12 & 10.274 & 1.00 & 1.07 & 0.190 & 0.570 $\pm$ 0.098\\
Par 1455 & 10.88 & 11.8 & 9.588 & 1.03 & 1.12 & 0.242 & 0.923 $\pm$ 0.056\\
Par 1646 & 9.72 & 10.19 & 8.853 & 0.89 & 0.93 & 0.000 & 1.158 $\pm$ 0.056\\
Par 1736 & 11.246 & 12.141 & 8.166 & 0.95 & 1.00 & 2.932 & 1.791 $\pm$ 0.042\\
Par 2441 & 10.741 & 11.430 & 9.432 & 1.08 & 1.19 & 0.168 & 0.890 $\pm$ 0.042\\
PX Vul & 11.83 & 12.57 & 9.324 & 0.35 & 0.41 & 2.955 & 1.639 $\pm$ 0.111\\
RY Ori & 10.80 & 12.76 & 9.444 & 0.95 & 1.00 & 0.503 & 0.918 $\pm$ 0.113\\
V 1000 Sco & 12.050 & 12.99 & 8.761 & 1.36 & 1.75 & 2.171 & 0.439 $\pm$ 0.093\\
V 1001 Sco & 11.59 & 12.82 & 8.981 & 1.26 & 1.52 & 0.476 & 0.313 $\pm$ 0.093\\
V 1002 Sco & 10.829 & 12.15 & 8.313 & 1.30 & 1.63 & 1.535 & 0.440 $\pm$ 0.048\\
V 1044 Ori & 11.51 & 12.123 & 9.701 & 1.16 & 1.33 & 0.674 & 0.813 $\pm$ 0.042\\
V 1149 Sco & 10.14 & 10.92 & 8.358 & 1.08 & 1.19 & 0.835 & 0.664 $\pm$ 0.066\\
V 1152 Sco & 11.4 & 12.6 & 9.324 & 1.38 & 1.83 & 0.348 & -0.044 $\pm$ 0.059\\
V 1156 Sco & 11.254 & 12.43 & 8.875 & 1.36 & 1.75 & 0.887 & 0.204 $\pm$ 0.051\\
\hline
\end{longtable}}
}

When trying to determine the bolometric luminosities of NGC 6530 SCB 7 and NGC 6530 SCB 739, we found abnormally high values (log(L/L$_{\odot})=2.55$ and 3.16, respectively). Because of these very high luminosities, these 2 stars are far above the birthline, which is impossible according to standard PMS theory. These errors might have the following causes: the first one being they are not pre-MS but post-MS stars, the second one being bad photometric measurements (for both of them, the $B$-$V$-$J$ magnitude measurements are reported with "C" or even "D" flags on SIMBAD, meaning we should be cautious when using them), and the third one being bad estimates of their distances. We favour the third scenario: there are no GAIA parallaxes for these 2 stars, and thus we adopted the distance of NGC 6530 from \citet{Sung2000}. As these 2 stars are poorly studied, we suspect NGC 6530 SCB 7 and NGC 6530 SCB 739 of actually being in front of NGC 6530, explaining why the bolometric luminosity of these 2 stars is highly overestimated \footnote{In the meanwhile, GAIA DR2 distances \citep{Gaia2018,Brown2018} have been released for the two stars. They confirm these stars are very much in the foreground.}. Despite their luminosities being undetermined, it does not affect our determination of their effective temperatures, rotational velocities, and magnetic fields reported in this paper.

 \begin{figure*}[ht!]
 \centering
 \includegraphics[angle=180,width=0.800\textwidth,clip,bb=18 0 680 590]{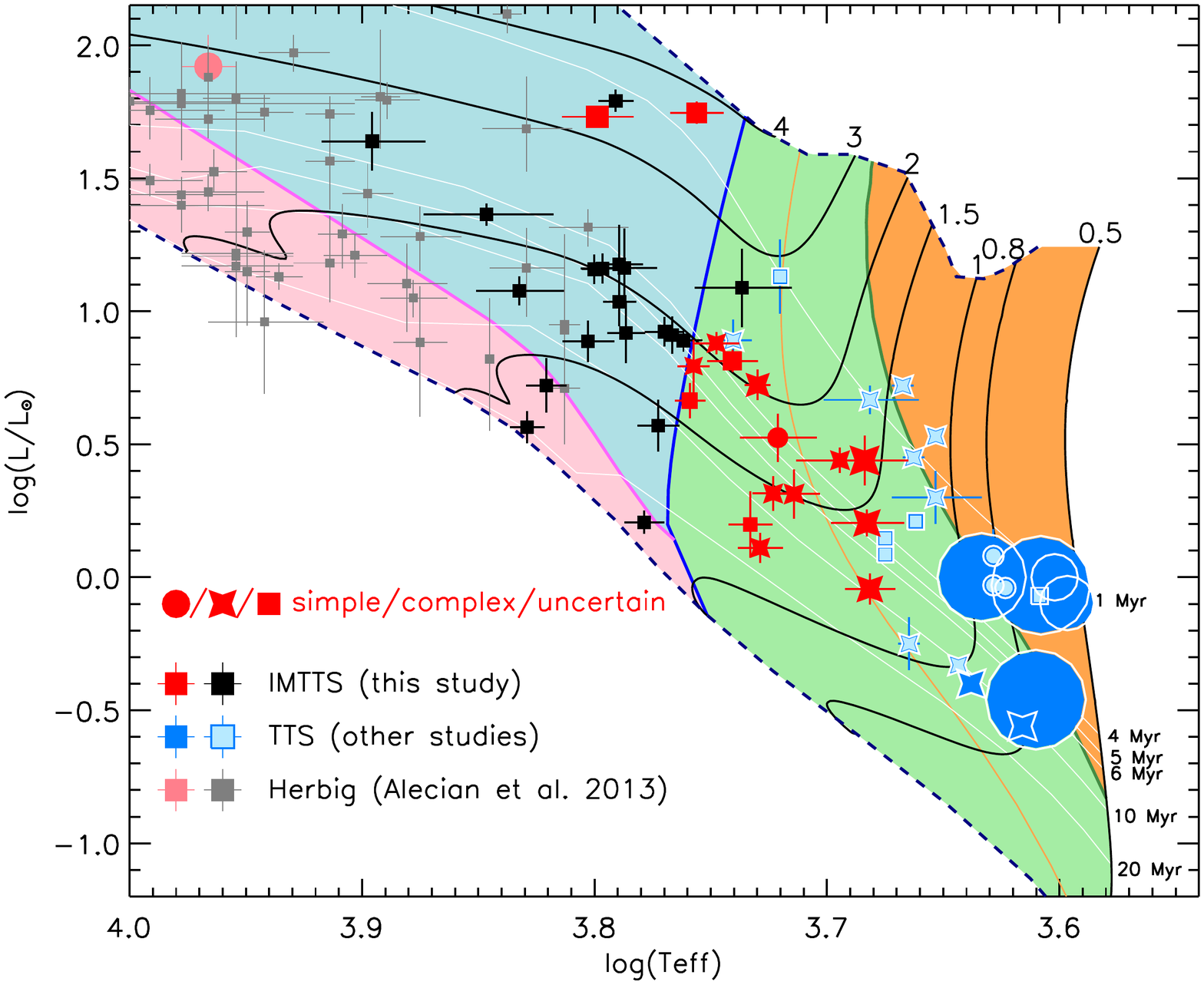}
 \hfill
\includegraphics[angle=180,width=0.186\textwidth,clip,bb=626 0 780 590]{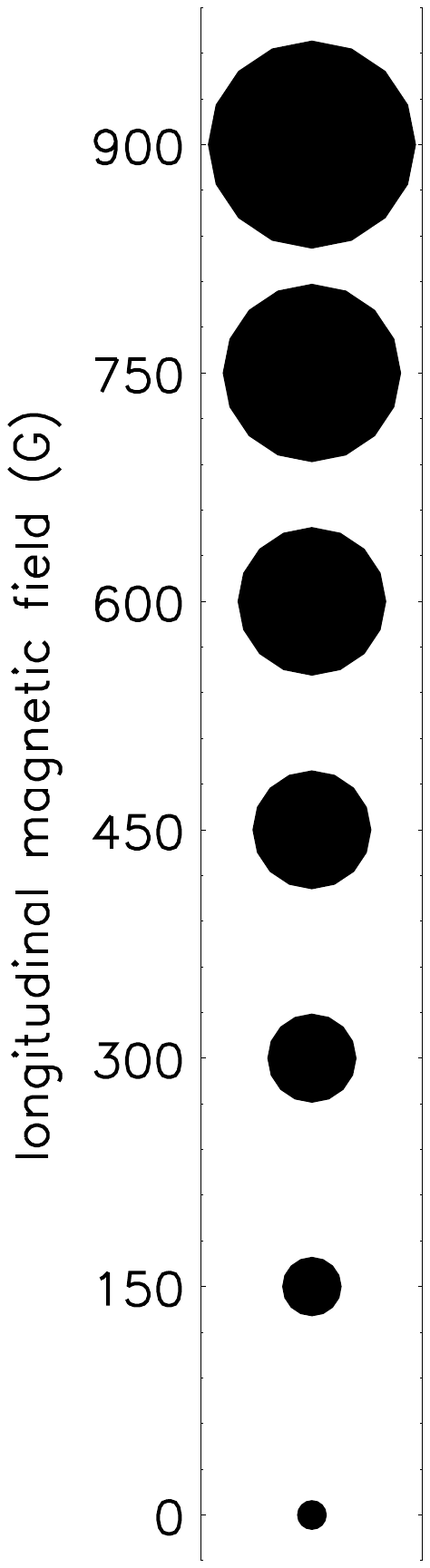}  
 \caption{HR diagram compiling the positions of IMTTS from this study (red and black symbols), and of IMTTS and LMTTS (blue symbols), and HAeBes (faded symbols) from other studies. Black and gray symbols are stars in which no magnetic field has been detected. Red and blue symbols are stars for which a magnetic field has been detected : stars represent complex magnetic topologies, circles represent simple magnetic topologies, and squares represent undefined magnetic topologies. The sizes of the red and dark blue symbols are proportional to the maximum absolute value of the longitudinal magnetic field measurements for each star. A light blue symbol means no such measurement is available. The youngest magnetic HAeBe star (HD 190073) is indicated on the top-left corner as an isolated red circle. The shaded areas have the following meaning; orange: fully convective; green: radiative core + convective envelope; blue: fully radiative; and pink: convective core + radiative envelope. The evolutionary tracks (solid black line, ranging from 1.0 - 4.0 M$_{\odot}$), isochrones (solid white line) and ZAMS (lower dashed line) are from the CESAM code, while the birthline (upper dashed line) is from \citet{BehrendMaeder2001}. The numbers above each evolutionary track are the stellar mass in {\bf solar units}. The numbers beside each isochron is the stellar age. The thin orange line is the location where $R_{\rm conv.env.}/R_{\star}=40$\%.}
  \label{fig:hrd}
\end{figure*}

\subsubsection{Mass, radius, convective turnover time, and internal structure}

We now want to determine other stellar parameters using PMS evolutionary models. To estimate the uncertainties introduced by these models, we compared 3 different grids of models that can be applied to the IMTTS mass range and calculated with 3 different stellar evolution codes: the CESAM code \citep{Morel2008,Marques2013}, the STAREVOL code \citep{Palacios2003,Lagarde2012,Amard2016}, and the Geneva code \citep[][Haemmerlé 2014 PhD thesis]{Eggenberger2008}. More details about these 3 sets of models and their comparison can be found in Appendix A and Fig A.1 at the end of this paper. We concluded that these 3 PMS models were similar enough -in the region of the HRD considered here and within the errors of our measurements of \teff\ and log($L$)- and confirm any of them can be used for our study. We decided to use the CESAM code, for practical reasons.

The resulting positions of the stars of our sample in the HR diagram are plotted in Fig. 1. From these, we interpolated stellar parameters that cannot be directly determined from spectropolarimetric observations: i.e. stellar mass, radius, convective turnover time $\tau_{c}$ (calculated at half a pressure scale height above the base of the convective envelope $H_{p}$/2, as described in \citet{Charbonnel2017}), and the mass and radius of the convective and radiative zones. We need to constrain all these parameters in order to test the magnetic field relaxation scenario. The procedure we used to interpolate these parameters is similar to the one described in \citet{Alecian2009}. All these results can be found in Table B.2.

\subsection{Magnetic properties}

We now want to determine which stars of our sample are magnetic, and for those that are we want to determine their magnetic field properties (strengths and topologies).

\subsubsection{Magnetic field detections}

From the Stokes $V$ profile of each spectrum, we computed the false alarm probability \citep[FAP,][]{Donati1997}. The detection of a magnetic field by the FAP is based on a comparison between the Stokes $V$ profile and the noise. We consider that a star hosts a magnetic field if we get at least one definite detection in one observation, as defined by \citet{Donati1997}. In addition, we checked that the null-spectra are flat for all our observations, and found no evidence for a spurious contribution to the Stokes $V$ profiles. We find that 18 stars out of our entire sample host a magnetic field, and we have no significant detections for the remaining 20 stars.

We then measured the line-of-sight component of the magnetic field averaged over the visible stellar hemisphere: the mean longitudinal magnetic field \citep[\bell, e.g.][]{Rees1979,Kochukhov2010}. \bell\ is measured by dividing the first moment of the Stokes $V$ profile by the equivalent width of the Stokes $I$ profile, and multiply this ratio by a factor that depends on the mean wavelength and mean Land\'e factor used for the normalisation of the LSD profile (Section 2). In Table B.5 are reported the $B_{l}$ and associated $\sigma_{B_{l}}$ of each observation, for both magnetic and non-magnetic stars. 

\begin{figure}[ht!]
 \centering
\includegraphics[width=0.52\textwidth,clip]{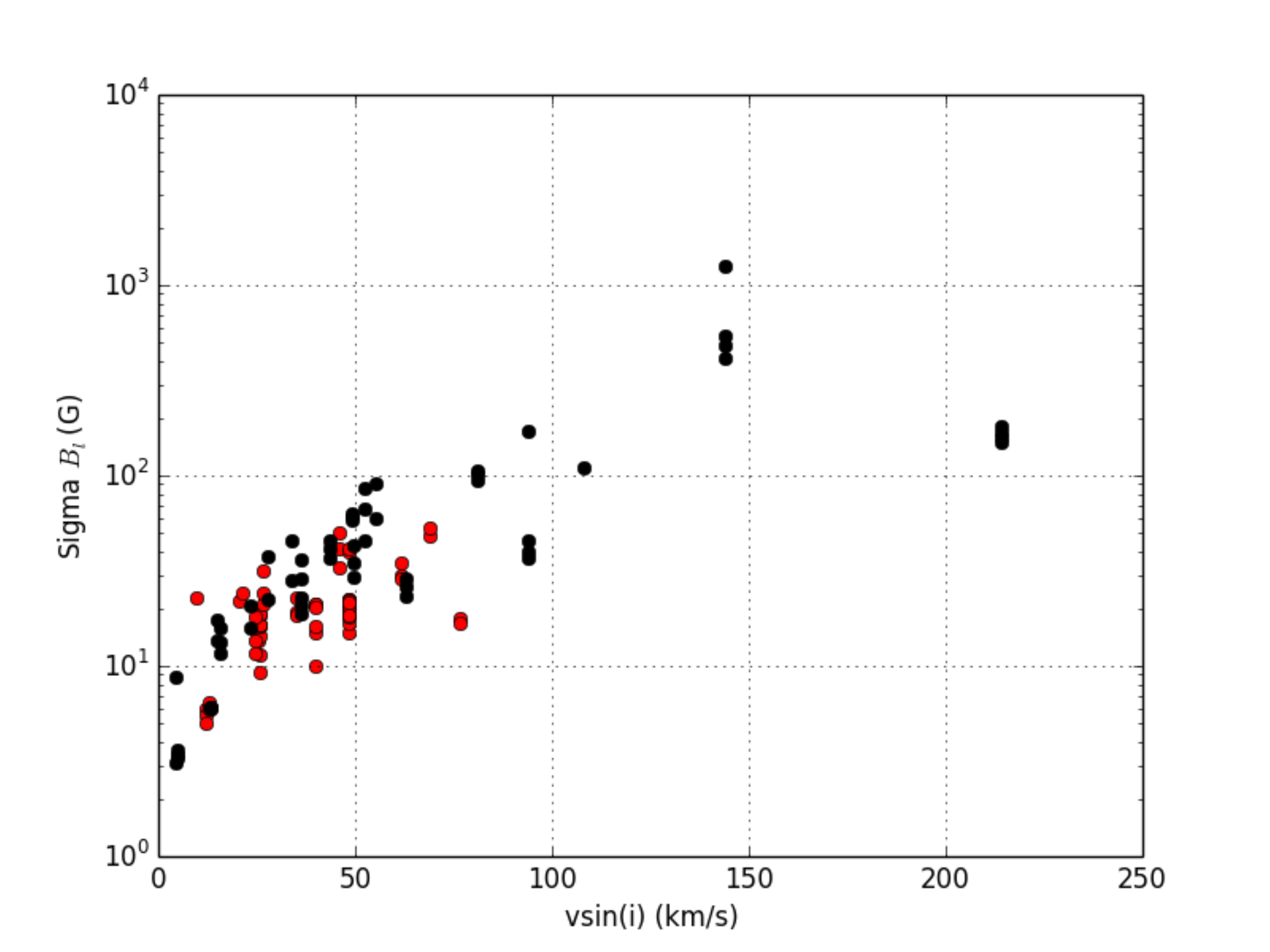}  
 \caption{$\sigma_{B_{l}}$ of each observation, with respect to the \vsini\ of its associated star. Red dots are for spectra coming from magnetic stars, black dots are for spectra coming from stars with no detected magnetic field.}
  \label{fig:sigbl}
\end{figure}

One might worry that, on average, all the stars for which we do not detect a magnetic field are just more difficult to detect a field on because they are rotating faster. To address this, we plotted the $\sigma_{B_{l}}$ of each spectrum with respect to the rotational velocity of its associated star in Fig.~\ref{fig:sigbl}. Red dots are for spectra coming from magnetic stars, and black dots are for spectra coming from stars with no magnetic detection. For \vsini\ up to 80 km/s (which concerns 33 stars out of 38), Fig.~\ref{fig:sigbl} shows that we can reliably detect magnetic fields with our data. Beyond this limit (concerning 5 stars) we did not detect any magnetic field: we therefore may be biased by fast rotation when above 80 km/s, but it also may be possible that none of these 5 stars actually host a strong large-scale magnetic field.

\longtab{
{\renewcommand{\arraystretch}{1.2}
\begin{longtable}{cccc}
\caption{Table compiling the extracted $B_{l}$ and associated $\sigma_{B_{l}}$ for each spectrum. Stars with a magnetic detection are written in bold, and so is the $B_{l}$ we retained for them (the maximum in absolute value among all spectra for one star).}\\
\hline\hline
Target & HJD (2450000+) & $B_{l}$ (G) & $\sigma_{B_{l}}$ (G) \\
\hline
BN Ori & 6196.01169 & 19.2 & 149.5 \\
- & 6272.05218 & -197.0 & 181.1 \\
- & 6286.08442 & -12.0 & 169.4 \\
- & 6293.85675 & 206.6 & 158.0 \\
\textbf{CHX 22} & 6140.48253 & \textbf{-35.3} & 23.0 \\
\textbf{CO Ori} & 6197.00549 & \textbf{-96.7} & 48.0 \\
- & 6284.80619 & -72.0 & 53.4 \\
\textbf{COUP 1350} & 6285.04068 & -1.5 & 34.5 \\
- & 6288.86765 & \textbf{23.4} & 29.8 \\
- & 6292.84744 & -20.9 & 28.7 \\
\textbf{CPD-43 7188} & 6128.60140 & \textbf{-30.5} & 13.7 \\
CQ Tau & 6284.02976 & 40.3 & 39.7 \\
- & 6287.06958 & -22.9 & 170.3 \\
- & 6289.89715 & 31.8 & 45.4 \\
- & 6290.84611 & -33.3 & 36.6 \\
\textbf{GW Ori} & 6195.99311 & 49.5 & 32.7 \\
- & 6258.12749 & 43.4 & 50.4 \\
- & 6272.07160 & \textbf{82.5} & 41.5 \\
\textbf{HBC 741} & 6100.12768 & \textbf{64.2} & 15.1 \\
- & 6102.12400 & 44.5 & 16.1 \\
- & 6104.11588 & -13.2 & 10.0 \\
- & 7879.12462 & 60.5 & 21.0 \\
- & 7886.11933 & 37.4 & 21.1 \\
- & 7886.13035 & 37.5 & 20.2 \\
\textbf{HD 133938} & 6140.56344 & \textbf{-28.7} & 22.0 \\
HD 135127 & 6131.52313 & -129.9 & 110.9 \\
HD 137059 & 6129.49163 & 42.3 & 37.3 \\
- & 6131.57721 & 33.3 & 22.3 \\
HD 143978 & 6129.70610 & -10.0 & 18.9 \\
- & 6130.55291 & -18.8 & 22.9 \\
- & 6130.63407 & -18.2 & 36.5 \\
\textbf{HD 147048} & 6131.65409 & \textbf{8.2} & 24.4 \\
\textbf{IRAS 22144} & 6100.10110 & \textbf{-14.4} & 19.2 \\
- & 6101.08220 & 4.0 & 22.9 \\
- & 6103.02717 & -3.0 & 18.5 \\
IRAS 22152 & 6100.10110 & -16.5 & 36.8 \\
- & 6101.08220 & 29.1 & 45.5 \\
- & 6103.02717 & -65.9 & 41.2 \\
NGC 2264 108 & 6270.15205 & 50.9 & 67.1 \\
- & 6285.13122 & 4.3 & 85.1 \\
- & 6289.03847 & 28.7 & 45.1 \\
NGC 2264 121 & 6267.16535 & 87.0 & 1266.2 \\
- & 6269.13658 & 152.5 & 481.8 \\
- & 6286.11356 & 516.6 & 411.9 \\
- & 6288.03579 & -91.6 & 539.2 \\
NGC 2264 84 & 6201.13037 & -64.9 & 28.6 \\
- & 6288.98198 & -16.0 & 20.8 \\
NGC 6530 SCB7 & 6108.00466 & 10.2 & 3.6 \\
- & 6110.84930 & 9.8 & 3.3 \\
- & 6111.85859 & 14.9 & 3.4 \\
NGC 6530 SCB739 & 6129.68867 & 0.9 & 3.1 \\
- & 6131.75909 & -2.0 & 8.8 \\
PAR 102 & 6200.03445 & 2.0 & 57.9 \\
- & 6284.82240 & -2.1 & 61.3 \\
- & 6288.76243 & 44.8 & 63.6 \\
PAR 1391 & 6265.58358 & 1.8 & 17.6 \\
- & 6267.76109 & 2.2 & 13.6 \\
PAR 1394 & 6284.94365 & -4.6 & 26.2 \\
- & 6288.79597 & 30.4 & 23.2 \\
- & 6200.08117 & -19.7 & 28.8 \\
PAR 1414 & 6266.71310 & -27.7 & 28.4 \\
- & 6268.84148 & 45.2 & 45.8 \\
PAR 1455 & 6265.64831 & -11.4 & 20.6 \\
- & 6267.63691 & 19.0 & 15.8 \\
PAR 1646 & 6265.86174 & -3.8 & 11.6 \\
- & 6268.58229 & 4.3 & 15.7 \\
- & 6266.86804 & 5.4 & 13.4 \\
PAR 1736 & 6265.72010 & -46.0 & 59.9 \\
- & 6268.65637& -30.9 & 91.5 \\
PAR 2441 & 6286.88901 & -4.6 & 6.1 \\
- & 6288.82923 & -4.7 & 6.0 \\
- & 6260.12791 & -6.0 & 6.1 \\
PX Vul & 6104.04050 & -120.9 & 106.8 \\
- & 6108.07777 & 1.0 & 102.3 \\
- & 6110.87961 & -112.1 & 94.7 \\
RY Ori & 6283.91923 & 20.5 & 43.2 \\
- & 6284.88690 & 38.7 & 34.7 \\
- & 6290.76770 & -67.9 & 29.4 \\
\textbf{V1000 Sco} & 5969.16094 & -38.9 & 9.3 \\
- & 6103.75536 & -55.4 & 11.5 \\
- & 7817.06600 & \textbf{-116.6} & 14.3 \\
- & 7879.09957 & 21.4 & 16.2 \\
- & 7879.83844 & -23.6 & 18.4 \\
- & 7881.97665 & 10.6 & 16.5 \\
- & 7883.05611 & 48.5 & 16.4 \\
- & 7886.89840 & 59.2 & 18.7 \\
\textbf{V1001 Sco} & 5968.12625& \textbf{-35.1} & 11.6 \\
- & 5972.11674 & -12.5 & 18.0 \\
- & 6103.78273 & -5.0 & 13.5 \\
\textbf{V1002 Sco} & 5969.10633 & 20.5 & 17.8 \\
- & 6109.77816 & \textbf{-27.2} & 16.7 \\
\textbf{V1044 Ori} & 6202.00591 & -24.7 & 21.1 \\
- & 6271.11957 & \textbf{-60.5} & 31.9 \\
- & 6290.93247 & -17.5 & 24.1 \\
\textbf{V1149 Sco} & 5967.09378 & 4.4 & 5.9 \\
- & 6099.78064 & 12.6 & 5.5 \\
- & 6101.78171 & \textbf{28.1} & 5.0 \\
\textbf{V1152 Sco} & 5968.16100 & \textbf{86.4} & 6.0 \\
- & 6102.75795 & 2.0 & 6.4 \\
\textbf{V1156 Sco} & 5968.09636 & -13.0 & 15.1 \\
- & 5972.09027 & -18.5 & 21.0 \\
- & 7881.83655 & 7.9 & 18.4 \\
- & 7882.00559 & -0.8 & 16.9 \\
- & 7882.85852 & 20.8 & 22.3 \\
- & 7883.08725 & -43.6 & 22.4 \\
- & 7905.81407 & 18.6 & 21.2 \\
- & 7906.03909 & 0.9 & 21.0 \\
- & 7906.75247 & 22.3 & 22.0 \\
- & 7906.98154 & 13.0 & 19.8 \\
- & 7907.75468 & -66.6 & 18.3 \\
- & 7907.99062 & -48.5 & 20.5 \\
- & 7908.87211 & \textbf{94.4} & 40.0 \\
- & 7908.90669 & -26.2 & 41.3 \\
- & 7909.87012 & -17.8 & 21.7 \\
- & 7910.81390 & -2.0 & 18.5 \\
\hline
\label{tab:bell}
\end{longtable}}
}

\subsubsection{Limits of detection for a dipolar configuration}

When we did not find any evidence of a magnetic field in a star, we computed an upper limit for the detection of a dipolar field configuration. Considering the case that the star hosts a dipolar magnetic field (i.e. the configuration commonly found in fossil field stars and in some cool stars possessing a field that is not fossil; e.g. \citealt{Petit2008,Morin2010,BoroSaikia2016}), we estimated the maximum strength below which the hypothetical dipolar field could not have been detected in our data, according to the SNR of our observations. The calculation of this limit is carried out by a Monte-Carlo simulation. 

We followed the same procedure as described in \citet{Alecian2016}: for 15 values of a hypothetical dipolar magnetic field strengths $B_{\rm D}$ ranging from 0 to 5000\,G, we created 1000 synthetic $V$ profiles, each of which was associated to a randomly generated oblique rotator configuration (random inclination angle to the observer $i$, random obliquity angle of the dipole $\beta$, and random rotational phase of the star $\phi$). For each of these configurations, we computed the FAP in the synthetic Stokes $V$ profile (after adding synthetic noise with respect to the quality of our observations) to decide if a magnetic field of strength $B_{\rm D}$ would have been detected, using the same detection criteria as in our observations and detailed in the previous section. After 1000 trials, we can therefore give a detection probability for a dipolar field of strength $B_{\rm D}$ at its pole. We set our detection threshold at 95\%, meaning that the value $B_{\rm D 95\%}$ mentionned in Table 6 is the value of $B_{\rm D}$ for which the synthetic magnetic field has been detected in 95\% of the trials. Having more than one observation for a star is beneficial as it gives us more chances to detect a magnetic field: we consider all the different observations we got as independant draws, and can therefore combine the detection probabilities to improve our limit of detection.

\begin{figure}[ht!]
 \centering
\includegraphics[width=0.52\textwidth,clip]{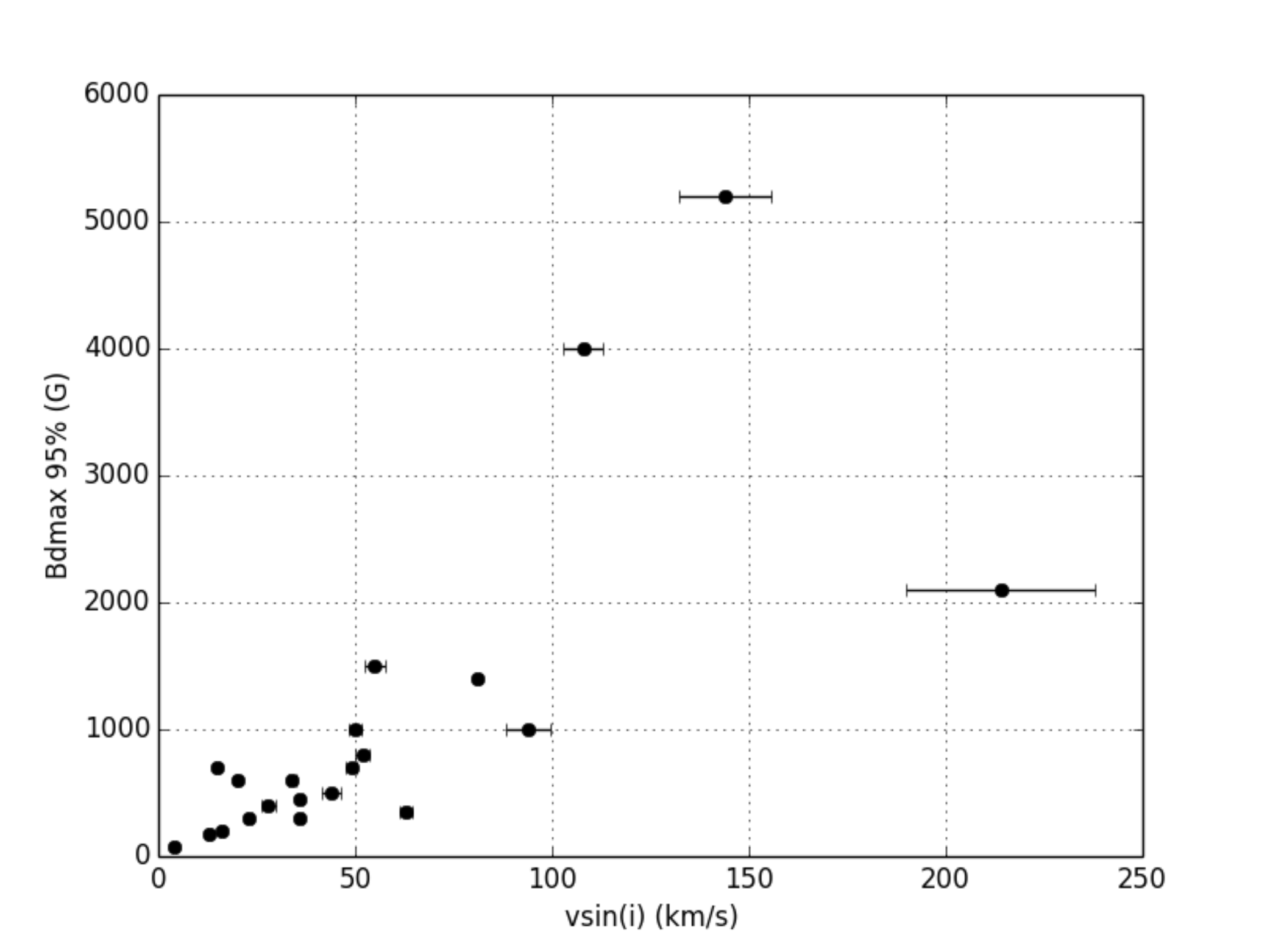}  
 \caption{Limit of detection for a dipolar magnetic field at the 95\% confidence for the 20 non-magnetic stars of our sample versus the projected rotational velocities of these stars.}
  \label{fig:blim}
\end{figure}

The values of $B_{\rm D 95\%}$ for the 20 non-magnetic stars are plotted in Fig.~\ref{fig:blim} : we observe that $B_{\rm D 95\%}$ is lower than 500\,G for half of our non-magnetic stars, and lower than 1\,kG for 80\% of them. We also observe a correlation between $B_{\rm D 95\%}$ and \vsini. Indeed, rotational broadening may hide magnetic signatures by spreading them over more spectral pixels, reducing the amplitude of the magnetic signature relative to the noise. For example, the very high rotational velocity of NGC 2264 121 ($v \sin i=144$\,\kms) is partly responsible for its very high detection limit (5200\,G). We would need to observe this target for much longer to increase significantly the SNR and give us a chance to detect a field of lower strength.


\subsubsection{Limits of detection for a multipolar configuration}

We initially computed the limits of detection in the case of a dipolar field for our stars in which there are no detections, as this is the typical configuration in fossil fields stars. However, the non-magnetic stars of our sample may also hide a multipolar field (or a multipolar+toroidal field, more typical of cool active stars). We investigated whether or not these stars could host complex, large-scale magnetic field signatures of the type found in the IMTTS CV Cha \citep{Hussain2009}. We used the published map of CV Cha to predict the Stokes $V$ signatures expected in each of the stars with non-detections. As the large-scale field is complex and multipolar, the main factors determining whether or not the signatures are likely to be detected are the stellar $v \sin i$ and the SNR levels of the LSD Stokes $V$ profile. By randomly sampling hundreds of phases for each star, we can assign the probability of detecting a CV Cha-type complex multipolar field at the 3-$\sigma$ level. The results are shown in the third column of Table 6, which shows the percentage probability of detecting a field, given the stellar parameters and the data quality. Our findings suggest that a complex multipolar field signature typical of partly-convective T Tauri stars should definitely have been detected for NGC 6530 SCB 739 and Par 2441 (though we cannot exclude a significantly weaker field). For HD 143978, a robust field signature would be expected approximately 95\% of the time. For the rest of the sample, we cannot exclude the possibility that these stars host complex fields such as those found in the CV Cha and CR Cha.  

\begin{table}
\centering
\tiny
\caption{Stars with no-detected magnetic field: upper-limit of detection in the case of a dipolar field ($B_{\rm D 95\%}$, second row) and $3-\sigma$ level probability of detecting complex, large-scale magnetic field signatures of CV Cha-type ($P_{3-\sigma}$, third row).}
\begin{tabular}{|c|c|c|}
\hline
\hline
Target & $B_{\rm D 95\%}$ (G) & $P_{3-\sigma}$  \\
\hline
BN Ori & 2100 & 0 \\
CQ Tau & 1000 & 19 \\
HD 135127 & 4000 & 0 \\
HD 137059 & 400 & 82 \\
HD 143978 & 300 & 95 \\
IRAS 22152+6947 & 500 & 66 \\
NGC 2264 108 & 800 & 33 \\
NGC 2264 121 & 5200 & 0 \\
NGC 2264 84 & 450 & 84 \\
NGC 6530 SCB739 & 70 & 100 \\ 
PAR 102 & 700 & 11 \\
PAR 1391 & 700 & 88 \\
PAR 1394 & 350 & 86 \\
PAR 1414 & 600 & 58 \\
PAR 1455 & 300 & 88 \\
PAR 1646 & 200 & 90 \\
PAR 1736 & 1500 & 5 \\
PAR 2441 & 180 & 100 \\
PX Vul & 1400 & 0 \\
RY Ori & 1000 & 62 \\
\hline
\end{tabular}
\label{cont-emission}
\end{table}

\section{Discussion}

\subsection{The origin of magnetic fields in Herbig and Ap stars}

We want to understand the evolution of the magnetic properties of intermediate-mass stars from the birthline to the Herbig phase, which necessitates a knowledge of the magnetic properties of the intermediate-mass T Tauri stars, the thus far poorly studied evolutionary precursors of the Herbig Ae/Be stars. We have performed an observational campaign of a specific sample of IMTTS evolving from a fully convective structure to a fully radiative structure. The stars of our sample are plotted in Fig.~\ref{fig:hrd} with red symbols when we obtained a magnetic detection (16 stars in Fig.~\ref{fig:hrd} + 1 falling out of the HRD, see Sect. 3.1.2.), and black symbols when no magnetic field has been detected (19 stars in Fig.~\ref{fig:hrd} + 1 falling out of the HRD, see Sect. 3.1.2.). The sizes of the symbols depend on the maximum strength of our \bell\ measurements (the minimum sizes being for the non-magnetic stars).

We have divided the HRD in 4 different zones. In Fig 1., the orange zone indicates the position of the stars evolving with a fully convective interior. The green zone indicates partly-convective stars, i.e. stars with a radiative core larger than 1\% of the stellar radius \rstar\ (equivalent to 1\% of the stellar mass \mstar), and lower than 75\%\,\rstar\ (or 98\%\,\mstar). The blue zone, called the fully-radiative zone, contains stars with a radiative core containing more than 98\%\ of the stellar mass, and the pink region are for stars in which a convective core is growing and has a mass larger than 1\%\,\mstar.

In order to complete the global picture, we included stars from other spectro-polarimetric studies in Fig.~\ref{fig:hrd}: Herbig Ae/Be stars from \citet{Alecian2013} -- using similar symbols as the IMTTS but with faded colours; as well as other IMTTS and LMTTS -- using blue symbols 
\citep[][]{Donati2007,Donati2008,Donati2010a,Donati2011b,Donati2011c,Donati2013,Donati2015,Hussain2009,Kochukhov2015,Hill2017,Lavail2017,Yu2017}. In the latter case, all stars are magnetic, and the light-blue (blue) colour indicates a lack (or not) of information on the \bell\ values. We observe that, while the fully-convective zone above $\sim$1\,\msun\ is almost empty, our sample bridges well the gap between the fully convective and fully-radiative zones at intermediate-mass.

In the fully radiative part of the HR diagram (blue region in Fig. \ref{fig:hrd}), our IMTTS sample overlaps with the Herbig sample well, bridging the gap between the convective-radiative (CR) limit (thick blue line) and the radiative part of the HR diagram probed by the Herbig sample. Except for a small gap near the CR limit around 3 M$_{\odot}$, the connection between partly convective IMTTS (green zone) and fully radiative IMTTS (blue zone) is fairly complete in our sample. In this region of the HR diagram, our study provides evidence of a very clear trend: convective stars quickly lose their magnetic field. Indeed, almost 100\% (14 out of 15) of partly convective IMTTS host a magnetic field, while only $\sim$10\% (2 out of 18) of fully radiative IMTTS are magnetic, which is an occurence similar to the fossil fields in the HAeBe and Ap/Bp stars. Furthermore, the non-magnetic star in the partly convective zone is NGC~2264~121, the one with a detection limit much larger than typical TTS magnetic fields. This star may well host a relatively strong magnetic field that we are not able to detect with our data due to its very rapid rotation.

The magnetic/non-magnetic boundary appears obvious in Fig. \ref{fig:hrd} and may define the boundary between ubiquitous dynamo-generated magnetic fields in convective stars and the rarer fossil magnetic fields in fully radiative stars. Stellar evolutionary grids from the CESAM code reveal that stars evolve across this boundary very quickly so that the transition between the two types of fields must occur within a few times 0.1 Myr. This timescale is similar to that found with the two other evolutionary PMS grids considered in this study. We thus propose that 0.1~Myr is the typical time-scale for dynamo fields to dissipate in IMTTS. 
In addition, we find that this transition occurs when $M_{\rm conv. env.} / M_{*} \sim 2$\%, or in terms of volume when $R_{\rm con. env.} / R_{*}\sim25$\%.

Typical fossil fields detected in Ap/Bp stars have predominantly dipolar configurations and have dipole field strengths of a few kG \citep[e.g.][]{Donati2009}. Most of the detection limits we measured in our non-magnetic sample are lower than 1\,kG, with the majority having a limit lower than 0.5\,kG. Our data are therefore sensitive enough for detecting typical fossil fields.

This is true if we ignore the dilution effect of magnetic flux due to the radius variation during stellar evolution. The fossil field hypothesis assumes magnetic flux conservation implying that the magnetic strength at the stellar surface varies as $1/R_{*}^{2}$. As a consequence, magnetic fields are weaker at large radii. Between the birthline and the Terminal Age Main Sequence (TAMS), the radius changes the most during the early-PMS phase. However, between the CR limit (after which only fossil field can be present at the surface of the stars) and the ZAMS, the radius only changes by a factor of about 2, i.e. a similar factor between the TAMS and the ZAMS. Therefore, the dilution effect affects Ap/Bp stars and radiative PMS stars the same way, and can be neglected in our interpretation. In the fossil field approximation, the magnetic strength and topology we expect to detect in the radiative IMTTS are therefore likely of Ap/Bp type.

If our data are sensitive enough to detect Ap/Bp fields, by extension it should also be sensitive enough to detect "typical" Herbig Ae/Be fossil fields. We observe that our IMTTS dataset is more sensitive to magnetic field than the HAeBe dataset \citep{Alecian2013}. Indeed, 80\% of our \bell\ measurements have uncertainties below 50~G (see Table \ref{tab:bell}), which is the case for only 20\% of the HAeBe dataset, and 80\% of it has uncertainties below 250~G. Nonetheless, fossil fields have been detected in the HAeBe sample \citep[e.g.][]{Alecian2013}, which emphasises that our dataset is sensitive enough to detect fossil fields.

Fossil magnetic fields are present in 1 to 10\% of the A/B stars, depending on the mass: while it is found to be 5-10\% above 2.5~\msun\, it drops quickly to 1\% below 2.5~\msun\ \citep{Power2007, Sikora2018}. According to the fossil field hypothesis, a similar incidence is expected in the PMS radiative stars. Two out of the 18 radiative IMTTS we observed are magnetic. This is in global agreement with the fossil field theory, regardless of the mass, and considering the small size of our sample. However, about 90\% of our sample has masses concentrated between $\sim$1~\msun\ and $\sim$2.5~\msun\ (Fig. \ref{fig:hrd}). Yet, the only magnetic fields we have detected in the radiative zone are in stars more massive than 3~\msun. The absence of magnetic field detection in radiative stars less massive than 2.5~\msun\ is therefore striking. Our observations suggest that the incidence drop observed during the MS, is already present during the PMS phase.

While below 2.5\,\msun\ there is a very distinct boundary between the magnetic and non-magnetic stars, above 3\,\msun\ the boundary is not that evident. First, our sample does not contain any stars more massive than 3\,\msun\ with a large outer convective envelope ($>$0.25\,\rstar). Secondly our sample contains in total only 3 stars more massive than 3\,\msun, which is not statistically significant. Nonetheless, if the two magnetic stars detected among those 3 have fossil fields, it would mean that we would have been lucky to detect them. It is probably more likely to assume that what we observe may be dynamo-fields created in the outer convective envelope, even if it contains less than 2\% of the stellar mass or is contained in less than 25\% of the stellar radius. The still open question would be to know whether it is possible to generate magnetic fields in such light but extended convective zones. Our data does not allow us to choose between simple and complex field for those stars. Additional observations are required to determine the origin of their magnetic fields, and to better understand the origin of fossil fields in stars more massive than 3\,\msun.

\subsection{Transition from fully convective to fully radiative}

To better understand the evolution of magnetic fields during the PMS phases, we need to know their strengths and topologies. This requires good quality monitoring that necessitates a large amount of telescope time. This is currently being done on a subsample of our stars as a follow-up of this study. 

In the mean time, with the spectropolarimetric snapshots we obtained, we studied the evolution of the mean longitudinal magnetic field \bell. Mean longitudinal magnetic field measurements are not ideal because they vary significantly over the rotation of the star, but we can lessen this effect by choosing the highest absolute value amongst all observations of a particular star. As the star rotates, the \bell\ values vary periodically, and the periodic curve depends on the field strength and topology. For a fixed topology the maximum absolute value (\blmax) increases with the magnetic strength. With only two to four observations we cannot draw the entire curve. We therefore chose the maximum measured absolute value as being the most representative value for the magnetic field's strength. In Table B.5, the \blmax value of each star is indicated in bold: this value has been used to plot the size of the symbols in Fig~\ref{fig:hrd}.

In addition, from the shape of the LSD Stokes $V$ profiles, we can begin to diagnose whether the Zeeman signature is most likely produced by a low-order field (simple bipolar signatures spread over the entire width of the $I$ profile) or whether it is most likely produced by a complex multipolar field: composite and complex Stokes $V$ signals often spread over only a small part of the width of the $I$ profile, suggestting localised magnetic spots. For 10 out of the 16 magnetic IMTTS plotted in the HR diagram, we are able to say confidently that their magnetic field is complex, while only one star of our sample seems to host a simple magnetic field. The simple, complex and uncertain field natures are expressed in Fig. \ref{fig:hrd} with different symbols.

We found magnetic fields in only two radiative stars: CO Ori and GW Ori. They are our best candidates for hosting a fossil field. In addition, their \bell\ are among the strongest detected in this sample (96.7 and 82.5 G, respectively) despite the relatively poor monitoring (only 2 spectra for CO Ori and 3 spectra for GW Ori). We tried to determine the complexity of their fields, but find their Stokes $V$ profile time-series are difficult to interpret (see Fig~\ref{fig:coori} for CO Ori and Fig~\ref{fig:gwori} for GW Ori). They are both excellent targets for future studies as they may host freshly formed fossil fields. 

On the opposite side of the HR diagram, we note that IMTTS with large convective envelopes {\bf ($R>0.40$\,\rstar)} follow the same trend: most of the time, they host a multipolar or a multipolar+toroidal field. According to previous studies, including the magnetic mapping of the IMTTS CR Cha \citep{Hussain2009}, those multipolar and toroidal fields are indeed expected once the fully-convective limit (between the orange and green zone, see Fig. \ref{fig:hrd}) is passed \citep{Gregory2012}.

The rest of the magnetic sample (all the partly convective stars with a weaker magnetic field, located closer to the convective radiative limit, with $R<0.40$\,\rstar) is much more difficult to interpret because these stars have all kind of field complexities, with no apparent correlation with their degree of convection. For example, the Stokes $V$ profiles of IRAS~22144 behave like a composite of multiple signals and could therefore hide a complex field (despite IRAS 22144 being almost entirely radiative) while the more convective star CHX~22 profile seems of simpler configuration. Moreover, the $B_{l}$ amplitudes of these stars do not show any correlation with the apparent complexity of their fields. It then appears to us that this region of the HR diagram (where $R_{\rm conv.env.}$ evolves from 40 to 25\% of $R_{\rm tot}$) is a transition region between two regimes: the regime of strong and complex fields in highly convective stars (at the right of the HRD, where $R_{\rm conv.env.} \geq$ 40\% of $R_{\rm tot}$) and the regime of radiative IMTTS (at the left of the HRD, where $R_{\rm conv.env.} \leq$ 25\% of $R_{\rm tot}$) where more than 90\% of the stars have either no field, or a field too faint to be detected with our observations.

Another interesting point can be mentioned: it may be possible that the magnetic IMTTS which do not evolve into Ap/Bp stars could be the progenitors of the weakly-magnetised "Vega-like" A stars \citep{Lignieres2009,Petit2010,Petit2011,Blazere2016a,Blazere2016b}.

\subsection{Low-mass T Tauri stars and the role of accretion}

Magnetic fields are systematically detected in T Tauri stars. The discussion in \citet{Hill2017} compiles and summarises the magnetic properties of 16 TTS with masses ranging from 0.5 to 1.9 M$_{\odot}$, and ages ranging from 1 to 10~Myr (8 classical TTSs from the MaPP programme, 6 weak-line TTSs from the MaTYSSE programme, and 2 additional weak-line TTSs from Hill et al. 2017). The 3 most massive stars in this study (Par 2244, Par 1379 and V410 Tau, at around 1.4 - 1.8 M$_{\odot}$) overlap with the lightest and most convective stars of our sample and display similar magnetic topologies.

\citet{Lavail2017} have measured the averaged modulus of the magnetic field over the visible stellar hemisphere ($<B>$) of 6 IMTTS from the Zeeman broadening of IR spectral lines. When compared to LMTTS in which similar measurements have been performed, \citet{Lavail2017} proposed that we cannot find as strong magnetic fields in IMTTS as in some LMTTS. In Fig. \ref{fig:hrd}, we observe that the stars with the strongest \bell\ (larger than 300~G) are the coolest TTS with effective temperatures lower than $\sim$4000~K. These stars also have simple fields and are mainly convective. On the contrary, the TTS with the lowest longitudinal field values are slightly more evolved, and have complex fields. However, it does not mean that their total magnetic energy is lower because the cancellation effect affects \bell\ measurements, even more in the case of complex fields. It may just be that the lowest order dipole field component is weaker. Such a decrease of the dipole component with the decrease of the size of the convective envelope has already been noted by \citet{Gregory2012}. In addition, we can compare \bell\ measurements of complex fields between our sample and the LMTTS. We observe they all have values of the same order of magnitude. Therefore our \bell\ data cannot confirm the trend found by \citet{Lavail2017} from mean field measurements: it is important to keep in mind that the two studies measured different kinds of magnetic field (global and local) that do not necessarily follow the same trend with mass.

\citet{Hill2017} also looked for correlation between magnetic topologies of LMTTS and their accretion state. While remaining cautious, they point out that wTTSs generally show a wider range of field topologies compared to cTTSs, with large-scale fields that can be more toroidal and non-axisymmetric. We looked for correlations between magnetic strengths / topologies and the accretion states of our sample of stars: from the accretion state (wTTS or cTTS) we compiled from the literature (Table B.2), we do not find any kind of correlation.

\section{Conclusions}

We have analysed spectropolarimetric data of a sample of 38 IMTTS probing the zone of the HRD where intermediate-mass stars evolve from a fully convective to a fully radiative internal structure. We have detected magnetic fields in about half of our sample. We find that the magnetic incidence goes from almost 100\% to $\sim$10\% as soon as the stars cross the limit over which the convective zone has a mass lower than $\sim$2\% of the stellar mass, which is equivalent to $\sim$25\% of the stellar radius. We argue that dynamo fields have to dissipate within about few 0.1\,Myr once this limit has been reached. We are however not able to constrain the time-scale over which they relax into fossil fields. Additional studies of the radiative magnetic IMTTS are required.

From the shape of the Zeeman signatures, we find that the stars of our sample with a convective envelope spreading over 40\% to 99\% of the stellar radius most likely host complex high-order magnetic fields, as observed among the low-mass TTS that are similarly convective. In addition, this may be evidence of a decrease in the low-order large-scale magnetic field component from the fully convective limit to the fully radiative limit, as in the LMTTS. We find no correlation between the magnetic signature shapes and strengths, and the accreting nature (wTTS or cTTS) of our IMTTS sample.

\section*{Acknowledgements}
We are grateful to the referee for his judicious comments and to Silvia Alencar for her help and contribution. 
We acknowledge financial support from "Programme National de Physique Stellaire" (PNPS) of CNRS/INSU, France. 
This work was based on observations made with ESO Telescopes at the La Silla Paranal Observatory under Programme IDs 089.C-0796 \& 090.C-0131, and is partly sponsored by the Swiss National Science Foundation (Project number 200020-172505).
FG acknowledges financial support from the Centre National d'Études Spatiales (CNES).
This work has made use of the VALD database, operated at Uppsala University, the Institute of Astronomy RAS in Moscow, and the University of Vienna, and of data from the European Space Agency (ESA) mission {\it Gaia} (\url{https://www.cosmos.esa.int/gaia}), processed by the {\it Gaia} Data Processing and Analysis Consortium (DPAC, \url{https://www.cosmos.esa.int/web/gaia/dpac/consortium}). Funding for the DPAC has been provided by national institutions, in particular the institutions participating in the {\it Gaia} Multilateral Agreement. 
Finally, we thank Aline Bellosguardo and the entire administrative team of IPAG for their excellent work and support.

\bibliographystyle{aa}  
\bibliography{article} 

\appendix
\section{PMS model influence}
When placing a sample of stars in the HR diagram and when trying to determine their internal structure through stellar evolution laws, the considered evolutionary model takes a crucial role: the interpretation may vary from one model to another. We thus need to test different PMS evolutionary models. We compared three grids of PMS evolution models computed with the CESAM code, the Geneva Code, and the STAREVOL code. These three codes can describe the evolution of pre-main sequence stars in the range of mass we are interested in, and can thus be used to derive the internal structure of our stars.

\subsection{The CESAM code}

The CESAM grid has been computed by one of us (YL) with the CESAM code \citep{Morel2008} using a standard set of input physics: OPAL96 opacity tables \citep{Iglesias1996} complemented at low temperatures by WICHITA data \citep{Ferguson2005}, OPAL2005 equation of state \citep{Rogers2002}, NACRE nuclear reaction rates \citep{Angulo1999} except for the $^{14}N(p,\gamma)^{15}O$ reaction, 
where the LUNA rate \citep{Formicola2004} was adopted, mixing-length theory of convection \citep{BohmVitense1958}, and outer boundary condition based on Eddington's grey atmosphere. Models do not include rotation, mass loss, element diffusion, convective core overshooting, or convective penetration of elements below the convective envelope. The AGSS09 solar mixture \citep{Asplund2009} has been adopted. The initial parameters of the models (initial helium abundance $Y_\mathrm{ini}$, metallicity $(Z/X)_\mathrm{ini}$, and mixing-length parameter $\alpha$) were fixed from the calibration of a solar model (i.e. a solar model that must have, at solar age, the observed solar luminosity, radius, and photospheric metallicity). This leads to $Y_\mathrm{ini}=0.2539, (Z/X)_\mathrm{ini}=0.01781, \alpha=1.622$.

\subsection{Geneva code}

The Geneva code grid \citep[][Haemmerlé 2014 PhD Thesis]{Eggenberger2008} has been computed by one of us (LH). Like the CESAM grid, that grid is calibrated on the sun, with no rotation, no mass loss, no diffusion, and no overshooting. The Geneva code grid is calculated with $Z_{\rm ini}$ = 0.0122, $Y_{\rm ini}$ = 0.2485, $\alpha$ = 1.6, the equation of state and the opacity come both from OPAL (1996). Abundances are from \citet{Asplund2005}, and the atmosphere model comes from \citet{MeynetMaeder1997}. 

\subsection{The STAREVOL code}

The STAREVOL grid has been computed by two of us (FG and LA). This grid is also calibrated on the sun, with no rotation, no mass loss, no diffusion, and no overshooting. They are calculated with $Z_{\rm ini}$ = 0.0134, $Y_{\rm ini}$ = 0.2676, $\alpha$ = 1.973. The basic input physics (equation of state, nuclear reactions, opacities) can be found in \citet{Lagarde2012}. The equation of state is based on the formalism developed by \citet{Eggleton1973} and extended by \citet{Pols1995}. OPAL opacities from \citet{Iglesias1996} are complemented at low temperatures by atomic and molecular opacities of \citet{Ferguson2005}. Abundances are from \citet{Asplund2005}, and the atmosphere model is a PHOENIX atmosphere \citep{Allard2011}. 

\subsection{Comparison of the models}

In Fig A.1. , we overplotted and compared isomass evolutionary tracks for 1.2, 1.5, 2.0, 3.0 and 4.0 solar masses for these three PMS grids of models. Green lines correspond to the Geneva Code, red lines correspond to the STAREVOL code, and blue lines correspond to the CESAM code. Red dots are stars with a definite magnetic detection (DD), while black dots are stars in which no magnetic field has been detected (ND). We tried to find a boundary between the population of magnetic stars (on the right) and the population of non-magnetic stars (on the left) in terms of stellar internal structure (radius and/or mass of the convective enveloppe compared to the radius and/or mass of the radiative core). We figured out that these two populations of stars can be splitted by a bound marking the moment when the radiative core fills more than 70\%-75\% of the total radius of the star, which also matches the moment when the radiative core contains more than 99\% of the total mass of the star. We propose to call this limit "the convective/radiative limit" (CR limit hereafter), i.e. the limit beyond which stars are almost entirely radiative and lose their dynamo field.

In the part of the HR diagram where our stars are located, we notice that the isomass evolutionary tracks are similar: the disparity from one model to another is always smaller than our smallest error bars on $T_{\rm eff}$ and $L$, meaning that the uncertainties on the stars' internal structure or characteristics are mainly due to our uncertainties on $T_{\rm eff}$ and $L$, and not on discrepancies between models. The same reasoning can be applied to the CR limits, all of them being located close to each other (the spread between them is smaller than the smallest error bars we have). We therefore can work with any of these three models, it wouldn't make any major difference in our interpretations. The biggest disagreement between these three models can be spotted at the very beginning of the evolutionary tracks, during the fully convective phase. This discordance comes from the atmosphere mode: the three models do not use the same outer boundary conditions. However, none of our stars are located is this region of the HR diagram, thus our interpretations shouldn't be impacted.

\begin{figure}[ht!]
 \centering
 \includegraphics[angle=270,width=0.5\textwidth,clip]{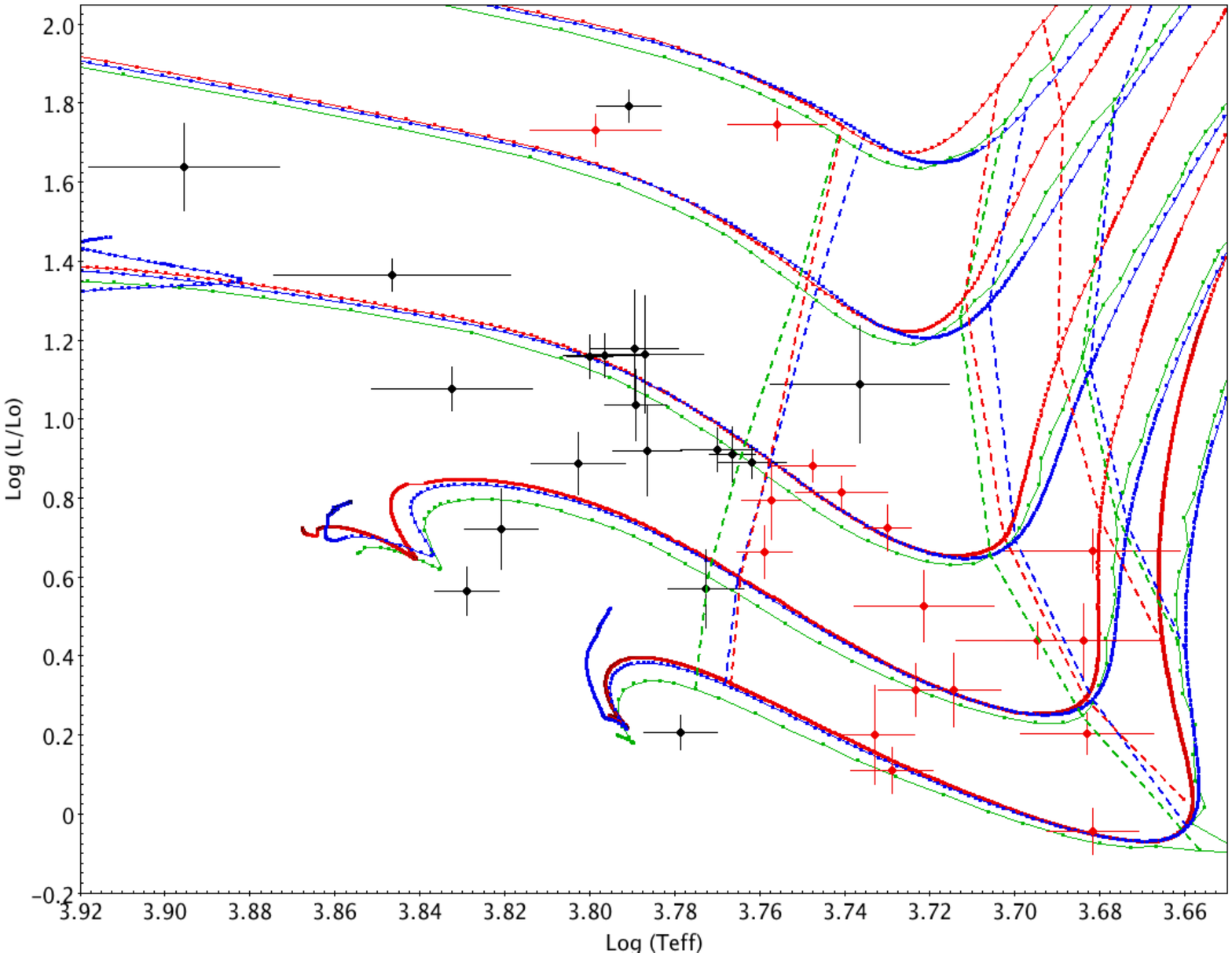}  
 \caption{Blue tracks correspond to CESAM evolutionary models, red tracks to STAREVOL, and green tracks to the Geneva Code. Red dots are magnetic stars, while black dot are stars with no magnetic detection. The solid lines are (from bottom to top) 1.2, 1.5, 2.0, 3.0 and 4.0\,M$_{\odot}$ isomass evolutionary tracks. Dashed lines are (from right to left) $M_{\rm conv.} / M_{\rm star}$ = 99\%, 50\% and 1\%.}
  \label{author1:fig1}
\end{figure}

\newpage

\section{LSD profiles}

\begin{figure}[ht!]
 \centering
 \includegraphics[width=0.45\textwidth,clip]{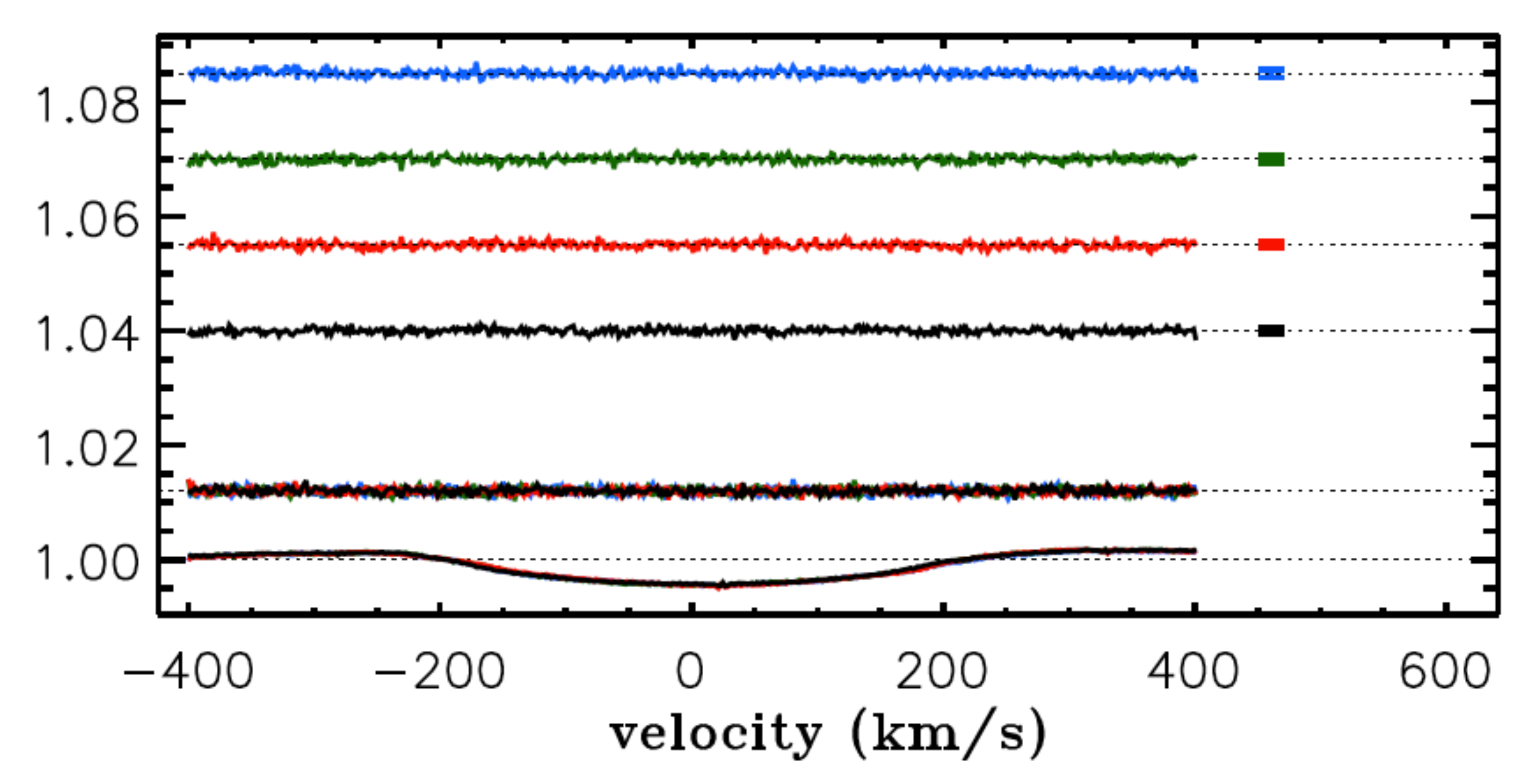}  
 \caption{LSD profiles for BN Ori (no magnetic detection).}
\end{figure}
 
\begin{figure}[ht!]
 \centering
 \includegraphics[width=0.45\textwidth,clip]{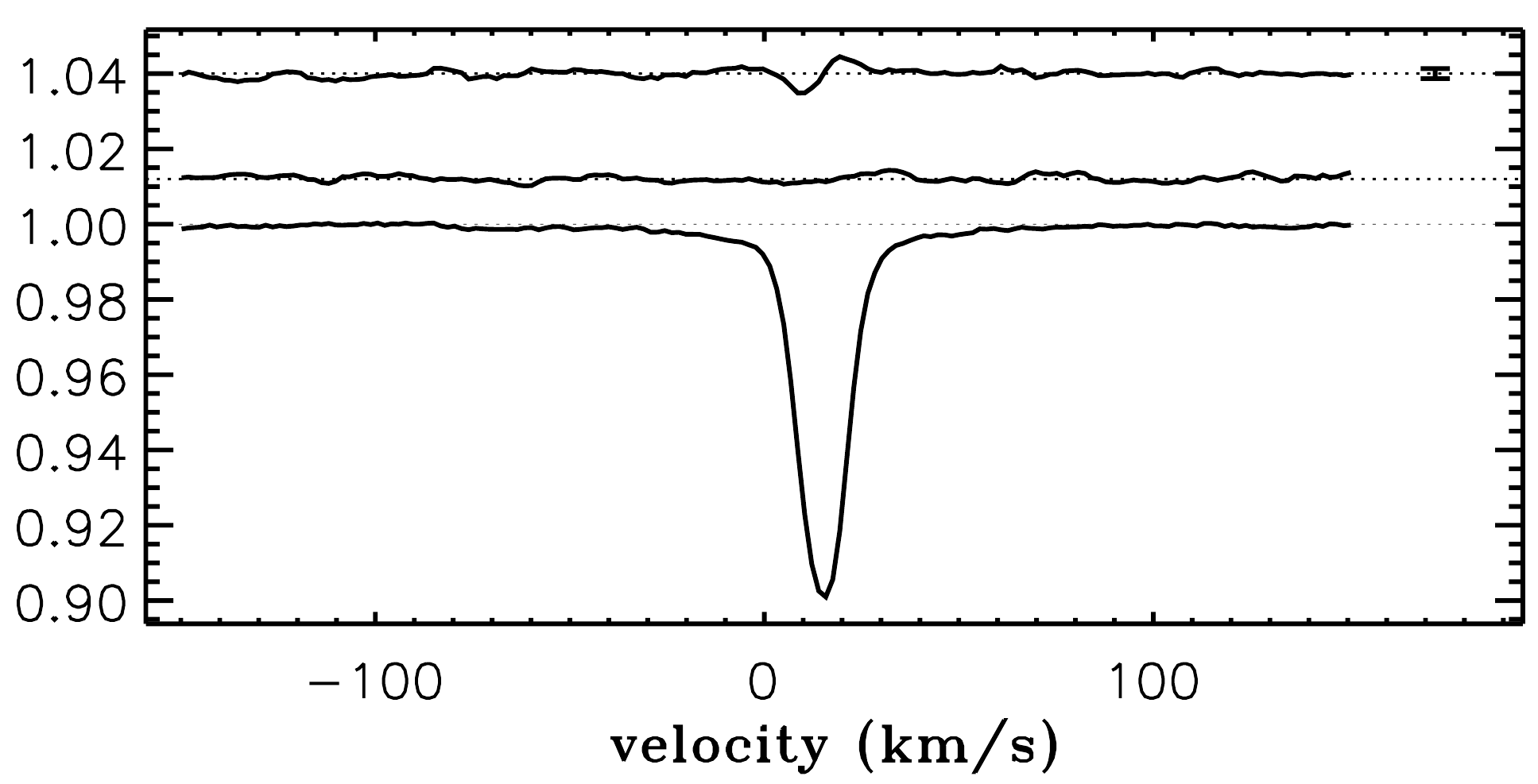}  
 \caption{LSD profiles for CHX 22 (magnetic field detected, simple topology).}
\end{figure}

\begin{figure}[ht!]
 \centering
 \includegraphics[width=0.45\textwidth,clip]{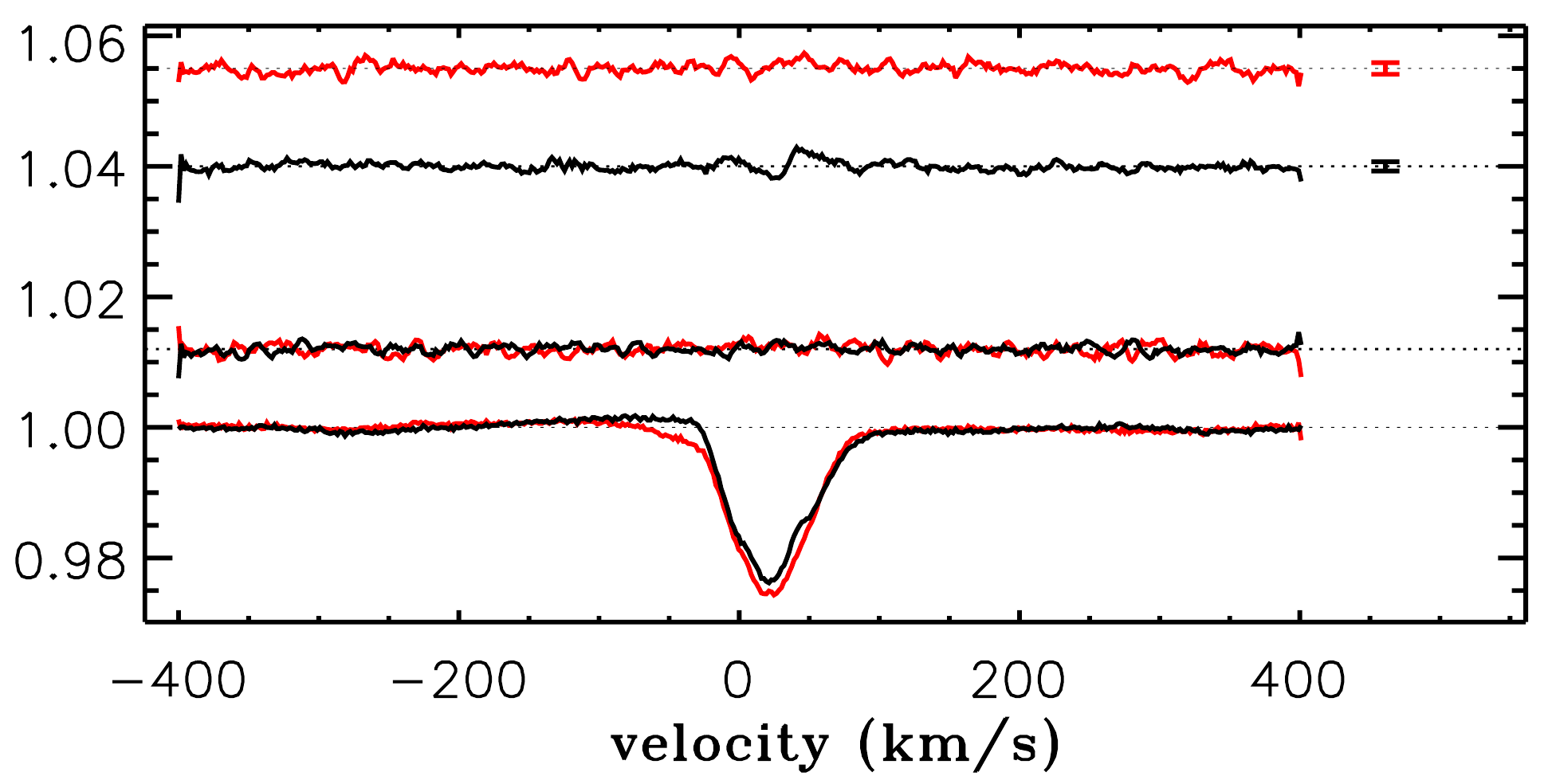}  
 \caption{LSD profiles for CO Ori (magnetic field detected, uncertain topology).}
 \label{fig:coori}
\end{figure}

\begin{figure}[ht!]
 \centering
 \includegraphics[width=0.45\textwidth,clip]{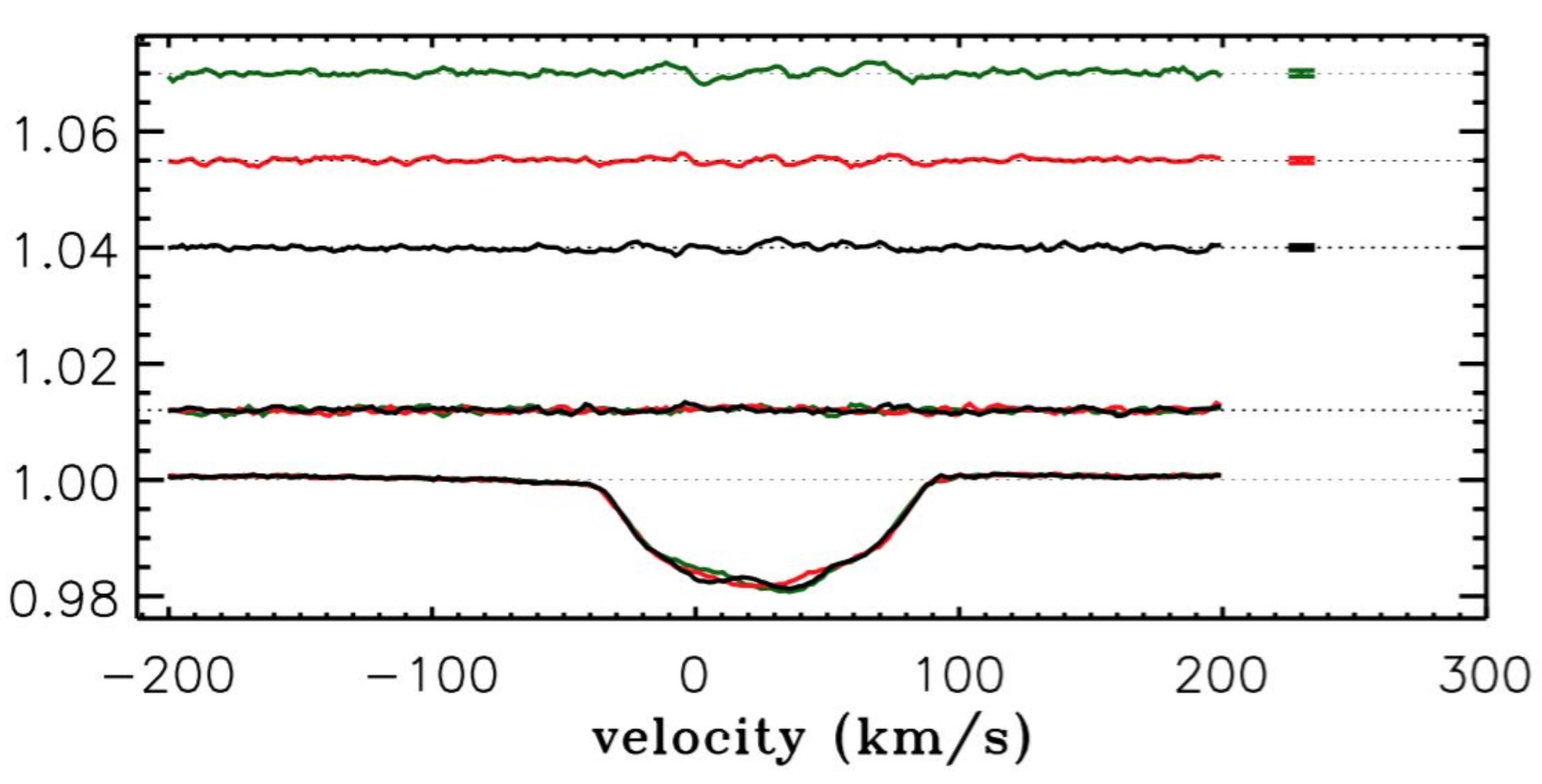}  
 \caption{LSD profiles for COUP 1350 (magnetic field detected, complex topology).}
\end{figure}

\begin{figure}[ht!]
 \centering
 \includegraphics[width=0.45\textwidth,clip]{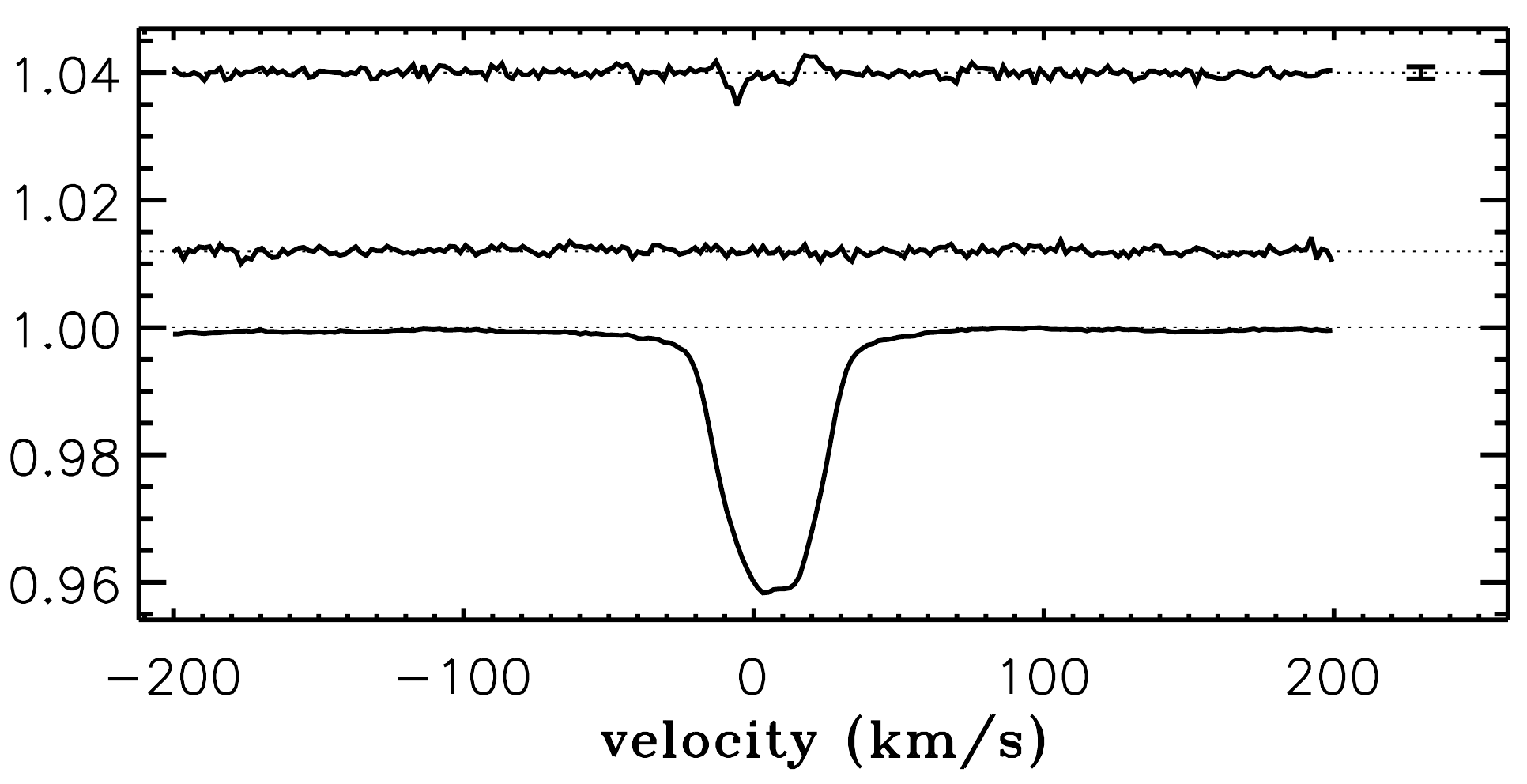}  
 \caption{LSD profiles for CPD -43 7188 (magnetic field detected, complex topology).}
\end{figure}

\begin{figure}[ht!]
 \centering
 \includegraphics[width=0.45\textwidth,clip]{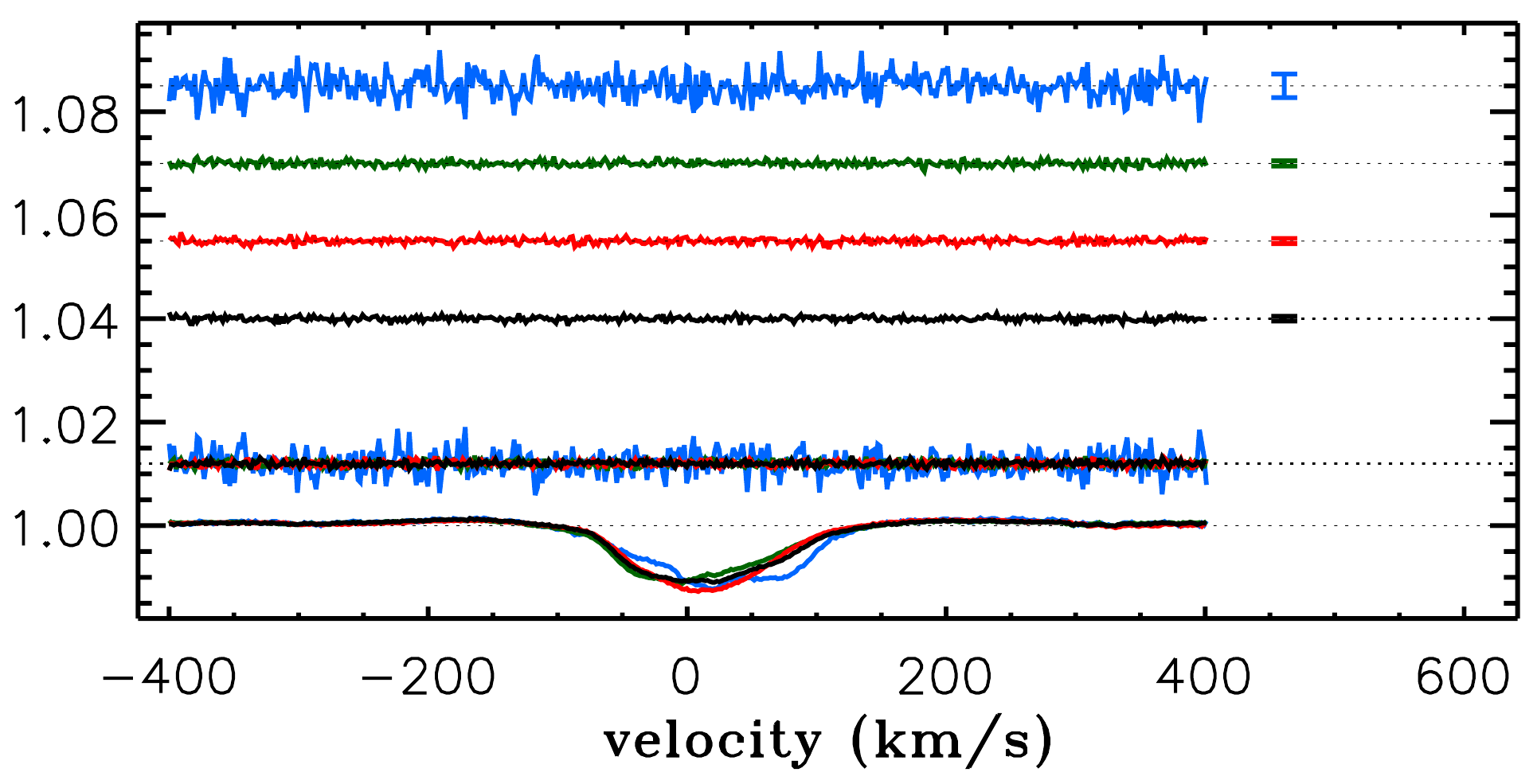}  
 \caption{LSD profiles for CQ Tau (no magnetic detection).}
\end{figure}

\begin{figure}[ht!]
 \centering
 \includegraphics[width=0.45\textwidth,clip]{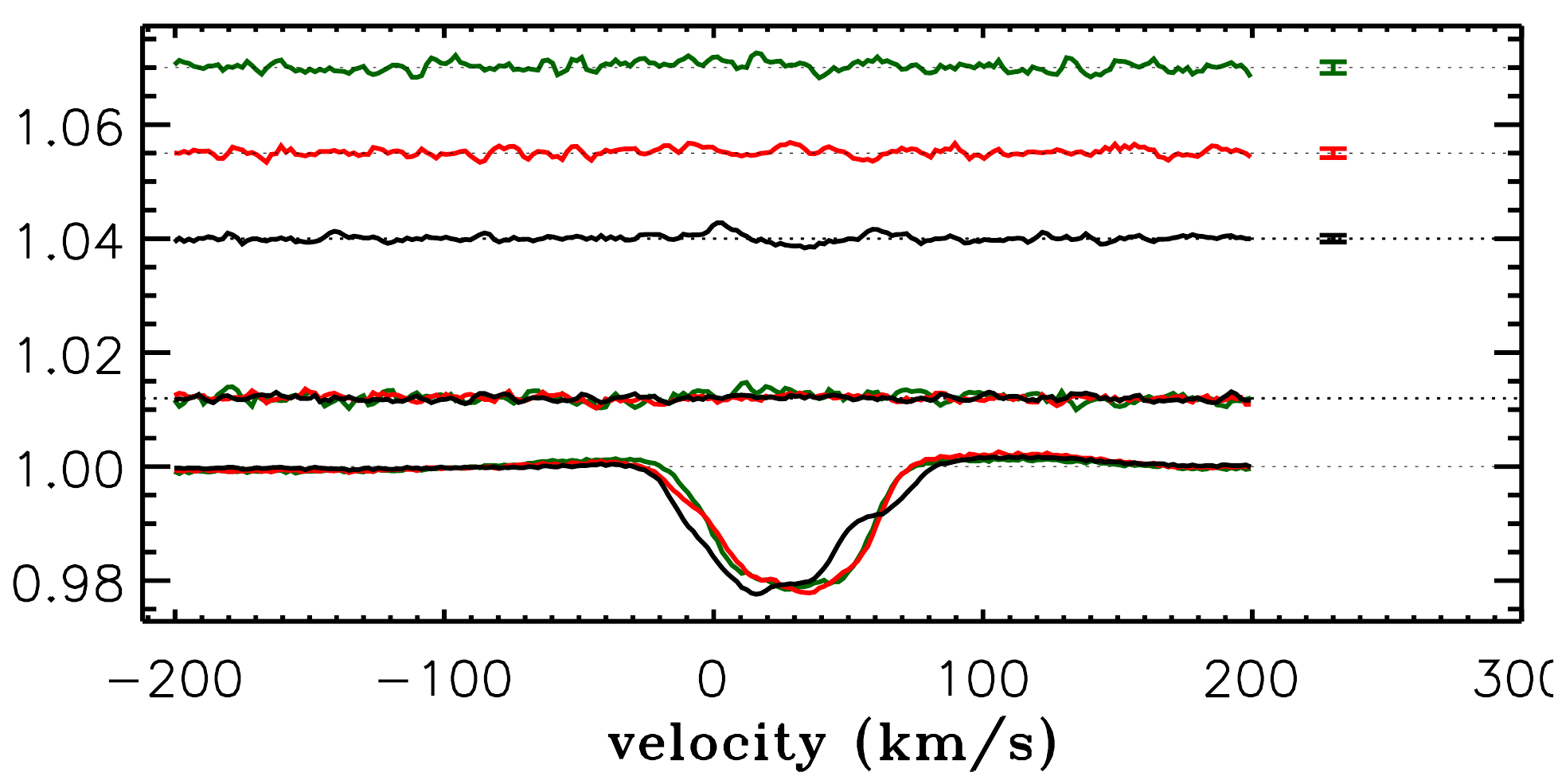}  
 \caption{LSD profiles for GW Ori (magnetic field detected, uncertain topology).}
 \label{fig:gwori}
\end{figure}

\begin{figure}[ht!]
 \centering
 \includegraphics[width=0.45\textwidth,clip]{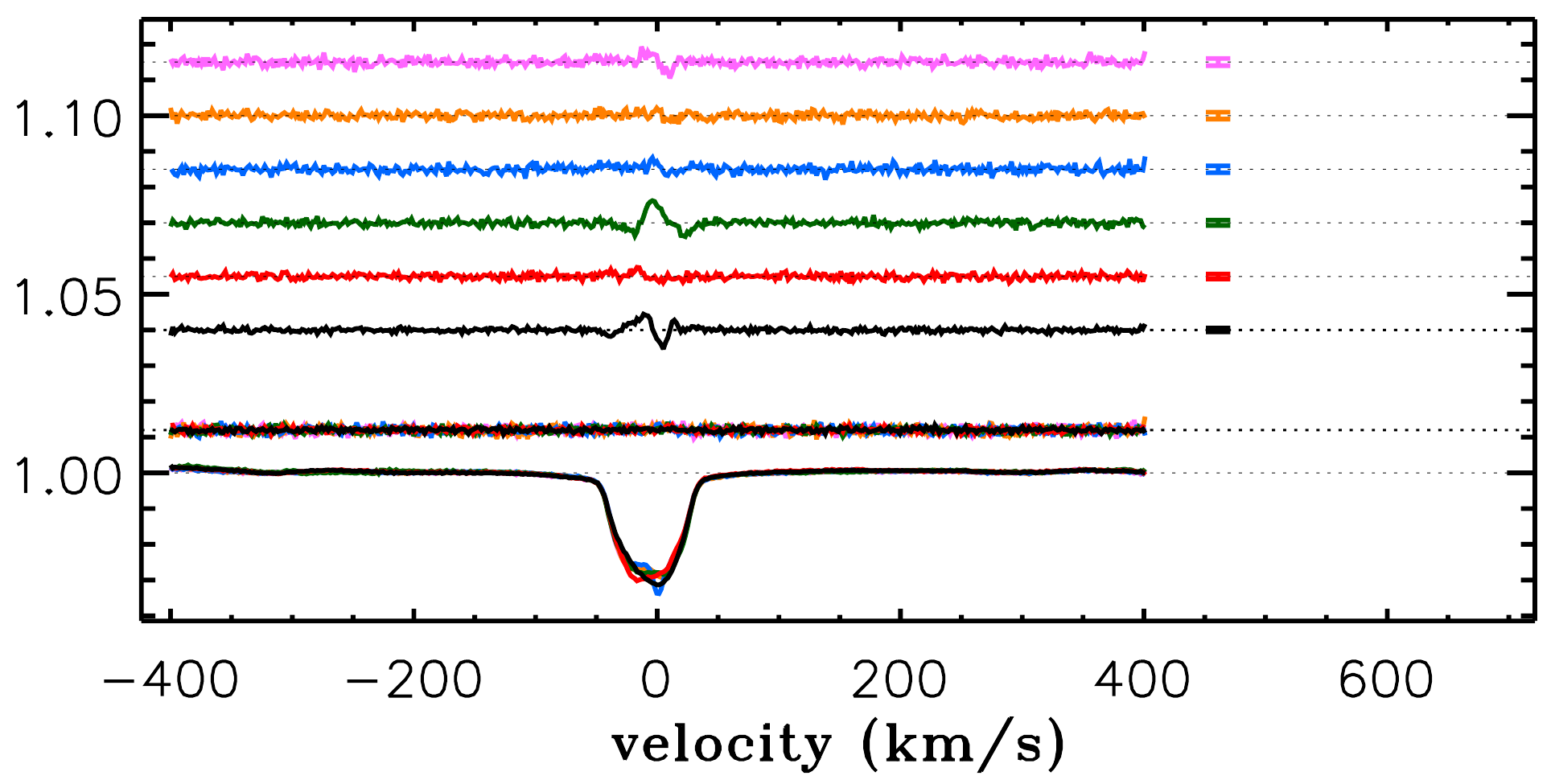}  
 \caption{LSD profiles for HBC 741 (magnetic field detected, complex topology).}
\end{figure}

\begin{figure}[ht!]
 \centering
 \includegraphics[width=0.45\textwidth,clip]{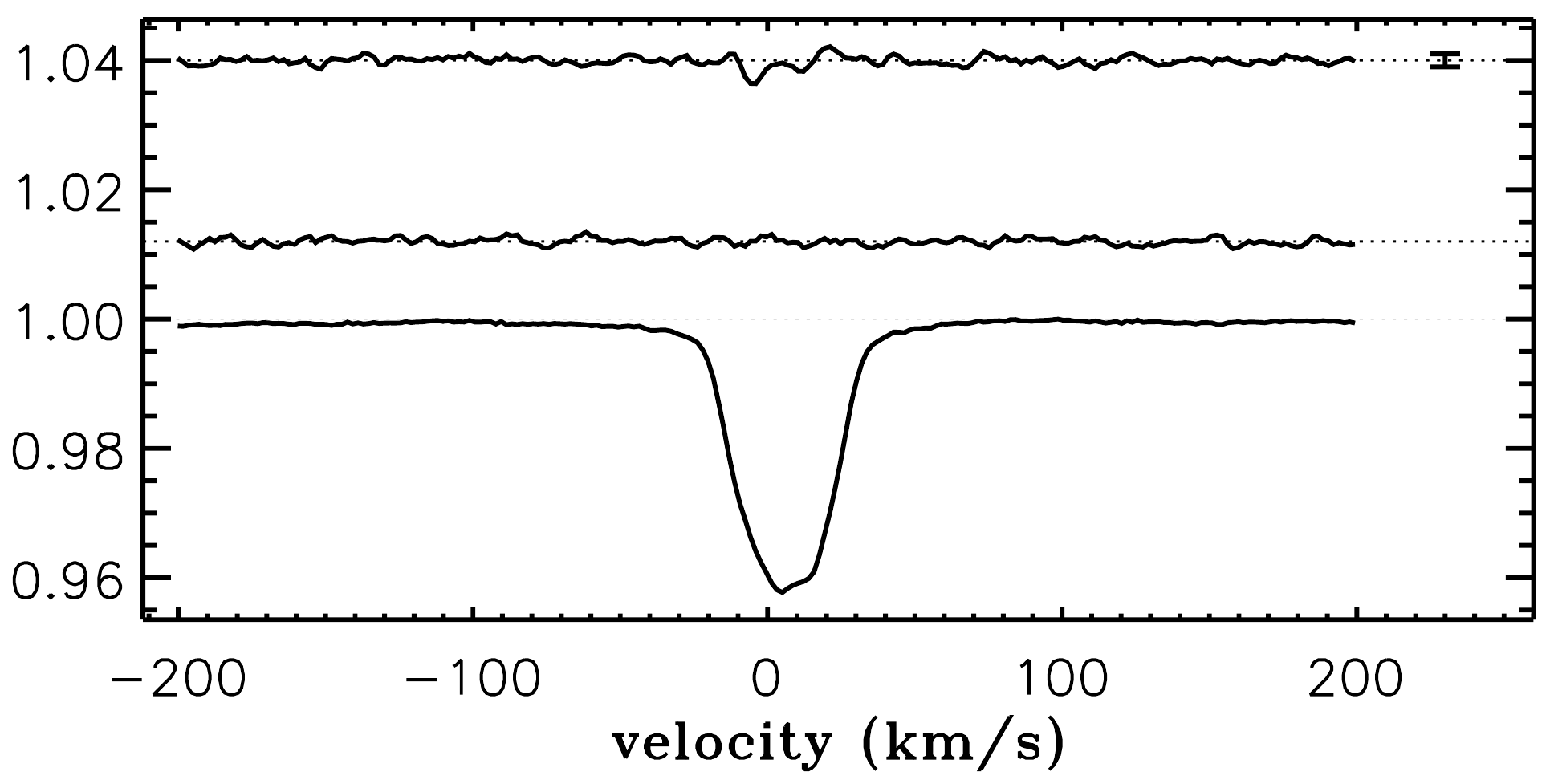}  
 \caption{LSD profiles for HD 133938 (magnetic field detected, complex topology).}
\end{figure}

\begin{figure}[ht!]
 \centering
 \includegraphics[width=0.45\textwidth,clip]{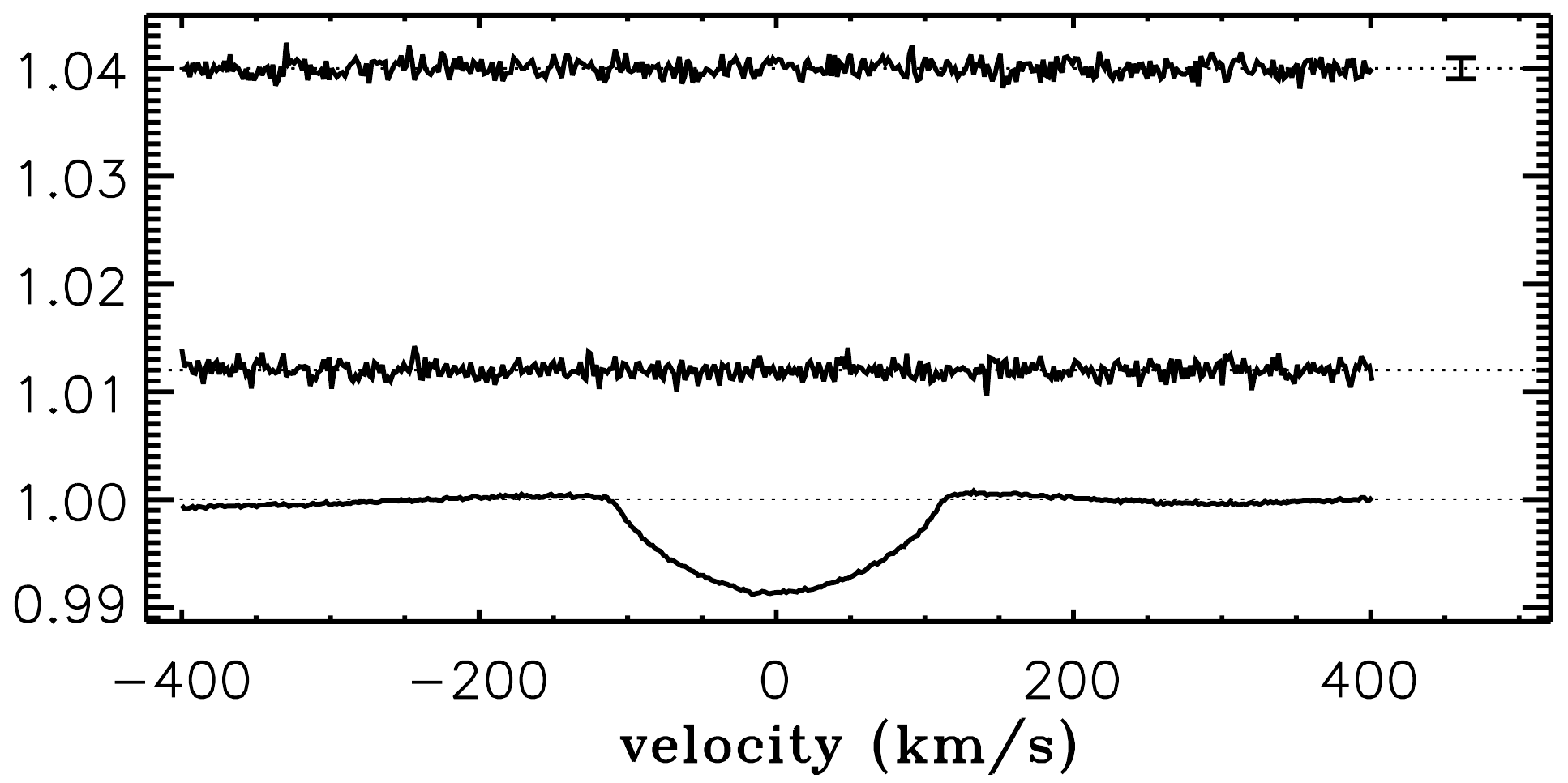}  
 \caption{LSD profiles for HD 135127 (no magnetic detection).}
\end{figure}

\begin{figure}[ht!]
 \centering
 \includegraphics[width=0.45\textwidth,clip]{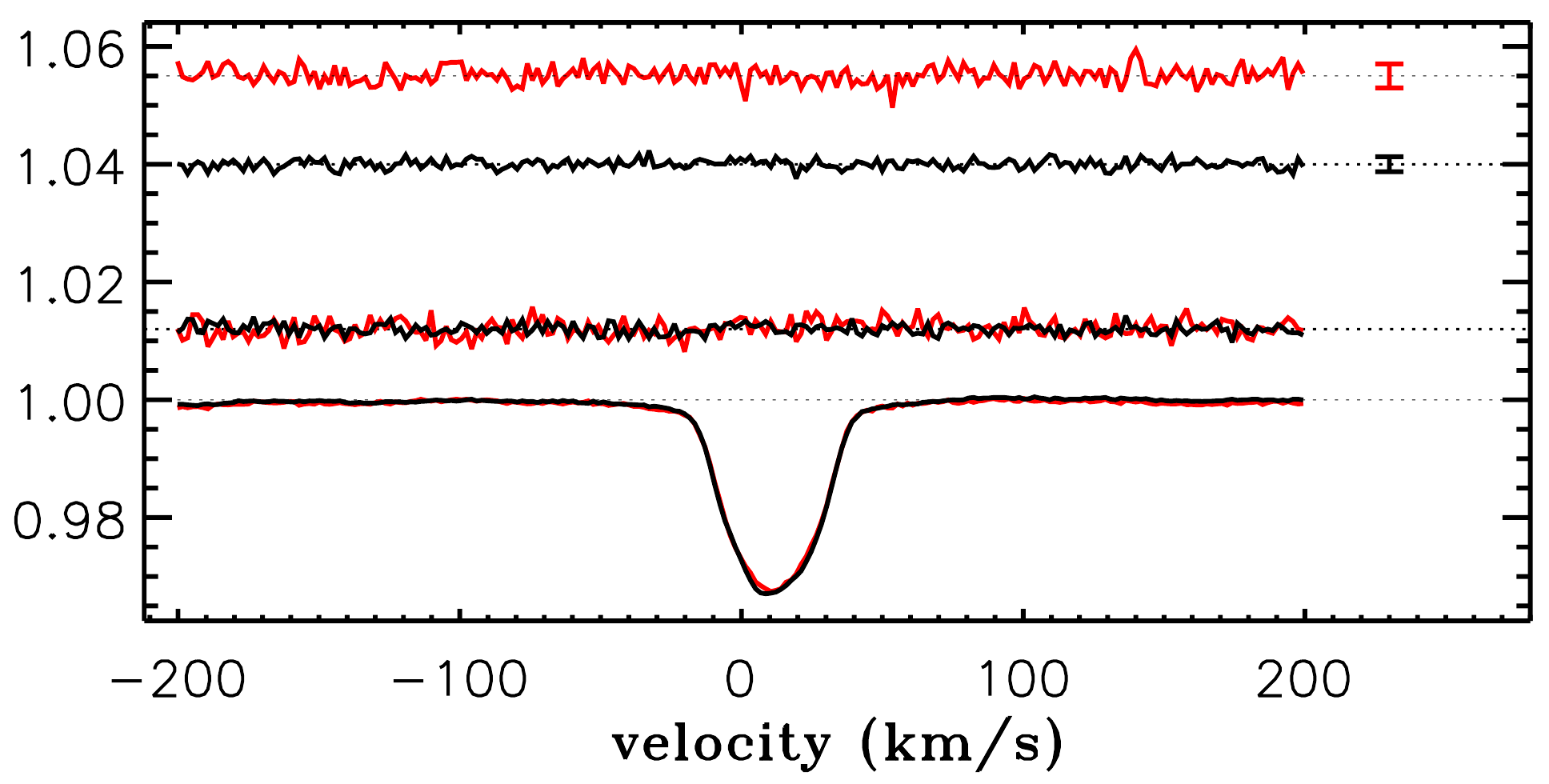}  
 \caption{LSD profiles for HD 137059 (no magnetic detection).}
\end{figure}

\begin{figure}[ht!]
 \centering
 \includegraphics[width=0.45\textwidth,clip]{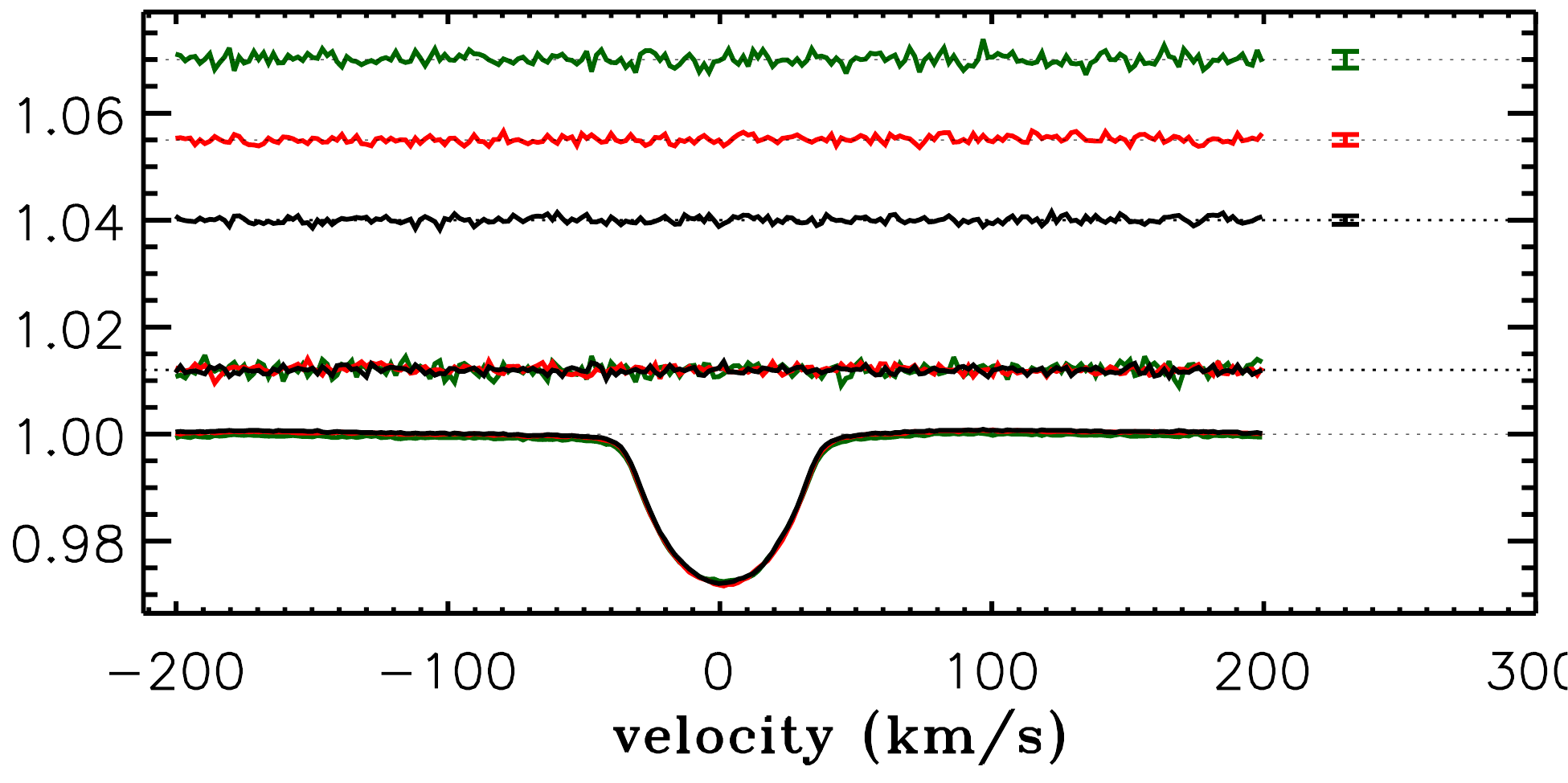}  
 \caption{LSD profiles for HD 143978 (no magnetic detection).}
\end{figure}

\begin{figure}[ht!]
 \centering
 \includegraphics[width=0.45\textwidth,clip]{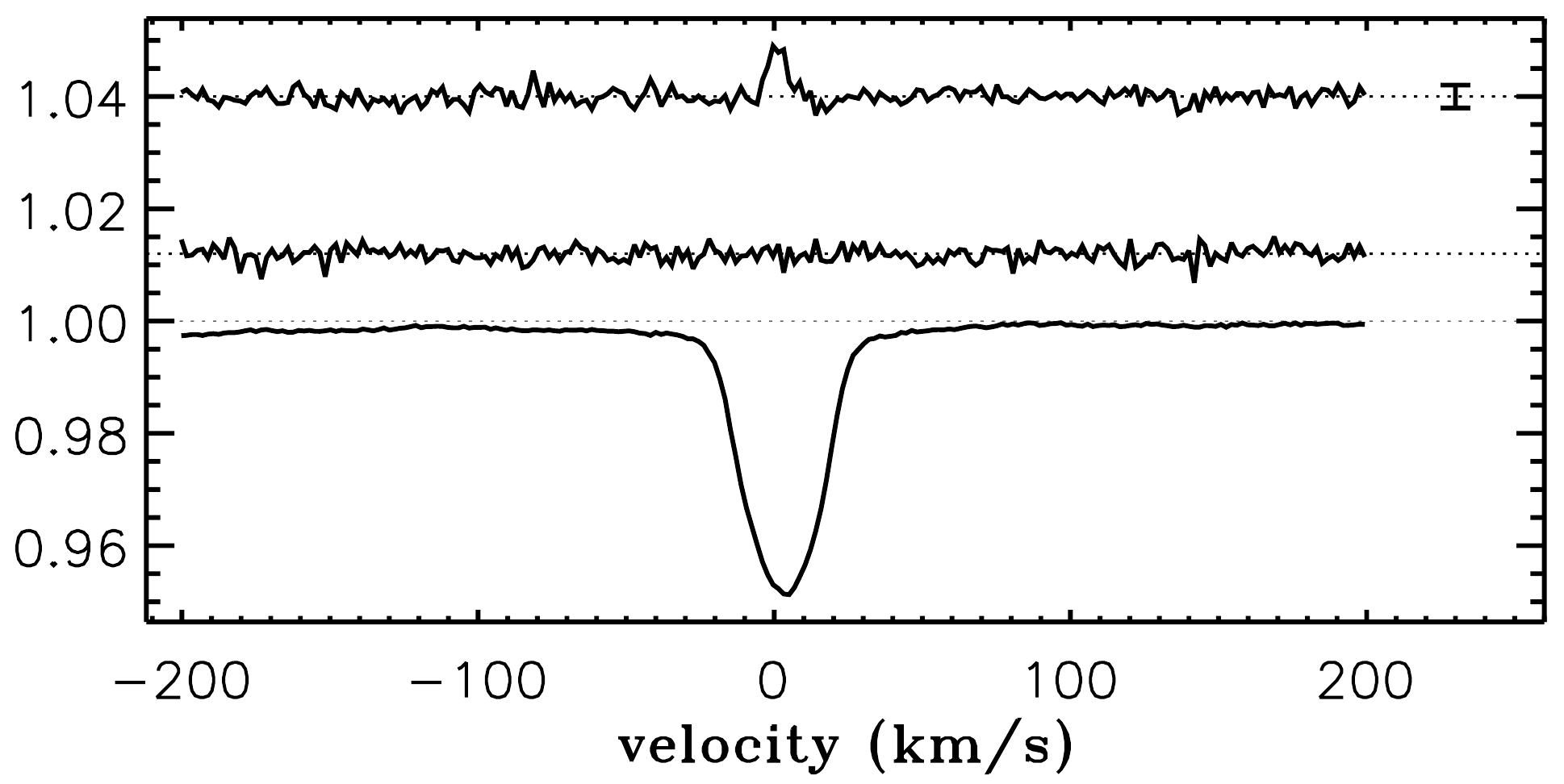}  
 \caption{LSD profiles for HD 147048 (magnetic field detected, uncertain topology).}
\end{figure}

\begin{figure}[ht!]
 \centering
 \includegraphics[width=0.45\textwidth,clip]{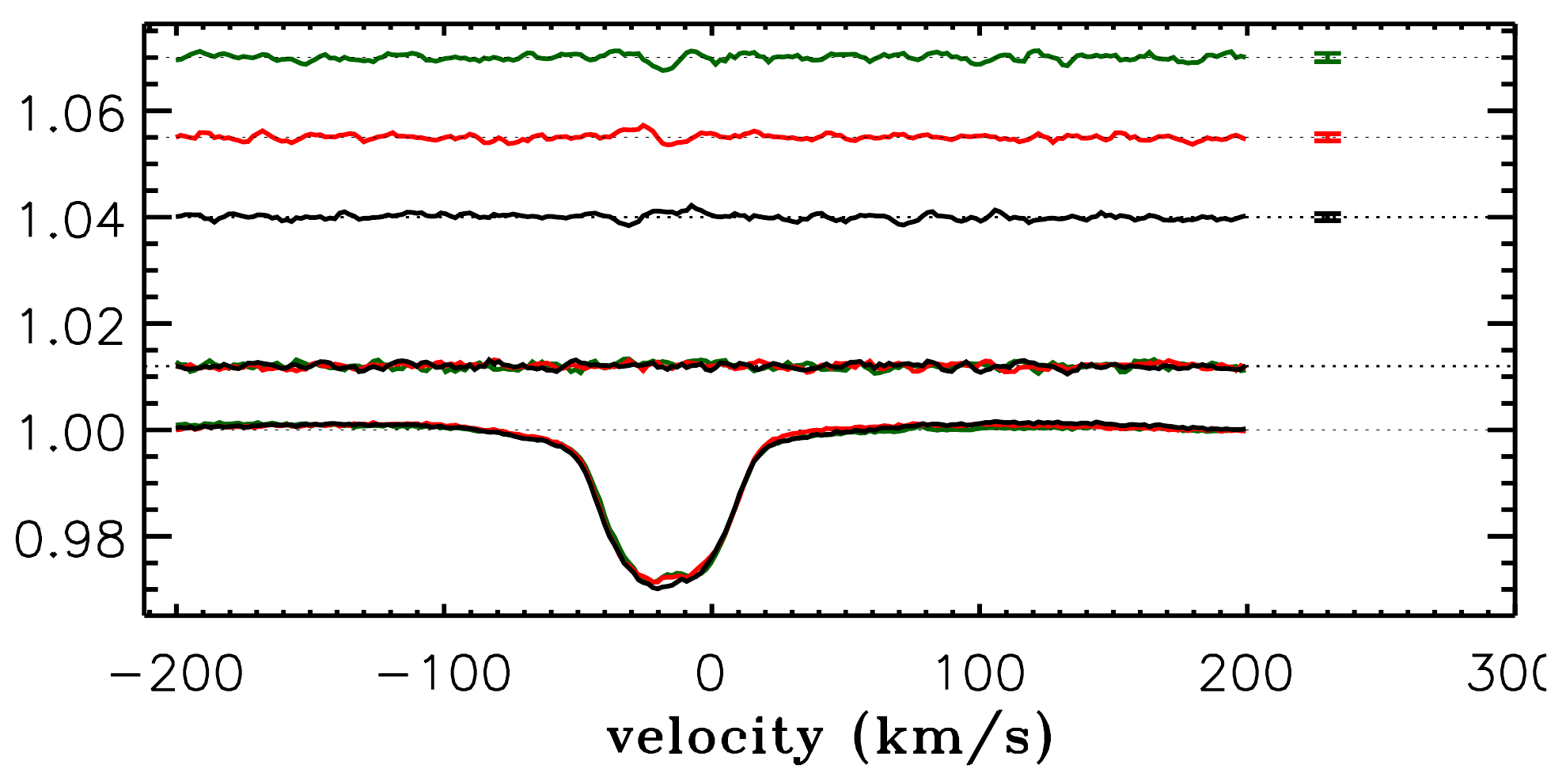}  
 \caption{LSD profiles for IRAS 22144 + 6923 (magnetic field detected, complex topology).}
\end{figure}

\begin{figure}[ht!]
 \centering
 \includegraphics[width=0.45\textwidth,clip]{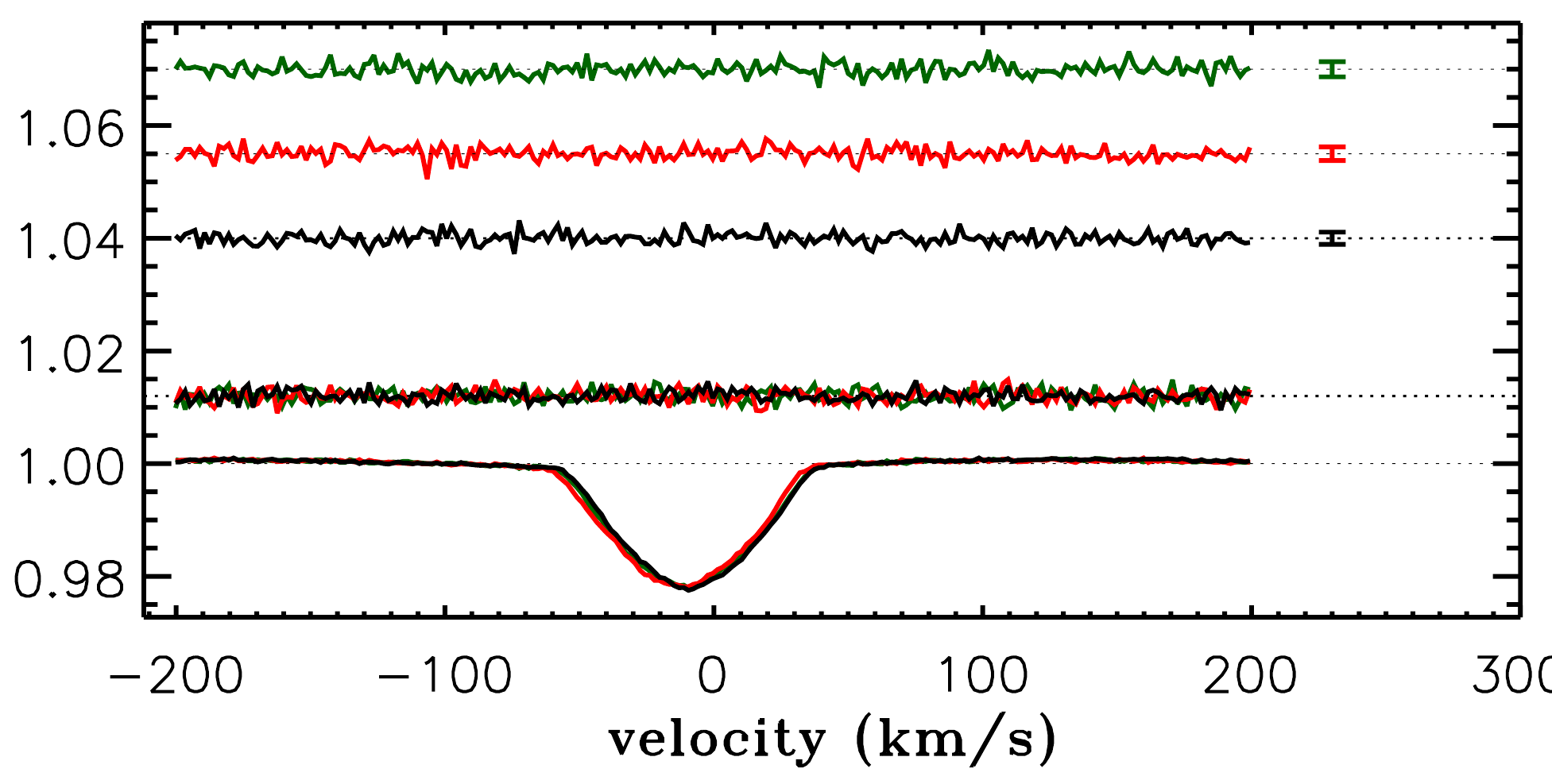}  
 \caption{LSD profiles for IRAS 22152 + 6947 (no magnetic detection).}
\end{figure}

\begin{figure}[ht!]
 \centering
 \includegraphics[width=0.45\textwidth,clip]{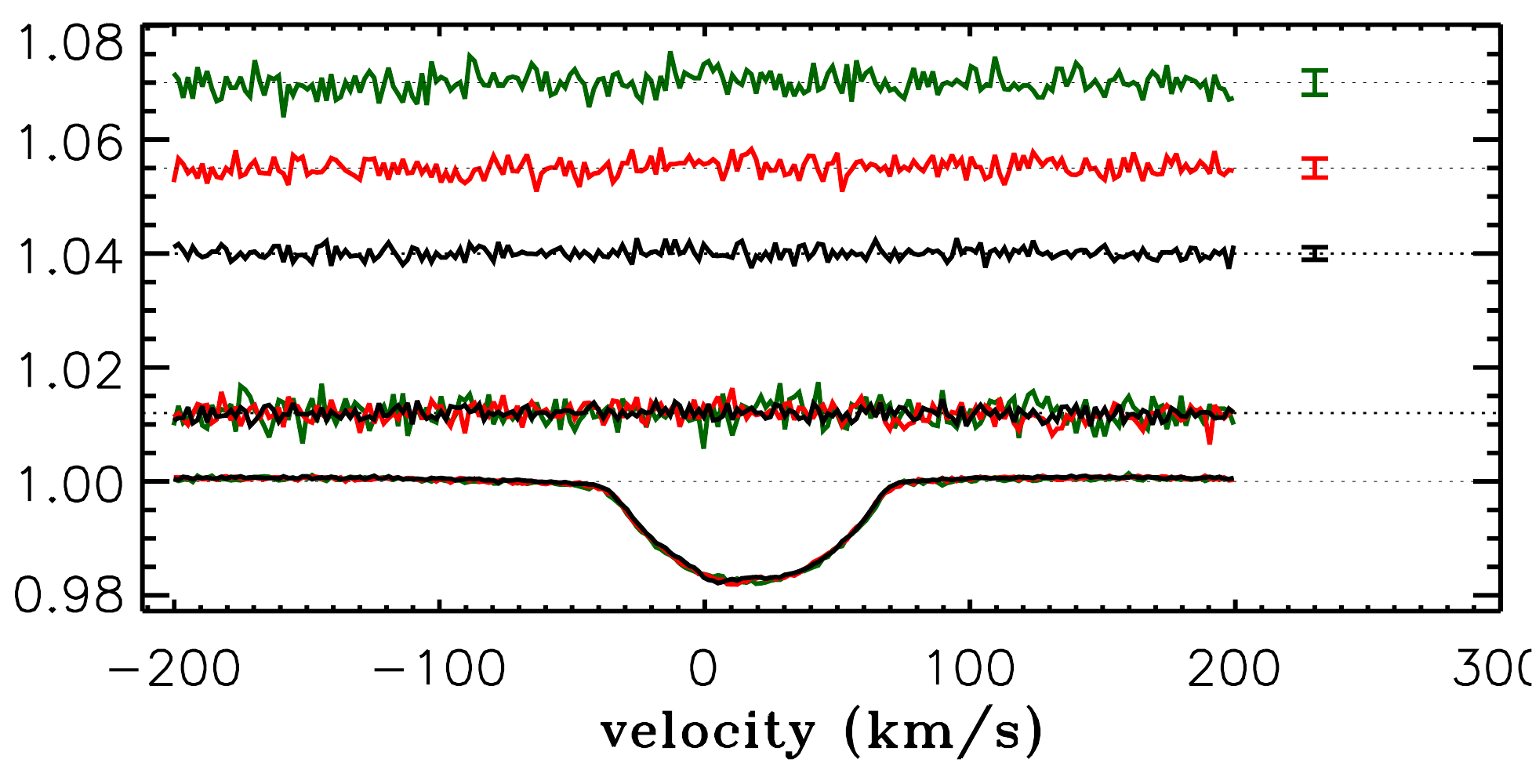}  
 \caption{LSD profiles for NGC 2264 108 (no magnetic detection).}
\end{figure}

\begin{figure}[ht!]
 \centering
 \includegraphics[width=0.45\textwidth,clip]{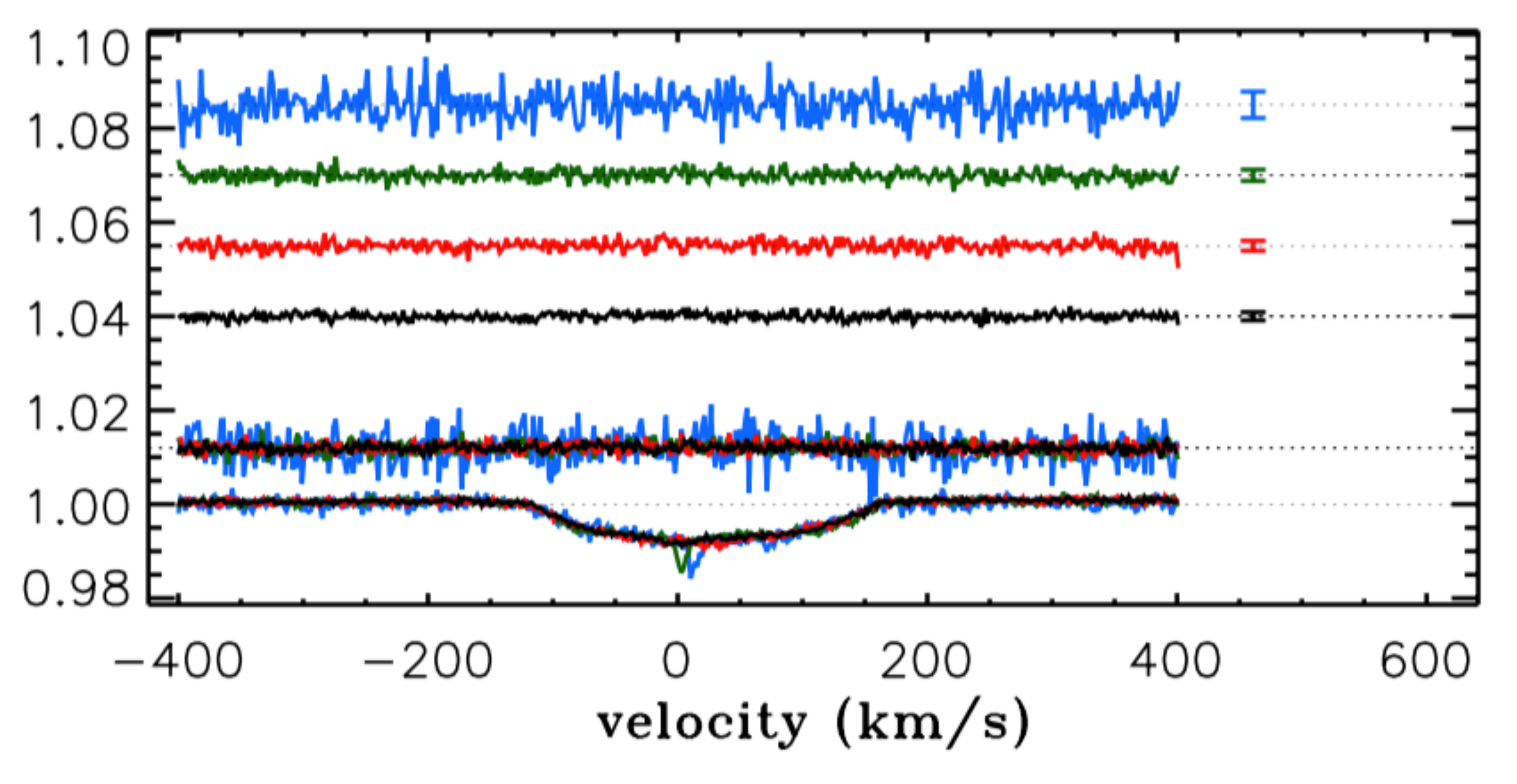}  
 \caption{LSD profiles for NGC 2264 121 (no magnetic detection).}
\end{figure}

\begin{figure}[ht!]
 \centering
 \includegraphics[width=0.45\textwidth,clip]{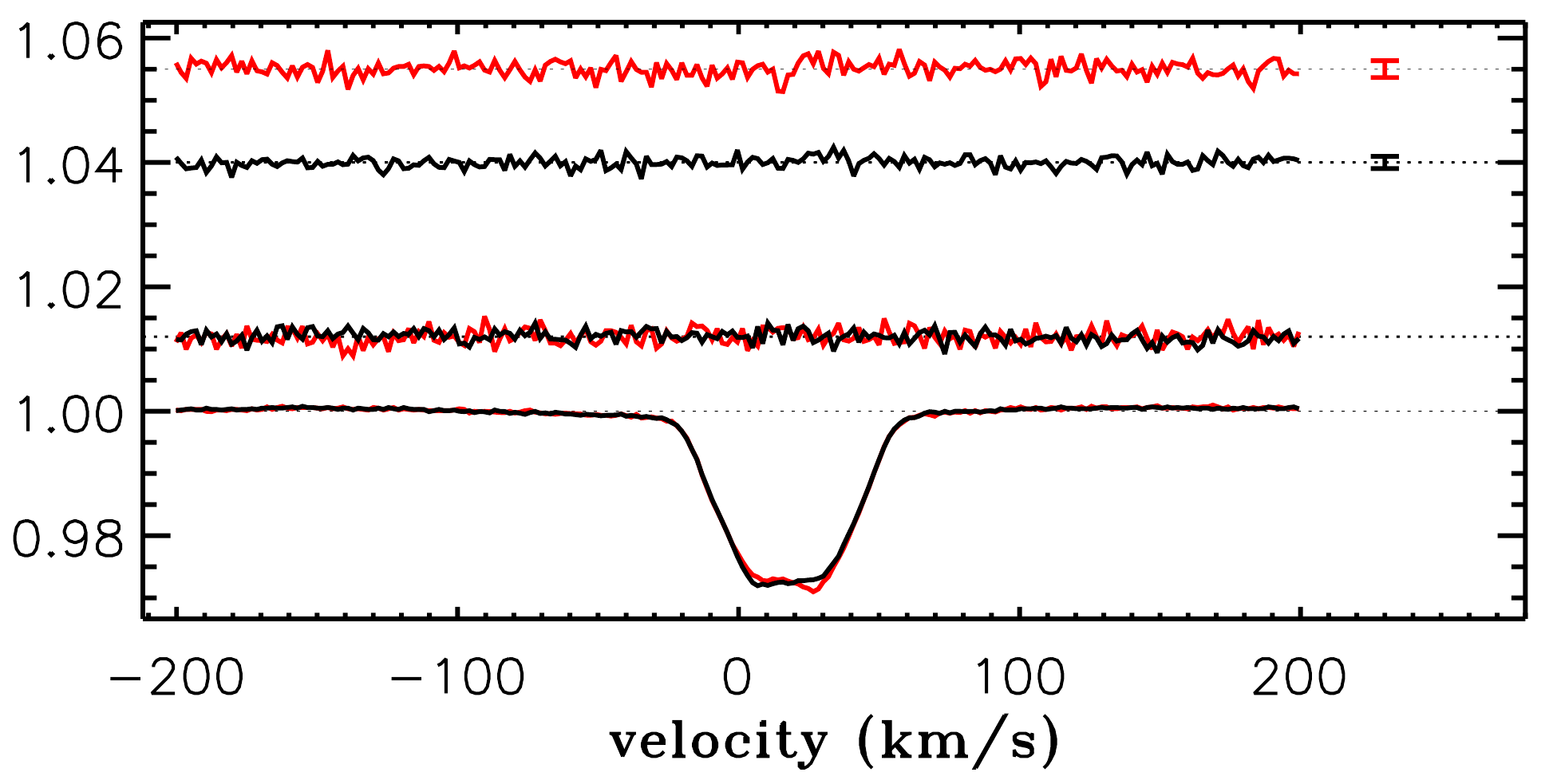}  
 \caption{LSD profiles for NGC 2264 84 (no magnetic detection).}
\end{figure}

\begin{figure}[ht!]
\centering
\includegraphics[width=0.45\textwidth,clip]{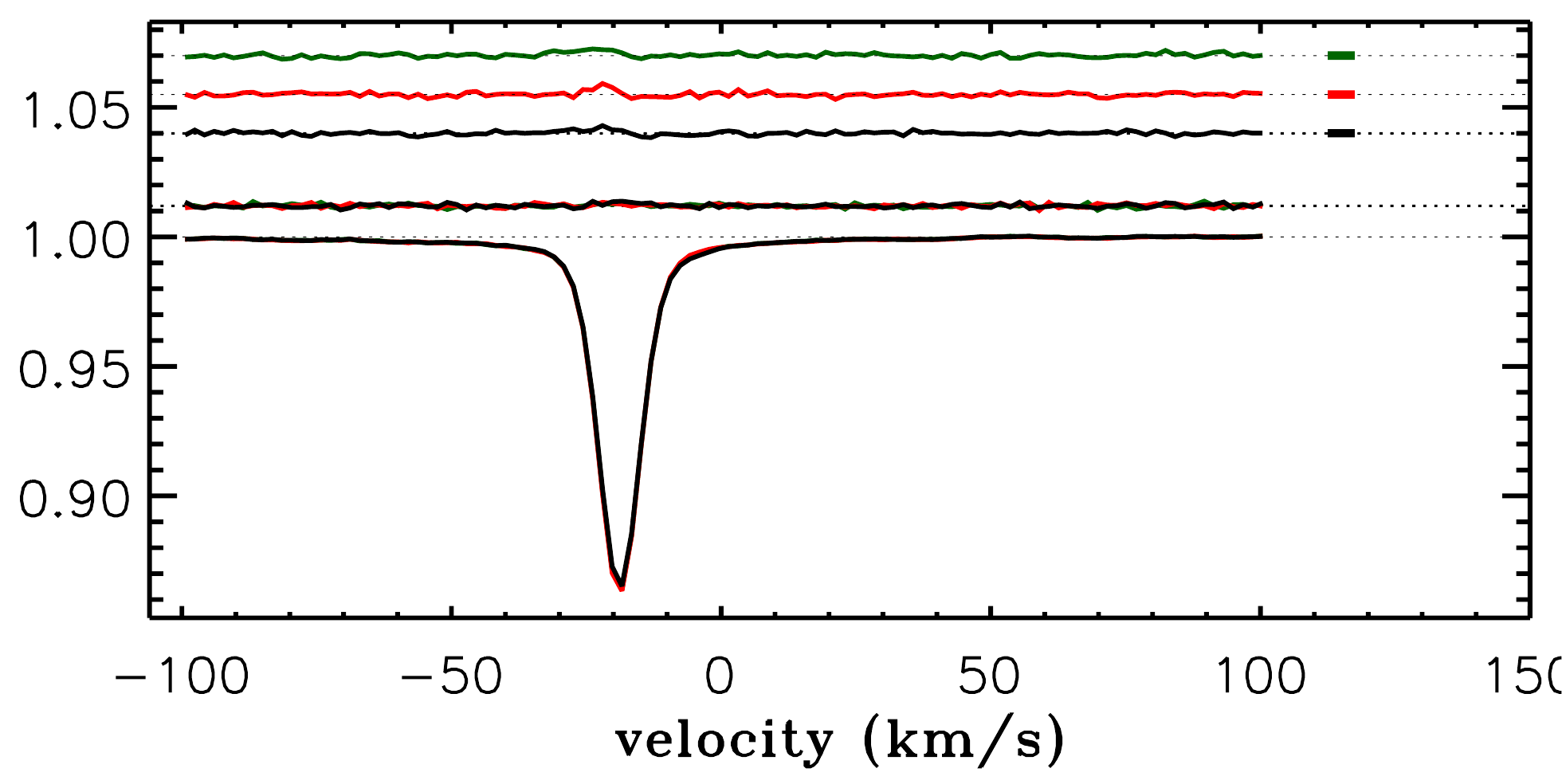}  
\caption{LSD profiles for NGC 6530 SCB 7 (magnetic field detected, uncertain topology).}
\end{figure}

\begin{figure}[ht!]
\centering
\includegraphics[width=0.45\textwidth,clip]{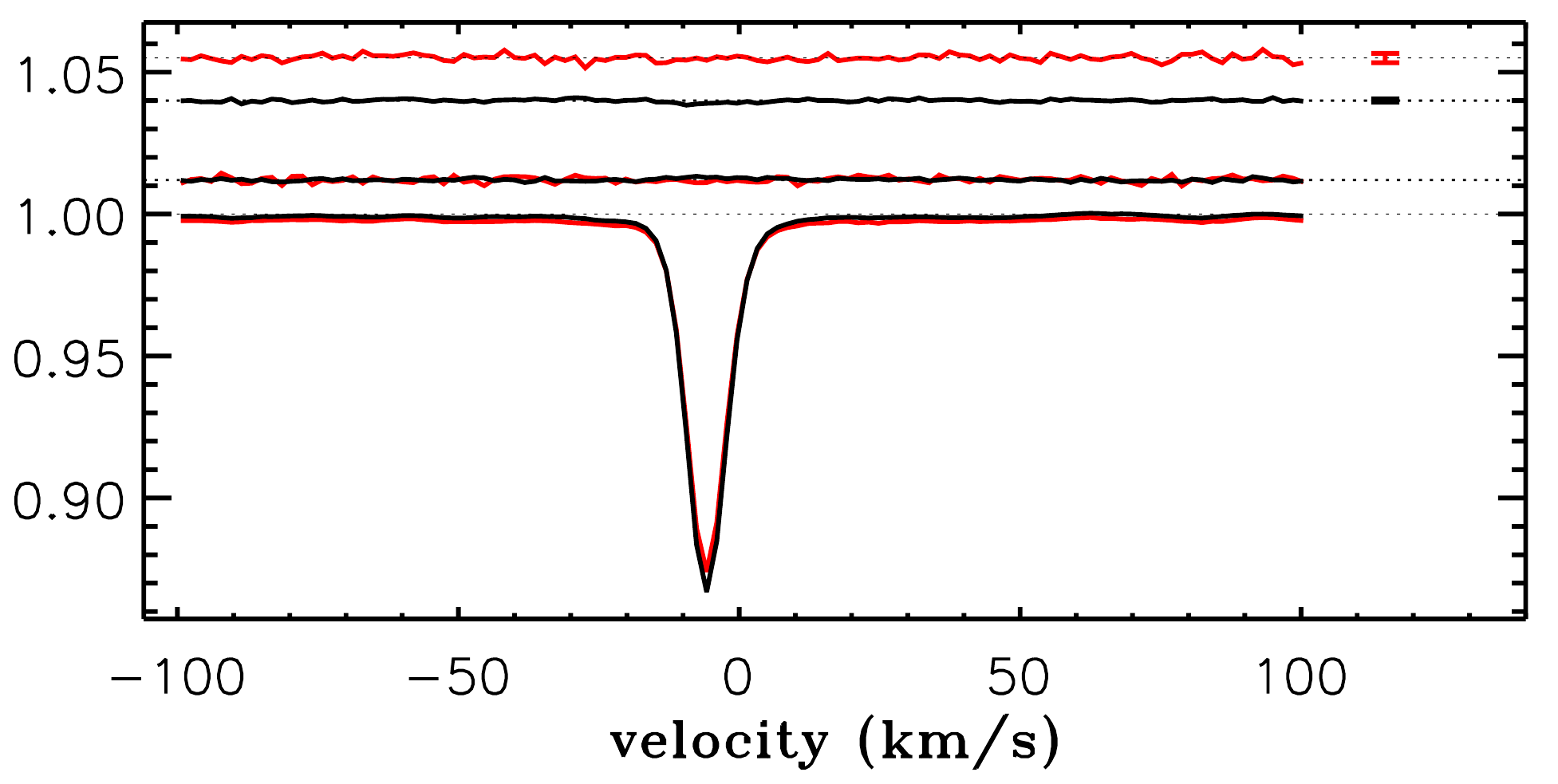}  
\caption{LSD profiles for NGC 6530 SCB 739 (no magnetic detection).}
\end{figure}

\begin{figure}[ht!]
\centering
\includegraphics[width=0.45\textwidth,clip]{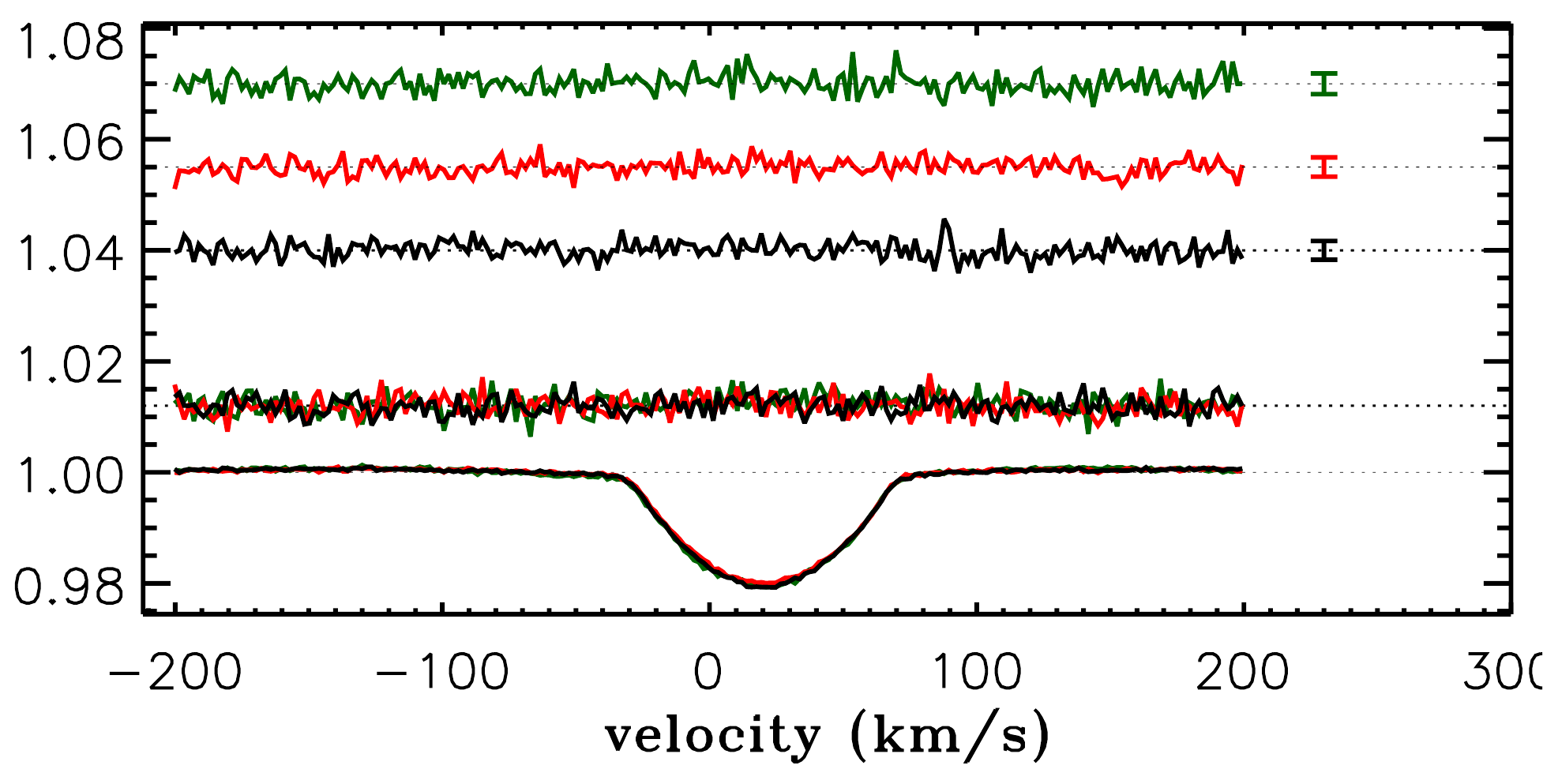}  
\caption{LSD profiles for Par 102 (no magnetic detection).}
\end{figure}

\begin{figure}[ht!]
\centering
\includegraphics[width=0.45\textwidth,clip]{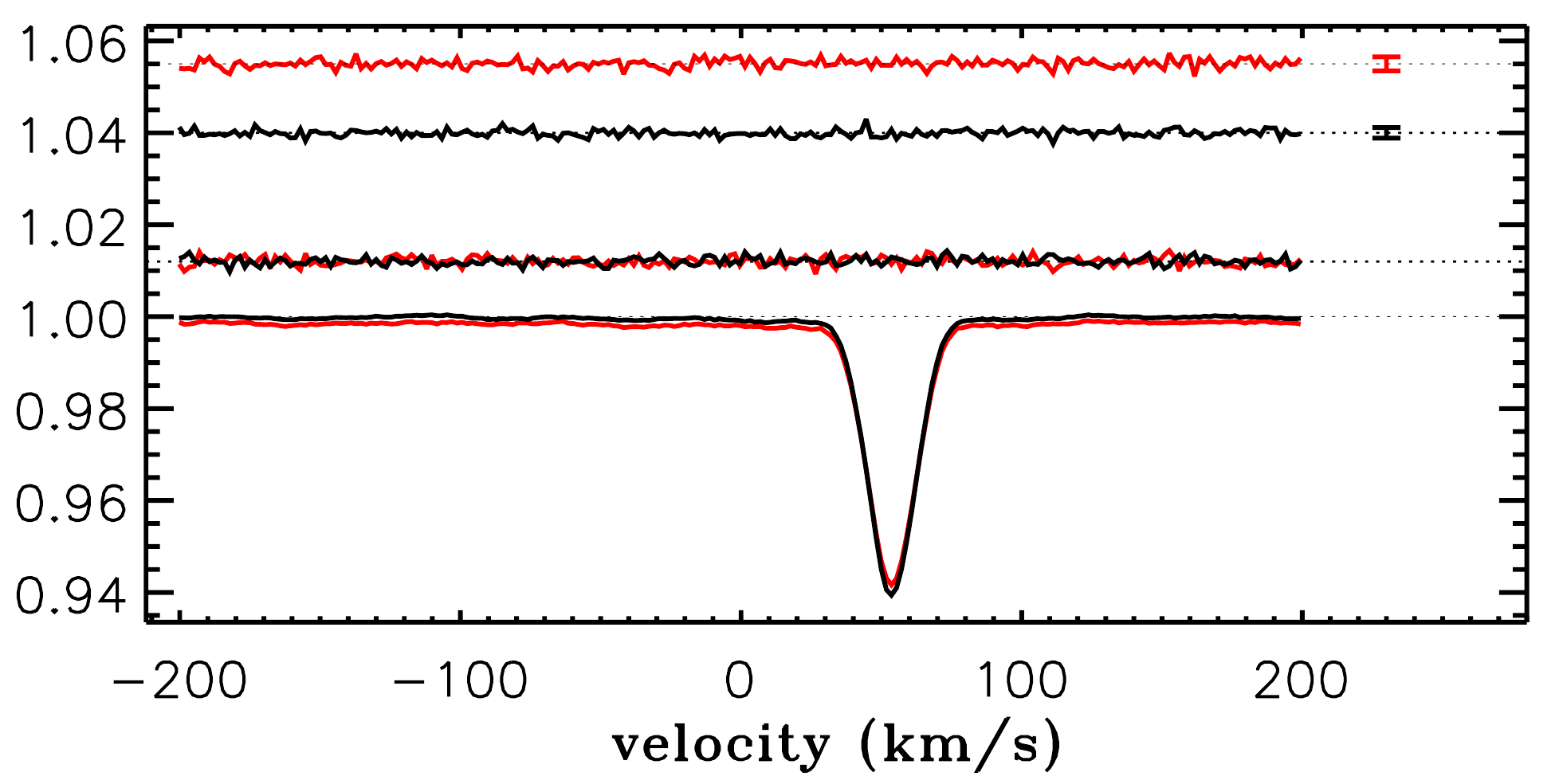}  
\caption{LSD profiles for Par 1391 (no magnetic detection).}
\end{figure}

\begin{figure}[ht!]
\centering
\includegraphics[width=0.45\textwidth,clip]{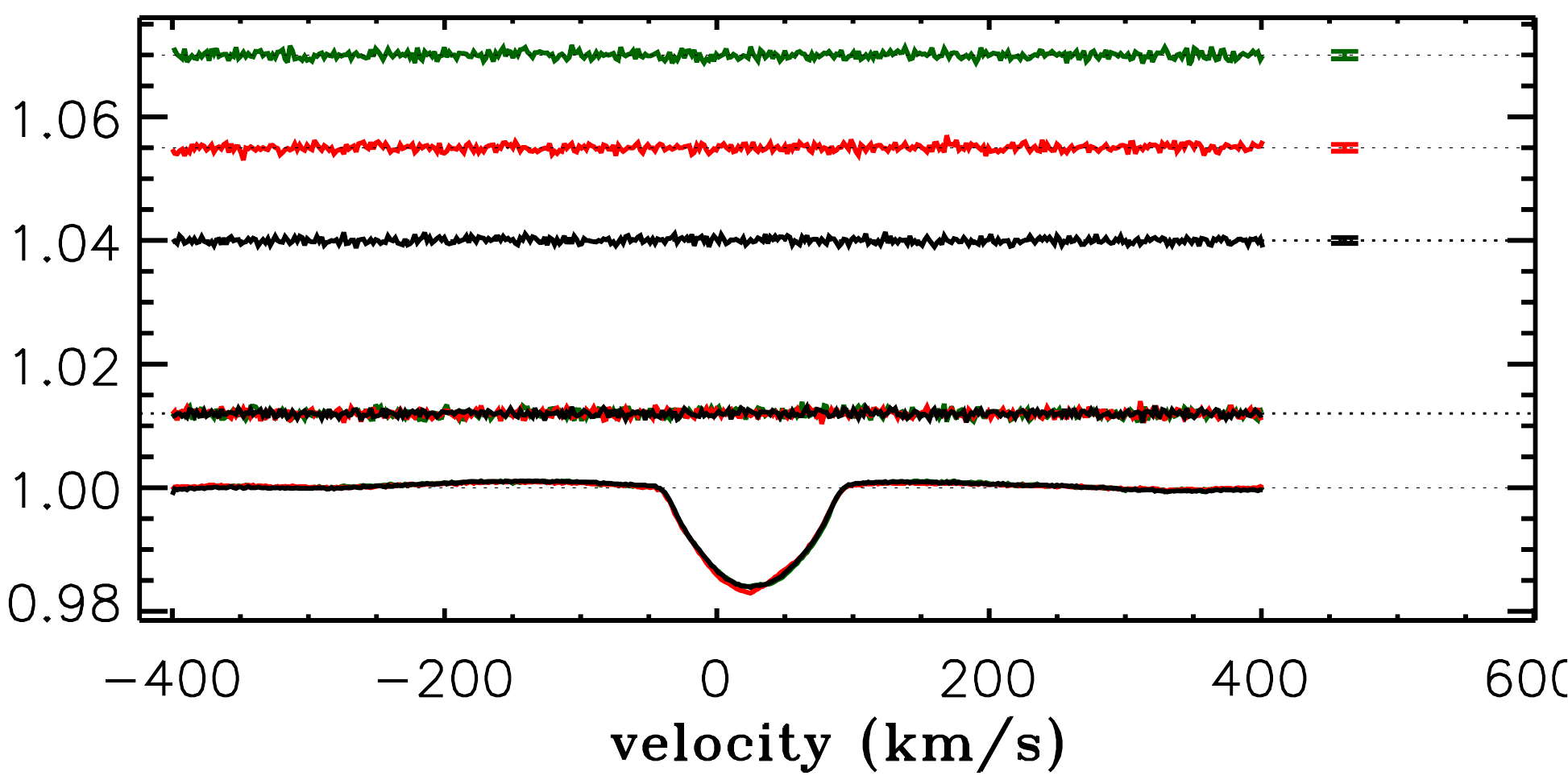}  
\caption{LSD profiles for Par 1394 (no magnetic detection).}
\end{figure}

\begin{figure}[ht!]
\centering
\includegraphics[width=0.45\textwidth,clip]{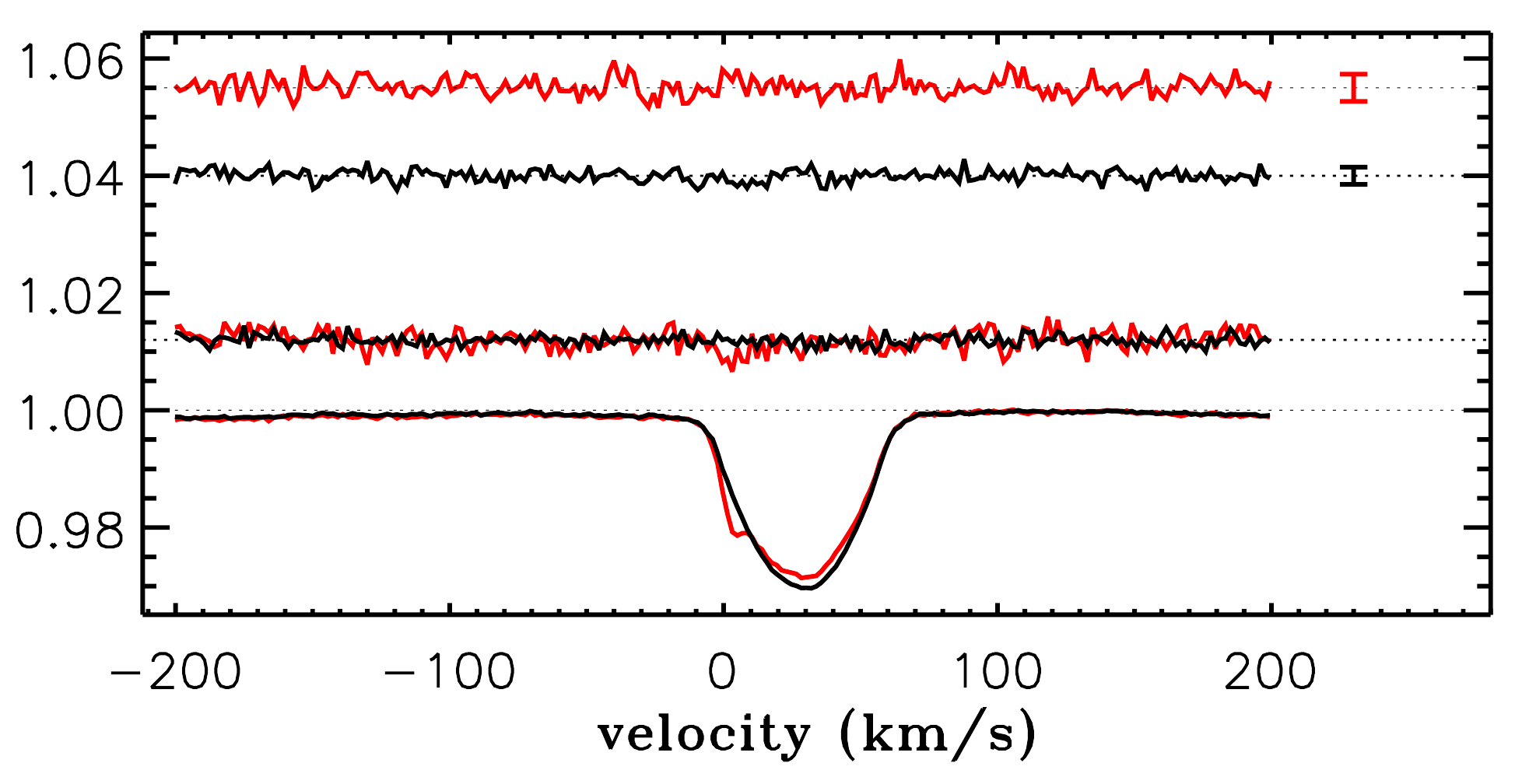}  
\caption{LSD profiles for Par 1414 (no magnetic detection).}
\end{figure}

\begin{figure}[ht!]
\centering
\includegraphics[width=0.45\textwidth,clip]{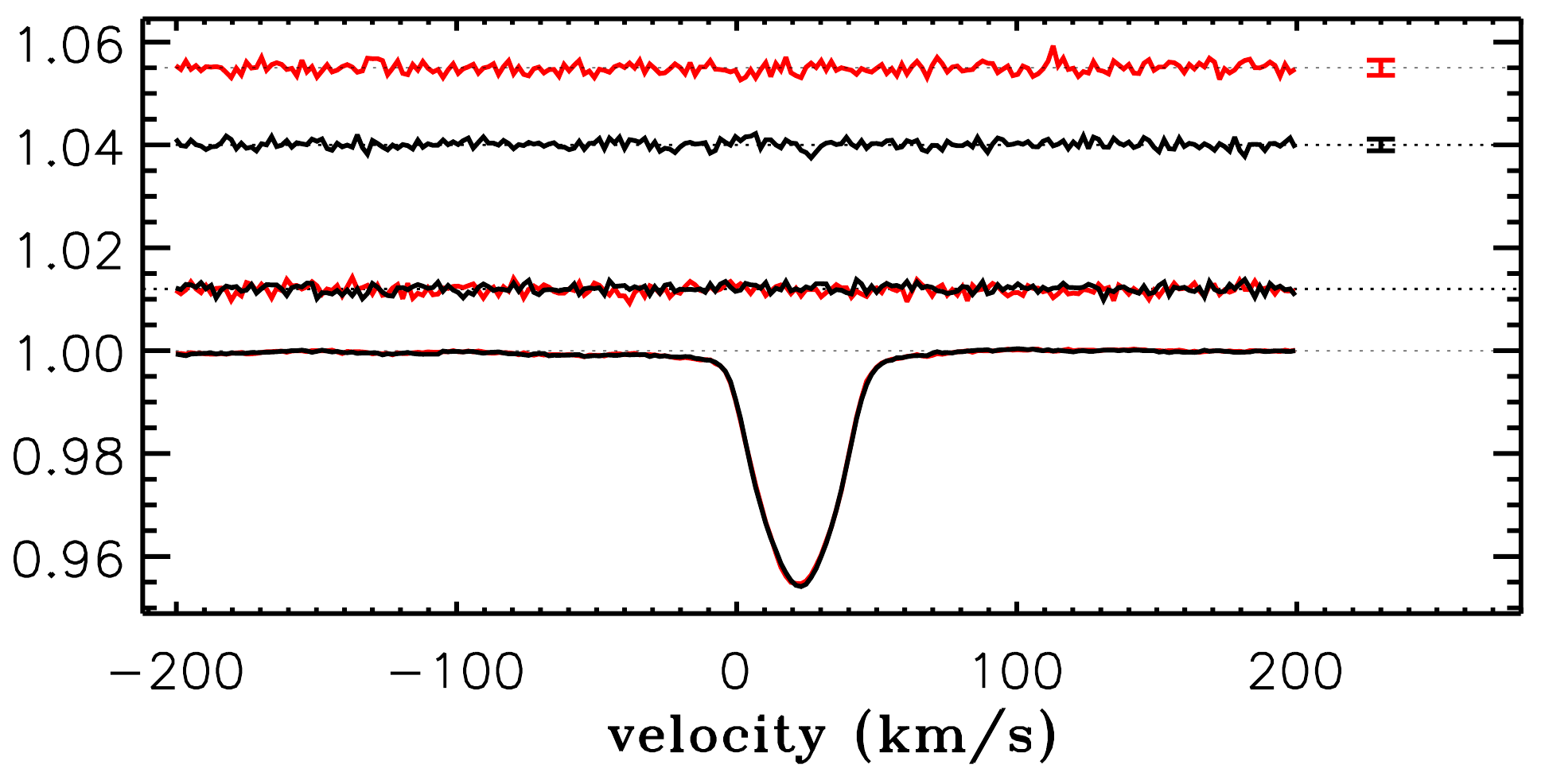}  
\caption{LSD profiles for Par 1455 (no magnetic detection).}
\end{figure}

\begin{figure}[ht!]
\centering
\includegraphics[width=0.45\textwidth,clip]{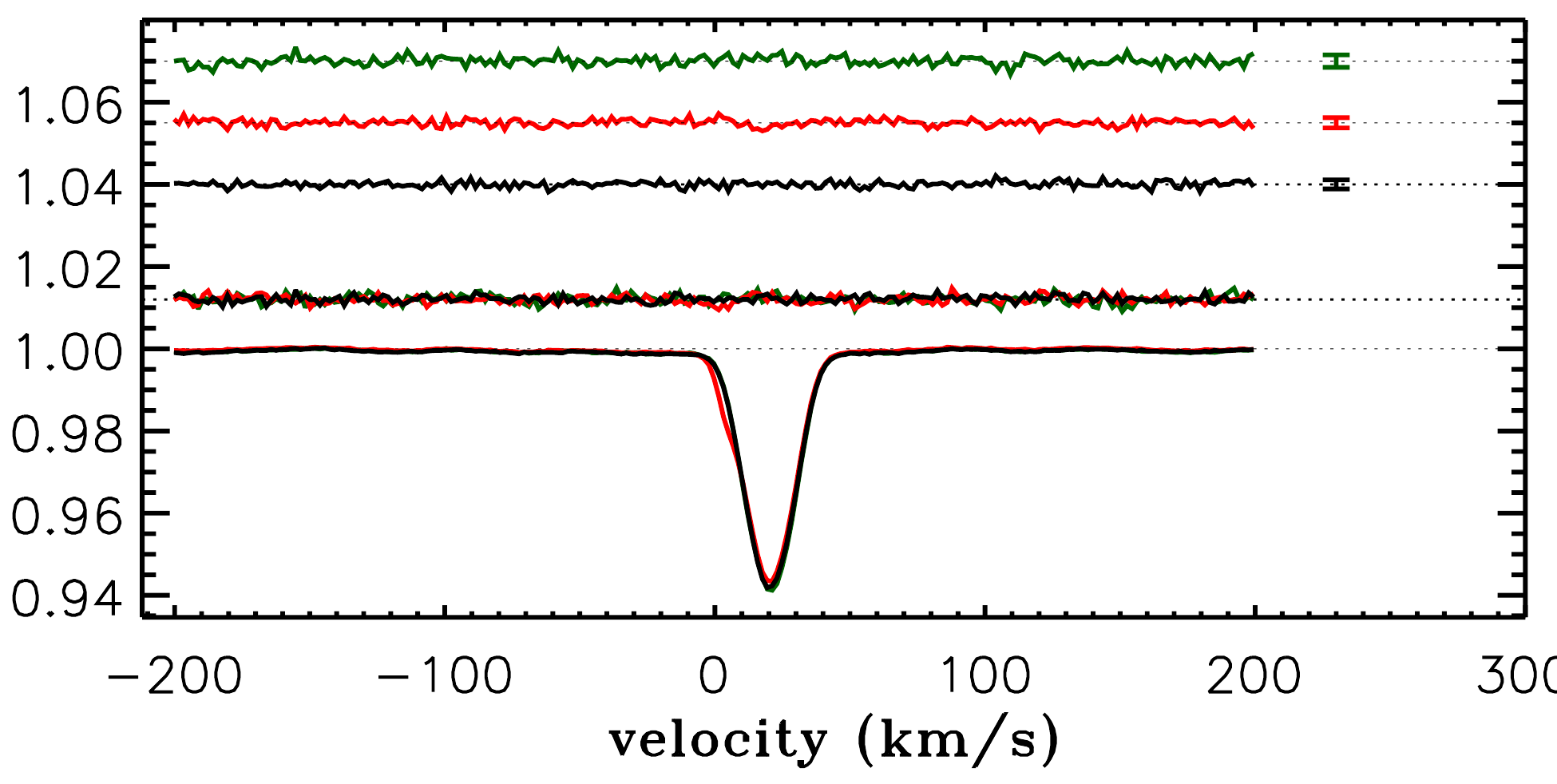}  
\caption{LSD profiles for Par 1646 (no magnetic detection).}
\end{figure}

\begin{figure}[ht!]
\centering
\includegraphics[width=0.45\textwidth,clip]{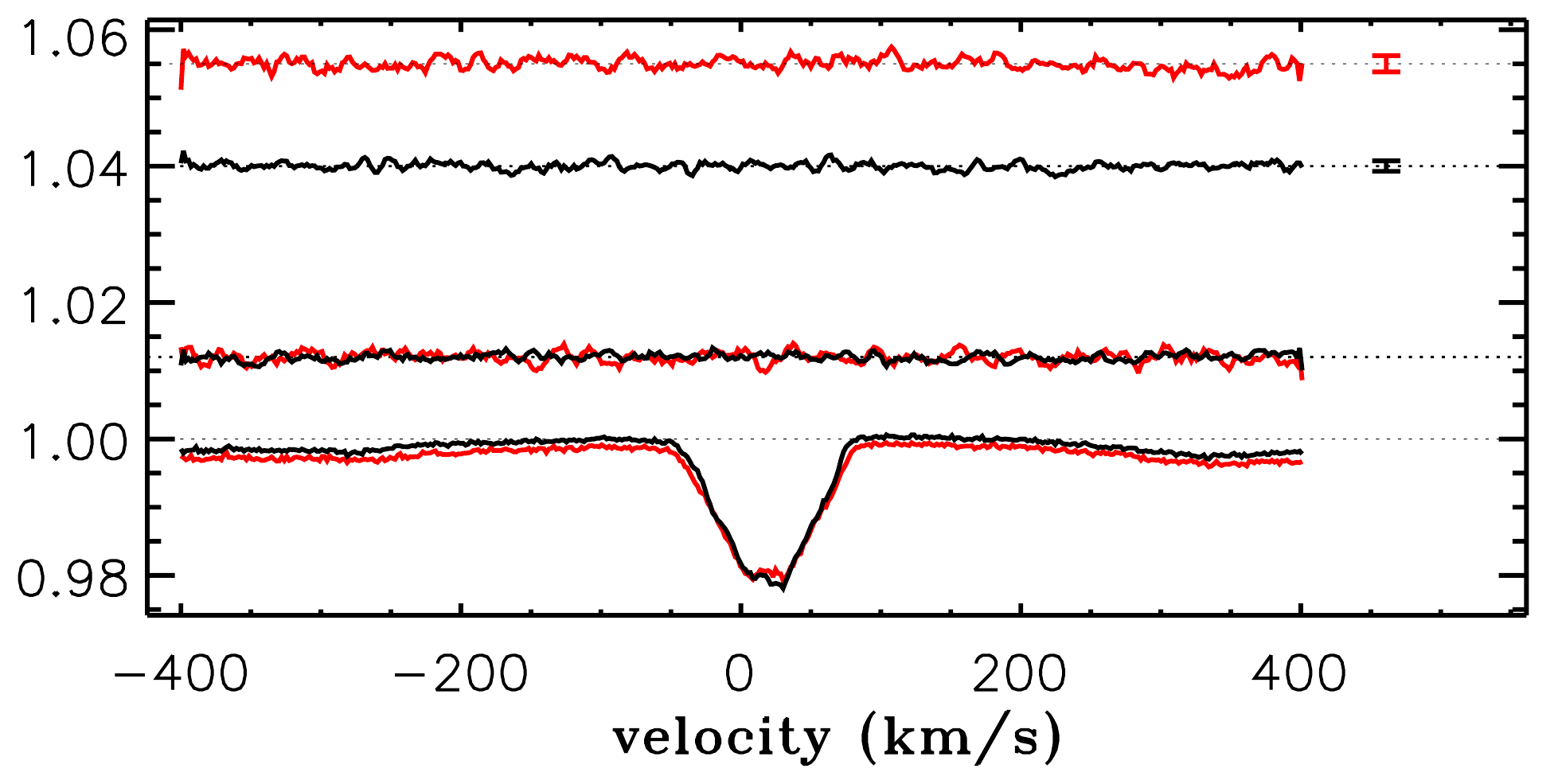}  
\caption{LSD profiles for Par 1736 (no magnetic detection).}
\end{figure}

\begin{figure}[ht!]
\centering
\includegraphics[width=0.45\textwidth,clip]{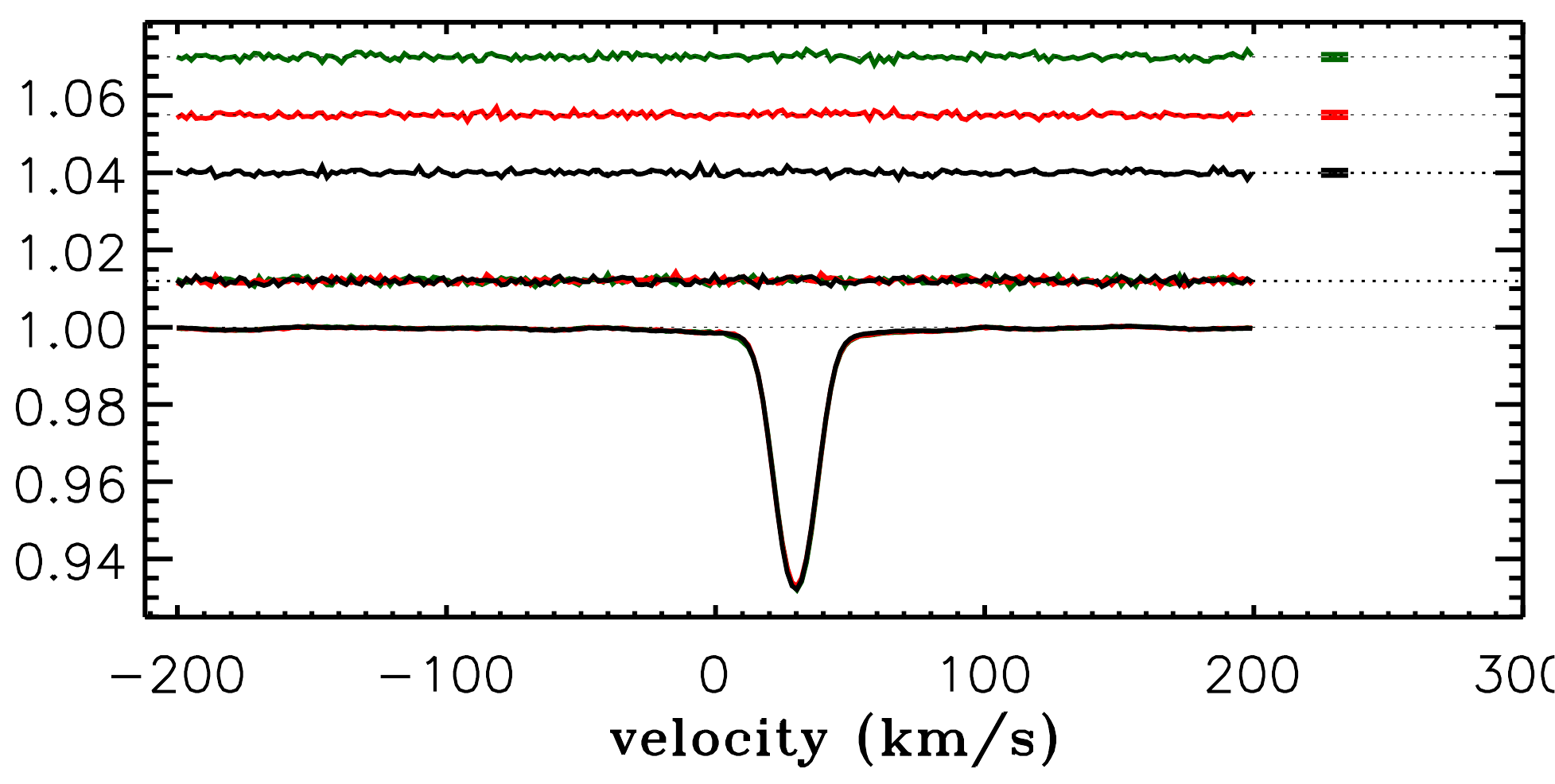}  
\caption{LSD profiles for Par 2441 (no magnetic detection).}
\end{figure}

\begin{figure}[ht!]
\centering
\includegraphics[width=0.45\textwidth,clip]{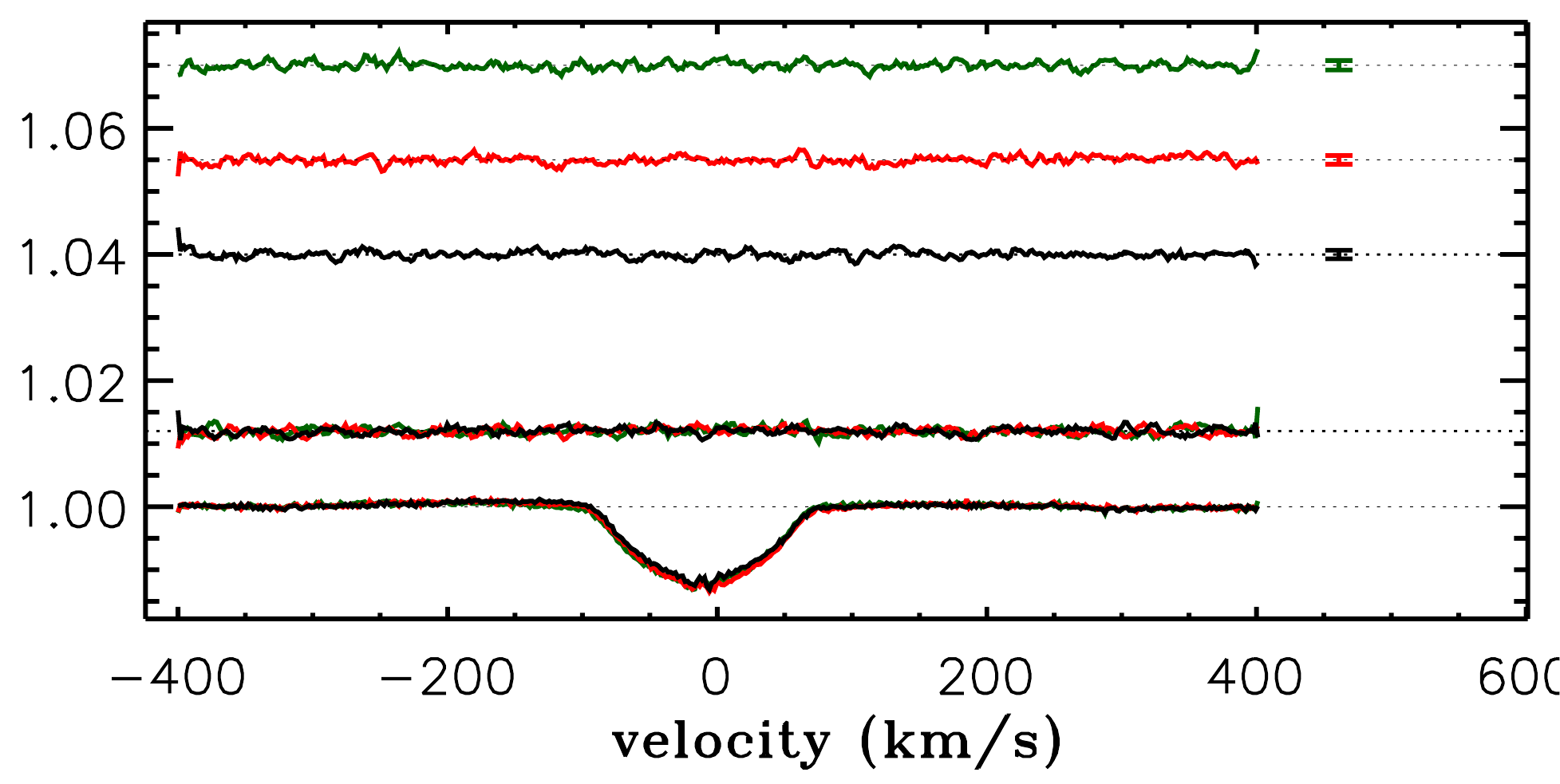}  
\caption{LSD profiles for PX Vul (no magnetic detection).}
\end{figure}

\begin{figure}[ht!]
\centering
\includegraphics[width=0.45\textwidth,clip]{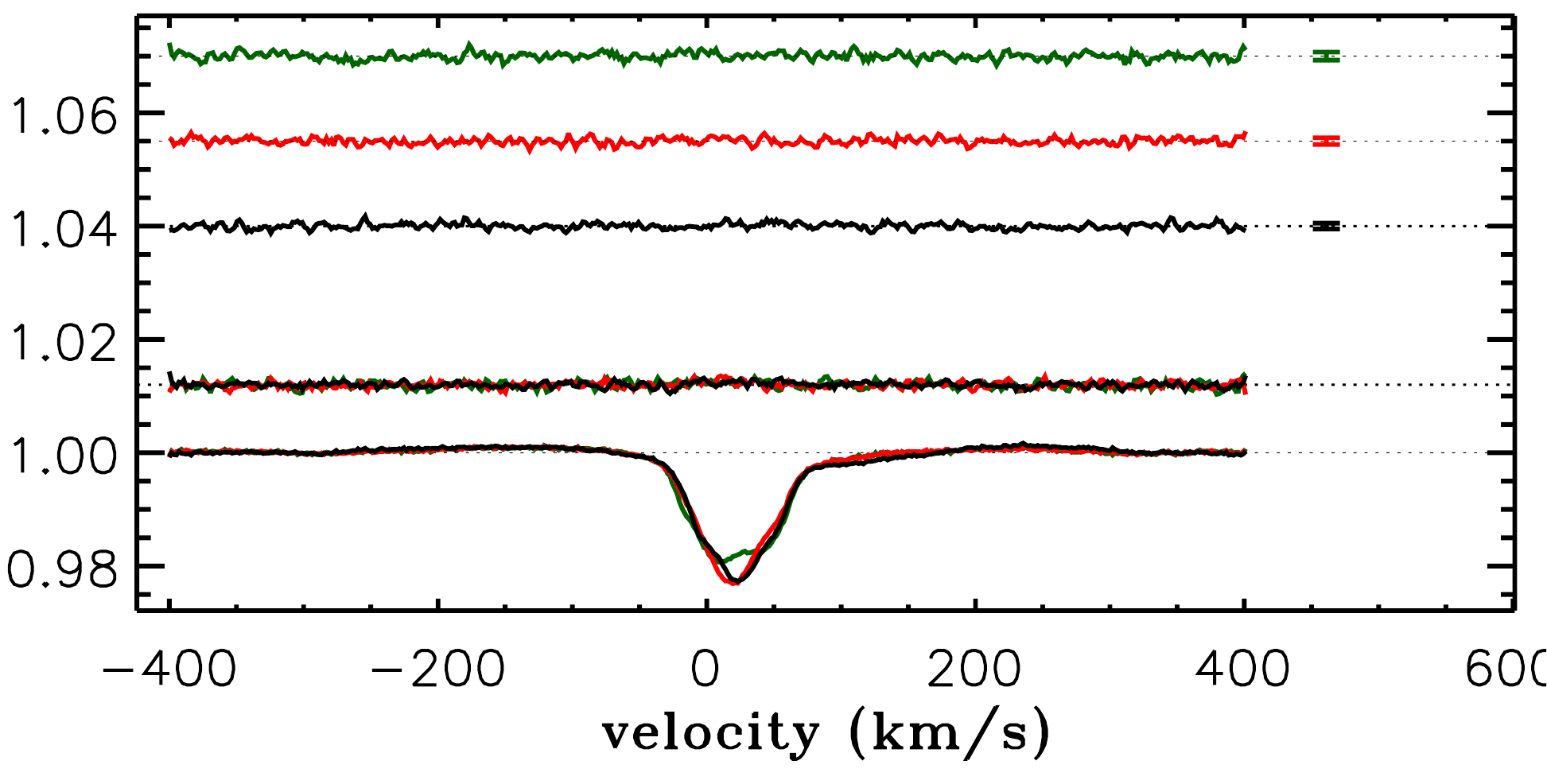}  
\caption{LSD profiles for RY Ori (no magnetic detection).}
\end{figure}

\begin{figure}[ht!]
\centering
\includegraphics[width=0.45\textwidth,clip]{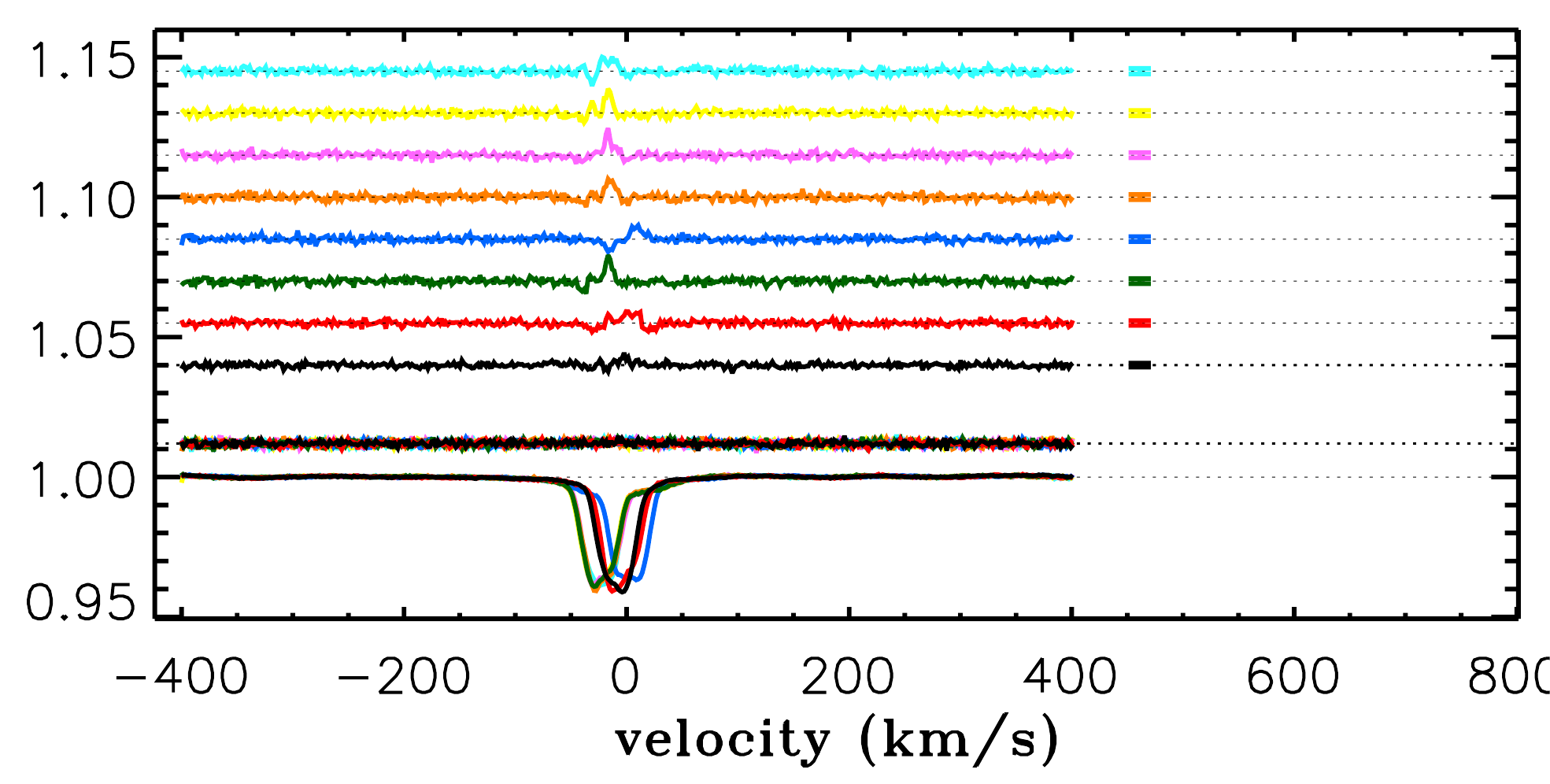}  
\caption{LSD profiles for V 1000 Sco (magnetic field detected, complex topology).}
\end{figure}

\begin{figure}[ht!]
\centering
\includegraphics[width=0.45\textwidth,clip]{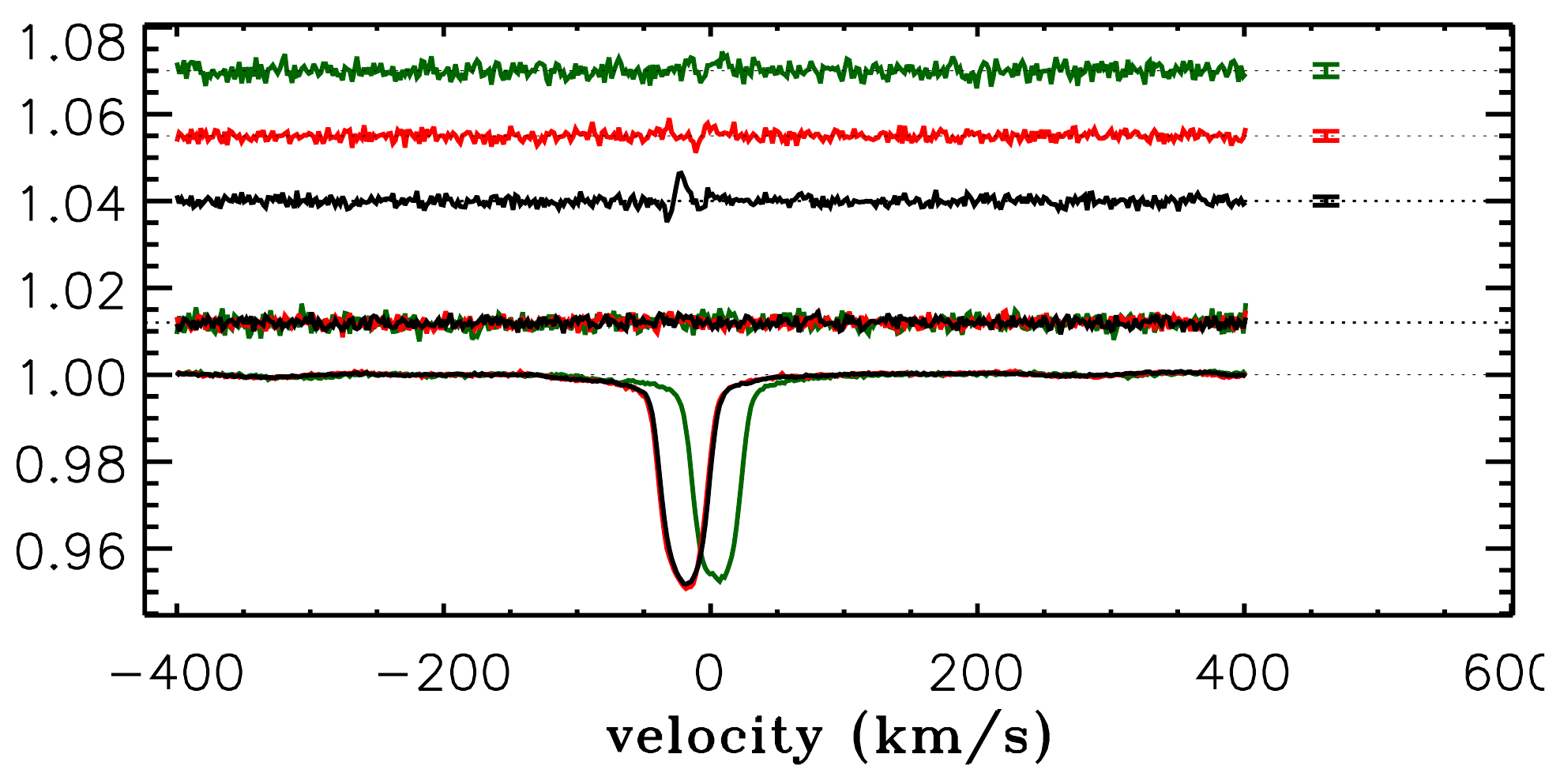}  
\caption{LSD profiles for V 1001 Sco (magnetic field detected, complex topology).}
\end{figure}
 
 
\begin{figure}[ht!]
\centering
\includegraphics[width=0.45\textwidth,clip]{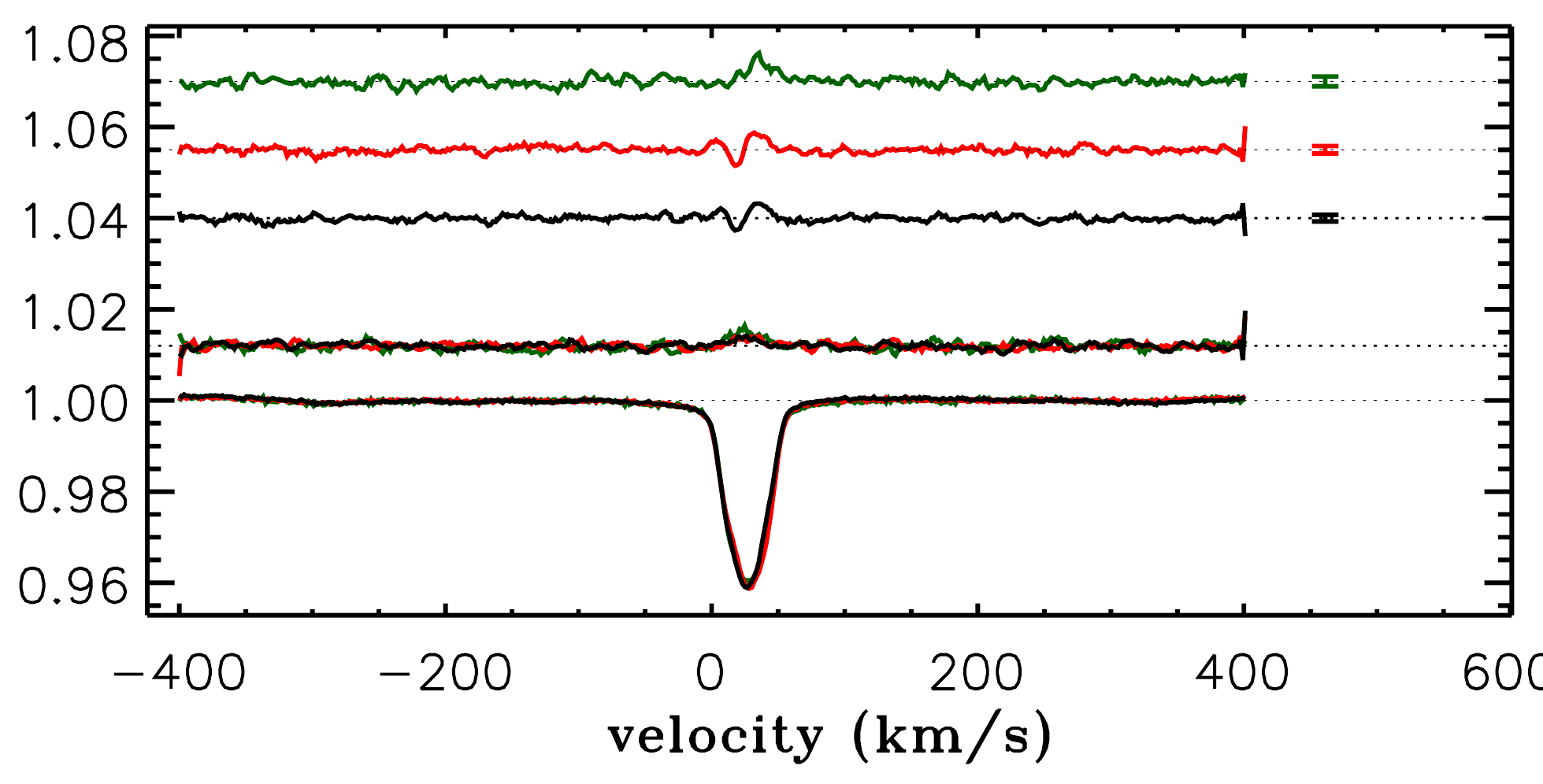}  
\caption{LSD profiles for V 1044 Ori (magnetic field detected, uncertain topology).}
\end{figure}

\begin{figure}[ht!]
\centering
\includegraphics[width=0.45\textwidth,clip]{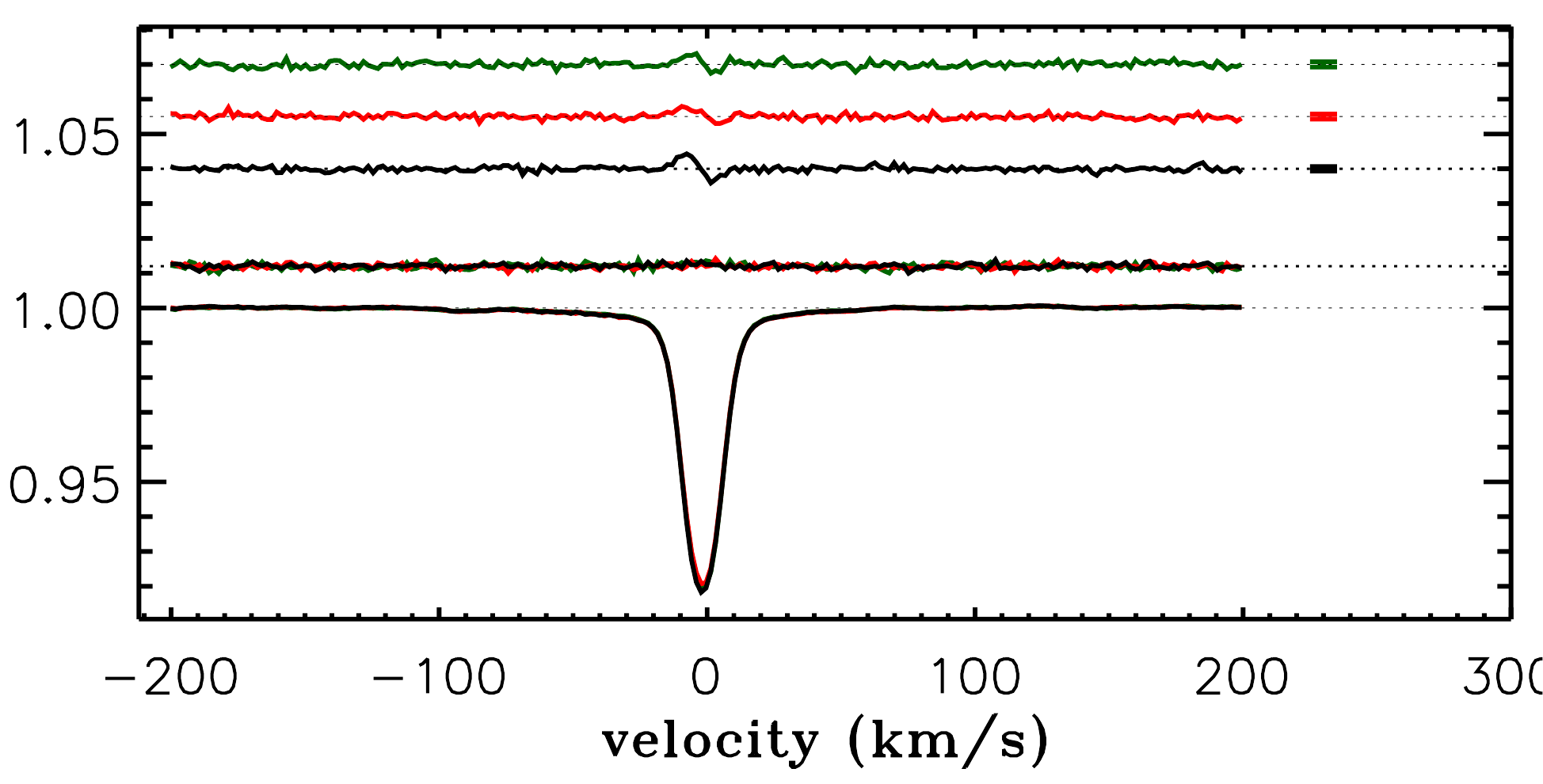}  
\caption{LSD profiles for V 1149 Sco (magnetic field detected, uncertain topology).}
\end{figure}

\begin{figure}[ht!]
\centering
\includegraphics[width=0.45\textwidth,clip]{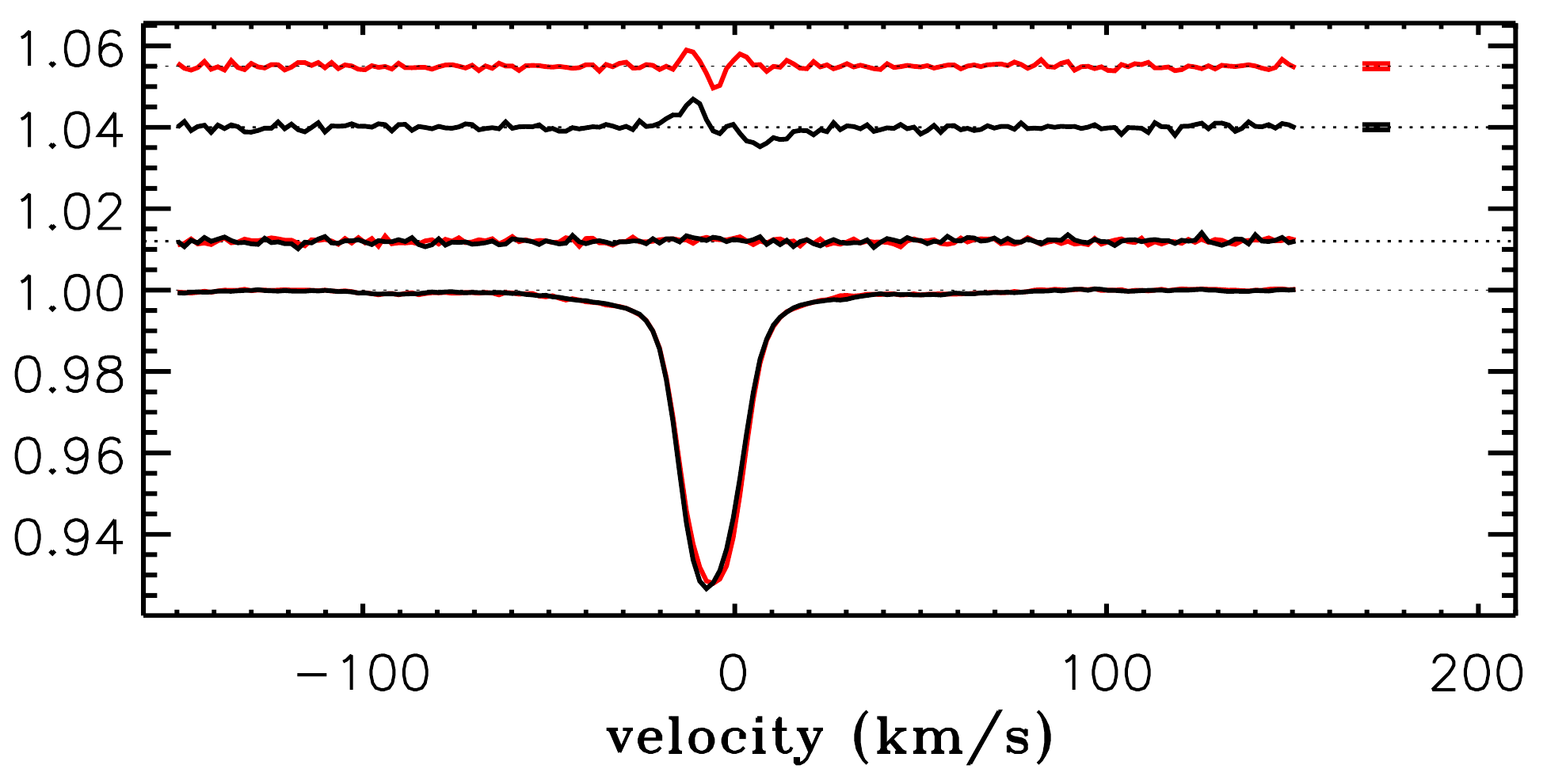}  
\caption{LSD profiles for V 1152 Sco (magnetic field detected, complex topology).}
\end{figure}

\begin{figure}[ht!]
\centering
\includegraphics[width=0.45\textwidth,clip]{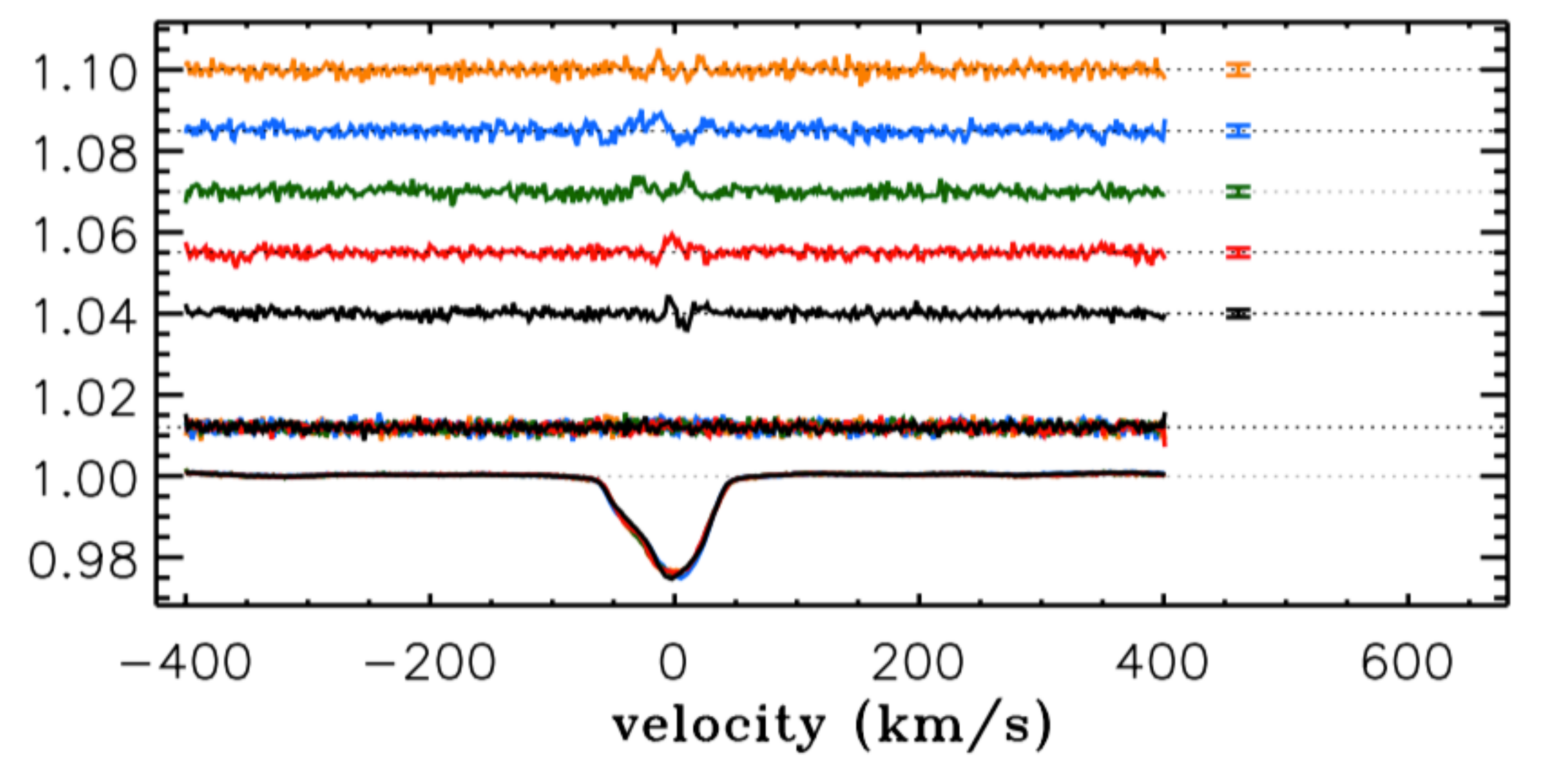}  
\caption{LSD profiles for V 1156 Sco (magnetic field detected, complex topology).}
\end{figure}

\begin{figure}[ht!]
\centering
\includegraphics[width=0.45\textwidth,clip]{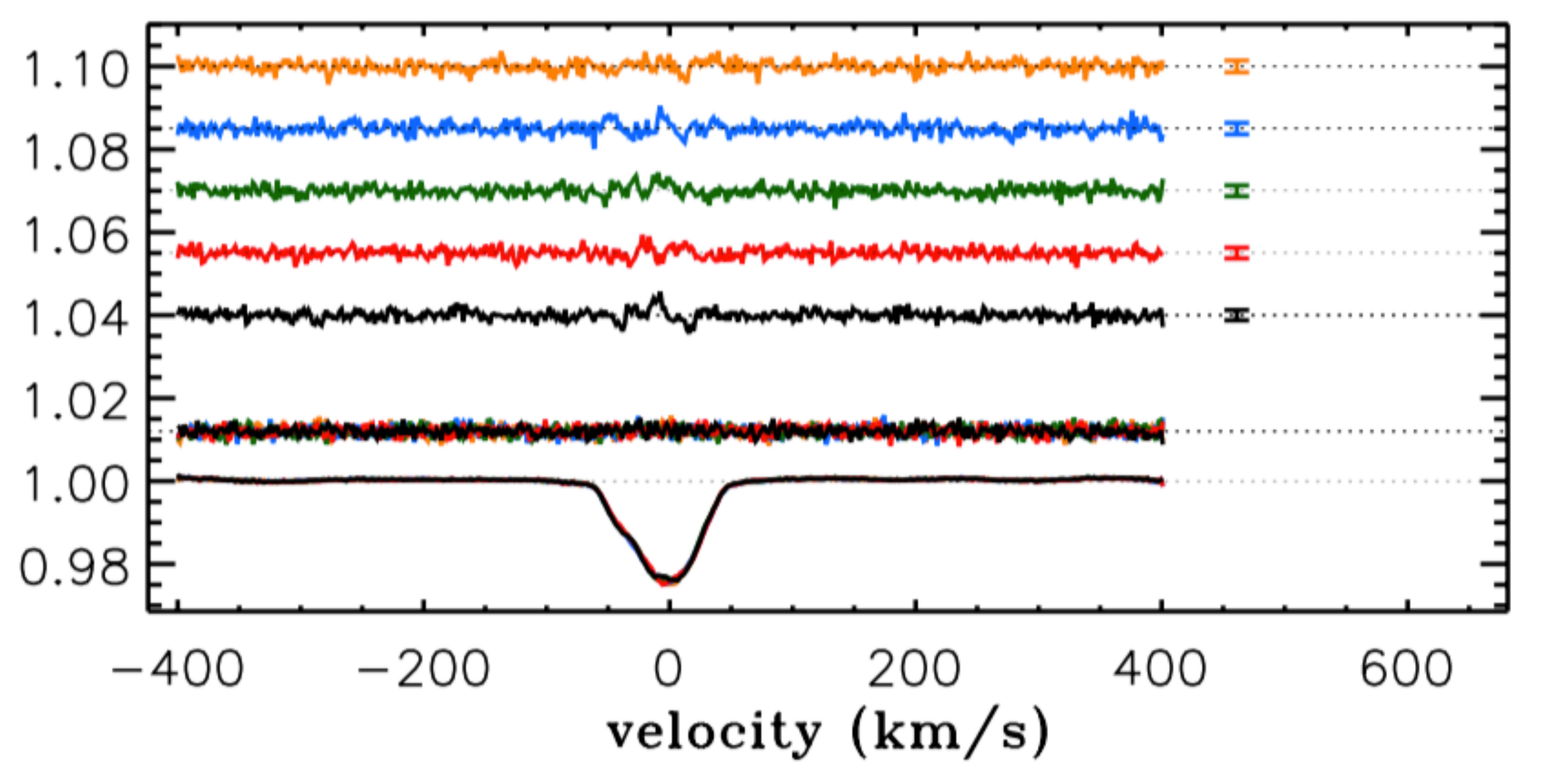}  
\caption{LSD profiles for V 1156 Sco (continued).}
\end{figure}

\begin{figure}[ht!]
\centering
\includegraphics[width=0.45\textwidth,clip]{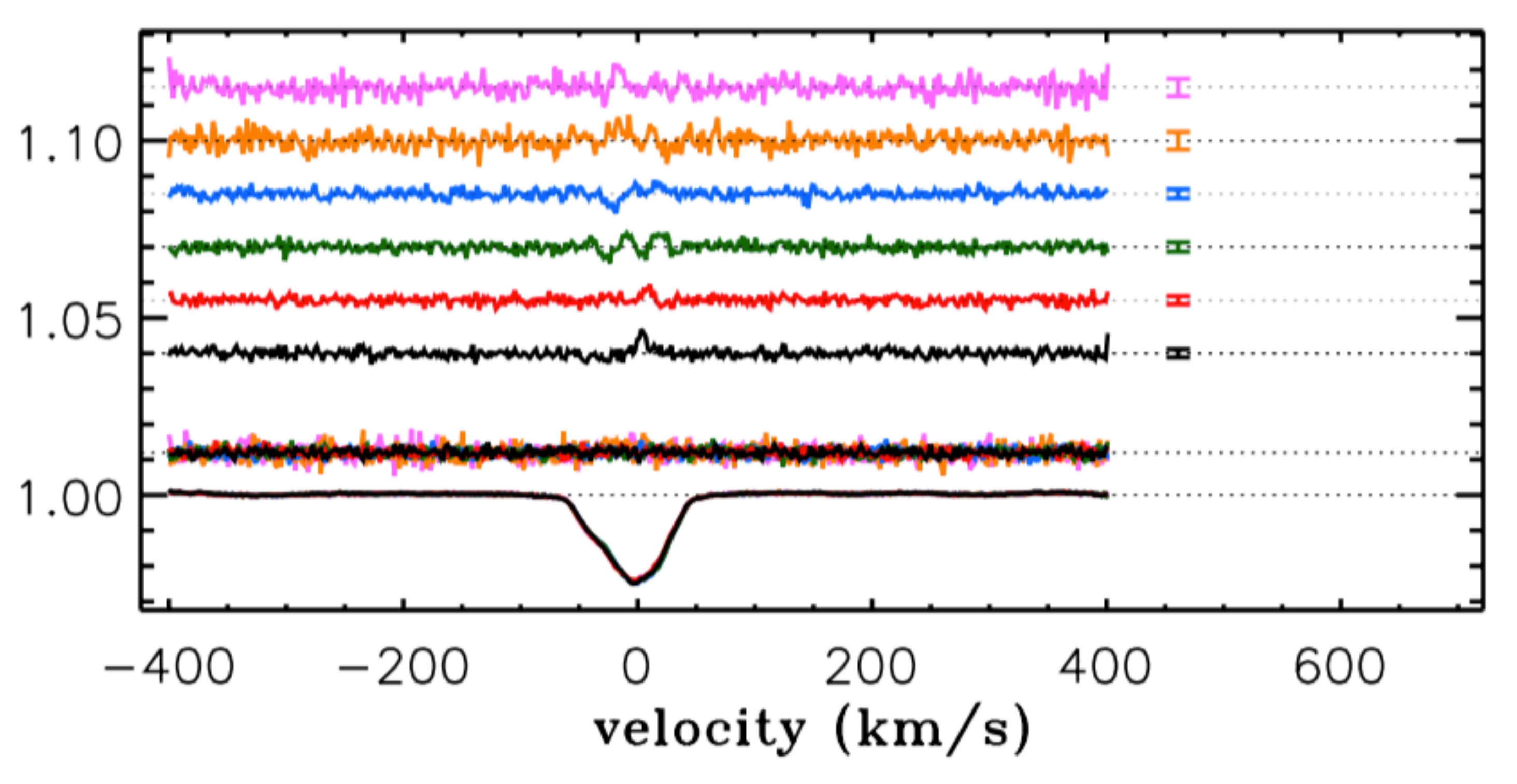}  
\caption{LSD profiles for V 1156 Sco (continued).}
\end{figure}

\end{document}